\documentclass[preprint2]{emulateapj}

\usepackage{xspace}
\usepackage{amsmath}
\usepackage{framed}
\usepackage{txfonts}
\usepackage{epstopdf}
\usepackage{xcolor}
\usepackage{rotating}
\usepackage{natbib}
\usepackage{ulem}
\usepackage{xspace}
\usepackage[colorlinks=true,urlcolor=blue,linkcolor=blue,citecolor=blue]{hyperref}

\newdimen\hssize
\newdimen\hdsize
\def\gcm3{\mathrm{g} / \mathrm{cm}^3}
\def\m200m{M_{\rm 200m}}

\def\gtsima{$\; \buildrel > \over \sim \;$}
\def\ltsima{$\; \buildrel < \over \sim \;$}
\def\prosima{$\; \buildrel \propto \over \sim \;$}
\def\gsim{\lower.7ex\hbox{\gtsima}}
\def\lsim{\lower.7ex\hbox{\ltsima}}
\def\simgt{\lower.7ex\hbox{\gtsima}}
\def\simlt{\lower.7ex\hbox{\ltsima}}
\def\simpr{\lower.7ex\hbox{\prosima}}

\def\bp{{\bf p}}

\usepackage{etoolbox}
\makeatletter
\patchcmd{\NAT@citex}
  {\@citea\NAT@hyper@{\NAT@nmfmt{\NAT@nm}\NAT@date}}
  {\@citea\NAT@nmfmt{\NAT@nm}\NAT@hyper@{\NAT@date}}
  {}
  {}
\patchcmd{\NAT@citex}
  {\@citea\NAT@hyper@{%
     \NAT@nmfmt{\NAT@nm}%
     \hyper@natlinkbreak{\NAT@aysep\NAT@spacechar}{\@citeb\@extra@b@citeb}%
     \NAT@date}}
  {\@citea\NAT@nmfmt{\NAT@nm}%
   \NAT@aysep\NAT@spacechar%
   \NAT@hyper@{\NAT@date}}
  {}
  {}
\patchcmd{\NAT@citex}
  {\@citea\NAT@hyper@{%
     \NAT@nmfmt{\NAT@nm}%
     \hyper@natlinkbreak{\NAT@spacechar\NAT@@open\if*#1*\else#1\NAT@spacechar\fi}%
       {\@citeb\@extra@b@citeb}%
     \NAT@date}}
  {\@citea\NAT@nmfmt{\NAT@nm}%
   \NAT@spacechar\NAT@@open\if*#1*\else#1\NAT@spacechar\fi%
   \NAT@hyper@{\NAT@date}}
  {}
  {}
\makeatother

\usepackage[figure,figure*]{hypcap}
\usepackage{ulem}
\usepackage{comment}

\def\avrg#1{\left\langle #1 \right\rangle}

\shorttitle{Dark Quest I.}
\shortauthors{Nishimichi et al.}

\begin{document}
\begin{flushright}
	\quad \\
	\quad \\
	\quad \\
	YITP-19-73\\
\end{flushright}

\title{
Dark Quest. I. Fast and Accurate Emulation of Halo Clustering Statistics and its Application to Galaxy Clustering
}

\author{Takahiro Nishimichi\altaffilmark{1,2}, Masahiro Takada\altaffilmark{1}, Ryuichi Takahashi\altaffilmark{3}, Ken Osato\altaffilmark{4}, Masato Shirasaki\altaffilmark{5}, Taira Oogi\altaffilmark{1}, Hironao Miyatake\altaffilmark{6,7,1}, Masamune Oguri\altaffilmark{4,8,1}, Ryoma Murata\altaffilmark{1,4}, Yosuke Kobayashi\altaffilmark{1,4}, and, Naoki Yoshida\altaffilmark{4,8,1}}

\altaffiltext{1}{Kavli Institute for the Physics and Mathematics of the Universe (WPI), The University of Tokyo Institutes for Advanced Study, The University of Tokyo, 5-1-5 Kashiwanoha, Kashiwa, Chiba 277-8583, Japan}
\altaffiltext{2}{Center for Gravitational Physics, Yukawa Institute for Theoretical Physics, Kyoto University,
Kyoto 606-8502, Japan}
\altaffiltext{3}{Faculty of Science and Technology, Hirosaki University, 3 Bunkyo-cho, Hirosaki, Aomori 036-8561, Japan}
\altaffiltext{4}{Department of Physics, The University of Tokyo, 7-3-1 Hongo, Bunkyo-ku, Tokyo 113-0033, Japan}
\altaffiltext{5}{Division of Theoretical Astronomy, National Astronomical Observatory of Japan, 2-21-1 Osawa, Mitaka, Tokyo 181-8588, Japan}
\altaffiltext{6}{Institute for Advanced Research, Nagoya University, Nagoya 464-8601, Japan}
\altaffiltext{7}{Division of Particle and Astrophysical Science, Graduate School of Science, Nagoya University, Nagoya 464-8602, Japan}
\altaffiltext{8}{Research Center for the Early Universe, The University of Tokyo, 7-3-1 Hongo, Bunkyo-ku, Tokyo 113-0033, Japan}

\begin{abstract}
We perform an ensemble of $N$-body simulations with $2048^3$ particles for 101 flat $w$CDM cosmological models sampled based on a maximin-distance Sliced Latin Hypercube Design. By using the halo catalogs extracted at multiple redshifts in the range of $z=[0,1.48]$, we develop \textsc{Dark Emulator}, which enables fast and accurate computations of the halo mass function, halo-matter cross-correlation, and halo auto-correlation as a function of halo masses, redshift, separations and cosmological models, based on the Principal Component Analysis and the Gaussian Process Regression for the large-dimensional input and output data vector. We assess the performance of the emulator using a validation set of $N$-body simulations that are not used in training the emulator. We show that, for typical halos hosting CMASS galaxies in the Sloan Digital Sky Survey, the emulator predicts the halo-matter cross correlation, relevant for galaxy-galaxy weak lensing, with an accuracy better than 2\% and the halo auto-correlation, relevant for galaxy clustering correlation, with an accuracy better than 4\%. We give several demonstrations of the emulator. It can be used to study properties of halo mass density profiles such as the mass-concentration relation and splashback radius for different cosmologies. The emulator outputs can be combined with an analytical prescription of halo-galaxy connection such as the halo occupation distribution at the equation level, instead of using the mock catalogs, to make accurate predictions of galaxy clustering statistics such as the galaxy-galaxy weak lensing and the projected correlation function for any model within the $w$CDM cosmologies, in a few CPU seconds.
\end{abstract}


\section{Introduction}

Cosmic large-scale structures
are promising avenues to fundamental questions in cosmology.
Various wide-area imaging or spectroscopic surveys of galaxies are ongoing and being planned, aimed at addressing
the nature of dark matter and dark energy in the universe.
These include the Subaru Hyper Suprime-Cam (HSC) Survey\footnote{\url{https://hsc.mtk.nao.ac.jp/ssp/}} \citep{2018PASJ...70S...4A}, the Dark Energy Survey\footnote{\url{https://www.darkenergysurvey.org}},
the Kilo-Degree Survey\footnote{\url{http://kids.strw.leidenuniv.nl}},
the Subaru Prime Focus Spectrograph (PFS) \citep{2014PASJ...66R...1T},
the Dark Energy Spectroscopic Instrument (DESI)\footnote{\url{https://www.desi.lbl.gov}},
the Large Synoptic Survey Telescope (LSST)\footnote{\url{https://www.lsst.org}},
the ESA satellite mission Euclid\footnote{\url{https://www.euclid-ec.org}},
and the NASA satellite mission WFIRST\footnote{\url{https://wfirst.gsfc.nasa.gov}}.
However, one of the most serious systematic effects in galaxy survey based cosmology lies in the galaxy bias that generally states an inevitable uncertainty in the relation between distributions of dark matter and large-scale structure tracers \citep[\citealt{kaiser84}, also see][for a recent review]{Desjacques18}.
Since physical processes involved in galaxy formation and evolution are
still impossible to solve from the first principles, it is of critical importance to explore a practical route to extracting cosmological information from observables of galaxy surveys, yet being least affected by the galaxy
bias uncertainty, in order to attain the full potential of ongoing and future galaxy surveys.

The growth of cosmic structures is driven mainly by the
spatial inhomogeneities of dark matter, which are easier to describe analytically on large scales
\citep{bernardeau02}
or via $N$-body numerical simulations down to small scales \citep{1975PASJ...27..333M,davis:1985fk}
than the variety of astrophysical processes where baryons play a major role in order to form galaxies \citep[e.g.,][]{2014Natur.509..177V}.
In practice, however, we can observe only the projected or three-dimensional distribution of galaxies from galaxy surveys from which we have to infer the dark matter distribution. This is not an easy task and a major challenge which all wide-area galaxy surveys must confront.
Nevertheless there is a theory-motivated working hypothesis that we can employ to make a connection between
galaxies and the dark matter distribution. Galaxies or galaxy clusters are believed to form inside dark matter halos,
which are self-gravitative systems and correspond to the peaks of the primordial mass density field \citep{kaiser84}. The distribution of halos with respect to the dark matter distribution, referred to as halo bias, and
its dependence on the halo mass and cosmological models can be
predicted in the cold dark matter dominated structure formation scenario using analytical models \citep{bardeen86,mo96,1999MNRAS.308..119S,2001MNRAS.323....1S} and/or using
$N$-body simulations \citep{Tinker10}. Here it is known that the large-scale bias of halos and therefore galaxies should approach to a constant value, known as ``linear bias'',
for the adiabatic initial Gaussian conditions of structure formation due
to the equivalence principle of gravity \citep[e.g.][]{Desjacques18} \citep[see][for a counter example such as the primordial non-Gaussian initial condition]{dalal08}. On small scales, the halo bias becomes scale-dependent and varies with cosmological models in a complex way due to nonlinearities of structure formation \citep{mcdonald06,2009JCAP...08..020M,taruya10,sato11,baldauf12,nishizawa13}. These distinct behaviors of halo bias over different scales have to be kept in mind in order not to have any bias in
cosmological parameter inference.

Observationally, there are promising
probes of galaxy surveys that help to link galaxies to the dark matter distribution
or halos, at least in a statistical manner.
Galaxy-galaxy or cluster-galaxy lensing, which can be measured by stacking shapes of background galaxies around the foreground tracers, allows us to probe the ``average'' projected matter (mostly dark matter) distribution around the tracers
\citep[e.g.][]{1996ApJ...466..623B,1996ApJ...473L..17D,2000AJ....120.1198F,2009ApJ...703.2217S}. The large-scale galaxy-galaxy lensing signal gives a direct estimate of the linear bias of the galaxies \citep[e.g.][]{2001ApJ...558L..11H,2004AJ....127.2544S}. However, the weak lensing signal is generally noisy. Although the small-scale lensing signal has a higher signal-to-noise ratio, it probes the dark matter distribution inside the same halo, which is generally difficult to predict accurately. Nevertheless, the integrated lensing signal within the projected
aperture of the virial radius
can be used to infer the average halo mass of galaxies in a sample \citep[e.g.][]{mandelbaum06},
 which can in turn be used to infer the linear bias at large scales with a help of theoretical model.
The auto-correlation function of galaxies' positions in the large-scale structure is another powerful probe of cosmology \citep[e.g.][]{peebles1980}. It can be measured from a wide-area spectroscopic sample, and is relatively easy to measure, i.e., with
high signal-to-noise ratios. If only the large-scale correlation signals are used and if the linear bias is {\it a priori} assumed, the cosmological information can be extracted from the shape information \citep[e.g.][]{2004PhRvD..69j3501T}. However, the small-scale correlations, which carry even higher signal-to-noise ratios, cannot be interpreted easily, and the correlations of galaxies in the same halo, the so-called one-halo term, add a significant contribution to the measured signal, which complicates the cosmological analysis.

Although each observable alone has its own pros and cons, combining different clustering observables enables us to perform a robust cosmological analysis, e.g., obtain tighter constraints on cosmological parameters, yet simultaneously calibrating
systematic errors such as the bias uncertainty that are otherwise difficult to calibrate with each observable alone \citep[e.g.][for similar discussion]{2011PhRvD..83b3008O,2012PhRvD..86h3504Y,2017PhRvD..95l3512S}. Implementations of joint-probes cosmology to actual data
can be found in various works \citep{2005PhRvD..71d3511S,2013MNRAS.432.1544M,hikage:2013kx,2014MNRAS.444..476R,2015ApJ...806....2M,2018PhRvD..98d3526A,2018MNRAS.474.4894J}.
Such analyses can be done by combining wide-area imaging and spectroscopic surveys over the same region of the sky;
for instance, this is the case for the Subaru HSC and PFS surveys.

Hence the purpose of this paper is to develop a software to make accurate model predictions for clustering observables
in preparation for high-precision cosmology achievable from ongoing and future wide-area galaxy surveys. Motivated by the fact that
dark matter halos are building blocks of the large-scale structure and the places hosting galaxies,
we build an ``emulator'', dubbed as \textsc{Dark Emulator}, that allows fast, accurate computations of ``halo'' clustering quantities; halo mass function, halo-matter cross-correlation function, and halo auto-correlation function as a function of halo mass, separation, redshift and cosmological models. To develop the emulator, we use a large number of $N$-body simulation realizations
and their halo catalogs at multiple output redshifts for different cosmological models that cover a sufficiently
broad range of
models within flat-geometry, time-varying dark energy and cold dark matter cosmologies (hereafter $w$CDM).
These halo clustering quantities include all the relevant physics such as the linear halo bias,
nonlinear bias and the halo exclusion effect.
Since we use a limited number of $N$-body simulation realizations for sparsely-sampled cosmological models in six-dimensional cosmological parameter space, we carefully propagate statistical
uncertainties in halo clustering quantities to the model predictions (emulator outputs) by using the Principal Component Analysis (PCA)
and the Gaussian Process regression (GPR) in a high dimensional space of input and output data vector.

The concept of our study is somewhat similar to emulators developed in previous studies, which interpolate various quantities measured from simulations over the cosmological parameter space \citep{Heitmann06,Habib07,Coyote1,Schneider08,Coyote3,MiraTitan1,MiraTitan2,2015PhRvD..91j3511P,2018arXiv180910747L,2012MNRAS.424.1409A,2014MNRAS.439.2102A,2018arXiv180405865D,2018arXiv180405866M,2018arXiv180405867Z,2018MNRAS.tmp.2206W,2018ApJS..236...43G,EuclidEmulator}. However, our study is
quite different from these works in the sense that we do {\it not} make a one-to-one mapping between the input cosmological parameters to the final statistical quantities with the emulation process.
We focus more on developing a machinery consisting of several building blocks, each of which works as a separate emulator, and combining them in an analytical manner to work together.
Specifically, we focus on halo clustering statistics and do not
employ any specific prescription to connect halos to galaxies such as the halo occupation distribution (HOD)
\citep{2005ApJ...633..791Z}. Hence, to obtain predictions of galaxy clustering quantities that can be compared with the measurements, a user needs to adopt a prescription
to model the galaxy-halo connection, especially the one-halo term contributions arising from galaxies in the same halo, and then combine the outputs of \textsc{Dark Emulator} to compute the desired statistical
quantities. As a working example, we show how to combine the outputs of \textsc{Dark Emulator} and the other small-scale physics prescriptions such as
the HOD model and the distribution of satellite galaxies inside a halo analytically at the equation level
(e.g., Fourier transform and numerical integration) to compute clustering quantities of galaxies such as galaxy-galaxy weak lensing and projected galaxy correlation
function for galaxies in a hypothetical sample. In this sense our approach might be regarded as a numerical-simulation version of the halo model approach \citep{seljak:2000uq,peacock:2000qy,ma:2000lr,scoccimarro:2001fj,Valageas11a}
\citep[also see][for a review]{Cooray02}.
Thus our emulator gives a flexibility that an user can decide how to use
the emulator outputs for his/her desired purpose. This study is the initial work of the \textsc{Dark Quest} campaign project, and the final goal is to use the \text{Dark Quest} products to achieve accurate and robust cosmological analysis with wide-area galaxy surveys. Therefore the requirements we impose for the \textsc{Dark Emulator} are
giving sufficiently accurate predictions for desired observables and
being sufficiently fast to allow
cosmological parameter inference such as a Markov-Carlo Monte Carlo analysis in a high-dimensional parameter space, e.g. 6-dimensional cosmological parameters plus various nuisance parameters including HOD parameters.
We demonstrate how well we achieve these requirements.

The structure of the paper is as follows. We start with a brief review of the halo approach to the galaxy clustering and the relevant observables in \S~\ref{sec:halo_cosmology}. In \S~\ref{sec:sims}, we summarize the details of the simulation setups including the parameter sampling scheme, initial conditions, time evolution and post processing. We then discuss the details of each module that constitutes our emulator in \S~\ref{sec:emu} including the cross-validation tests. We focus on typical halos which host CMASS galaxies observed by the Sloan Digital Sky Survey at $z \sim 0.5$ in this section. We demonstrate how these modules can be combined to make predictions of various halo and galaxy statistics in \S~\ref{sec:demo}. We summaries in \S~\ref{sec:summary} with comments on the actual situations where our codes can be applied. Convergence studies, our treatment on the massive neutrinos, the mass and redshift dependence of our modules and an example HOD prescription implemented in the current version of the emulator are shown in appendices. Readers who are interested only in the final accuracies of \textsc{Dark Emulator} may go directly to Appendix~\ref{sec:extra_dependence} for the results of our validation study.

\section{Halo cosmology}
\label{sec:halo_cosmology}

Before going to details of
our method to construct
\textsc{Dark Emulator}, we first describe the concept of our approach. In particular we describe why we focus on
statistical quantities of halos and how we can connect the halo statistics to observables
for galaxies and
galaxy clusters that can be used
to extract cosmological information.

\subsection{Galaxy observables}
\label{subsec:galaxy}

Our final goal is to make predictions for clustering observables
that are available from wide-area galaxy surveys. For example, the galaxy-galaxy weak lensing signal is measured
by cross-correlating the positions of foreground galaxies with the shapes of background galaxies and
probes the average excess mass density profile around the lensing
galaxies, $\Delta\Sigma_\mathrm{g}(R)$.
This signal reflects the three-dimensional galaxy-mass cross correlation function, $\xi_{\rm gm}(x)$, projected along the line-of-sight direction:
\begin{eqnarray}
&&\Delta\Sigma_\mathrm{g}(R)= \bar{\Sigma}_\mathrm{g}(<R) - {\Sigma}_\mathrm{g}(R),\label{eq:DeltaSigma}
\end{eqnarray}
where
\begin{eqnarray}
&&\Sigma_\mathrm{g}(R) = \bar{\rho}_\mathrm{m0} \int_{-\infty}^\infty
\xi_\mathrm{gm}\!\left(\sqrt{R^2+\pi^2}\right)\,\mathrm{d}\pi,\\
&&\bar{\Sigma}_\mathrm{g}(<R) = \frac{2}{R^2}\int_0^R\Sigma_\mathrm{g}(y)~ y \mathrm{d}y.
\end{eqnarray}
Here we denote by $\pi$ and $R$ separations
in the line-of-sight and transverse directions, respectively, and $\bar{\rho}_\mathrm{m0}$ is the
present-day mean matter density.
The use of $\bar{\rho}_{\rm m0}$ is due to the fact that we define the surface mass density in the comoving
coordinates rather than the physical coordinates.
Similarly, the projected galaxy auto correlation function is related to the three-dimensional
galaxy auto correlation function, $\xi_\mathrm{gg}(r)$, via
\begin{equation}
w_\mathrm{gg}(R) = 2\int_0^{\pi_\mathrm{max}} \xi_\mathrm{gg}\!\left(\sqrt{R^2+\pi^2}\right)\,\mathrm{d}\pi,\label{eq:w_gg}
\end{equation}
for the projection width $[-\pi_\mathrm{max},\pi_\mathrm{max}]$.

The simplest linear deterministic bias model,
which connects the matter density field $\delta_\mathrm{m}$ and the galaxy number density field $\delta_\mathrm{g}$ as
$\delta_\mathrm{g} = b_\mathrm{g} \delta_\mathrm{m}$, leads to
\begin{equation}
\xi_{\rm gm} = {b_\mathrm{g}} \,\xi_\mathrm{mm},\qquad \xi_{\rm gg} = b_\mathrm{g}^2 \,\xi_\mathrm{mm},
\end{equation}
with a free parameter $b_{\rm g}$, which is completely degenerate with the normalization of linear matter power spectrum,
$\sigma_8$. Having both the lensing and clustering signals, one can break this degeneracy and infer the
underlying matter correlation function $\xi_{\rm mm}(x)$ by combining the two correlation functions:
\begin{equation}
\xi_{\rm mm}(x) = \frac{\left[\xi_{\rm gm}(x)\right]^2}{\xi_{\rm gg}(x)}.\label{eq:linear}
\end{equation}
In reality, however, both nonlinear corrections and stochasticity can alter this relation.
To quantify this, we introduce the cross coefficient \citep{1998ApJ...500L..79T} defined by
\begin{equation}
r_{\rm gm}(x) \equiv \frac{\xi_{\rm gm}(x)}{\sqrt{\xi_{\rm gg}(x)\,\xi_{\rm mm}(x)}}.\label{eq:cross_correl_coeff}
\end{equation}
The departure of $r_\mathrm{gm}$ from unity characterizes the degree to which the linear deterministic relation is violated.

\subsection{Halo model approach to galaxy bias}
\label{subsec:halo_to_gal}

Dark matter halos are basic building blocks of large-scale structure and the sites harboring formation of galaxies and
galaxy clusters.
Since physical processes involved in galaxy formation are still difficult to resolve or simulate from the first principles,
dark matter halos could give us a practical route to connect theory and observations of galaxy surveys. Hence, in this paper,
we develop an emulator that primarily predicts statistical quantities of halos as a function of halo mass, redshift, clustering separation scale,
and cosmological parameters (model).
This halo model
allows us to compute galaxy clustering statistics, e.g., by using a halo occupation distribution (HOD) prescription.

\begin{figure*}[ht!]
\begin{center}
\includegraphics[height=5.1cm,angle=0]{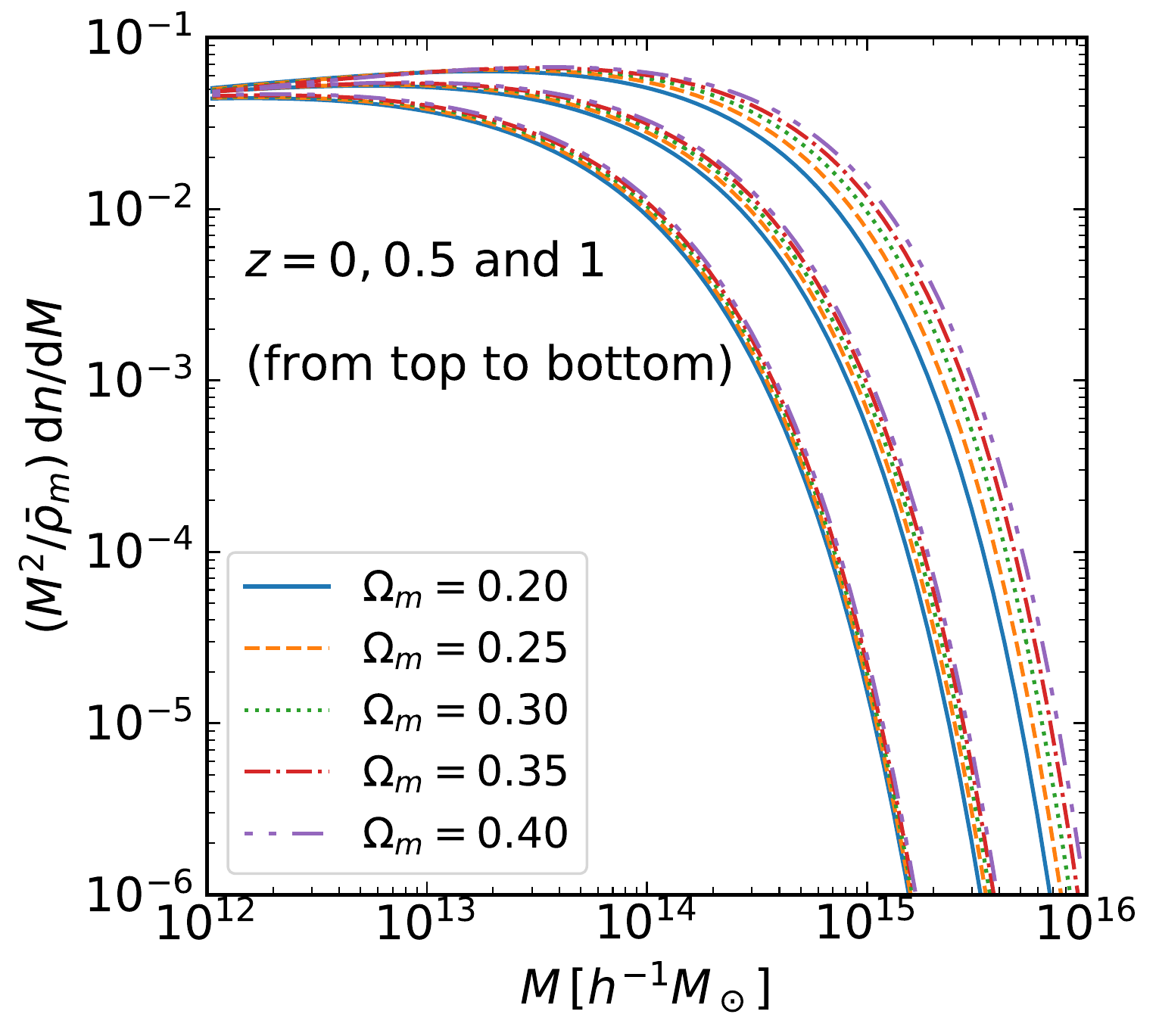}
\includegraphics[height=5cm,angle=0]{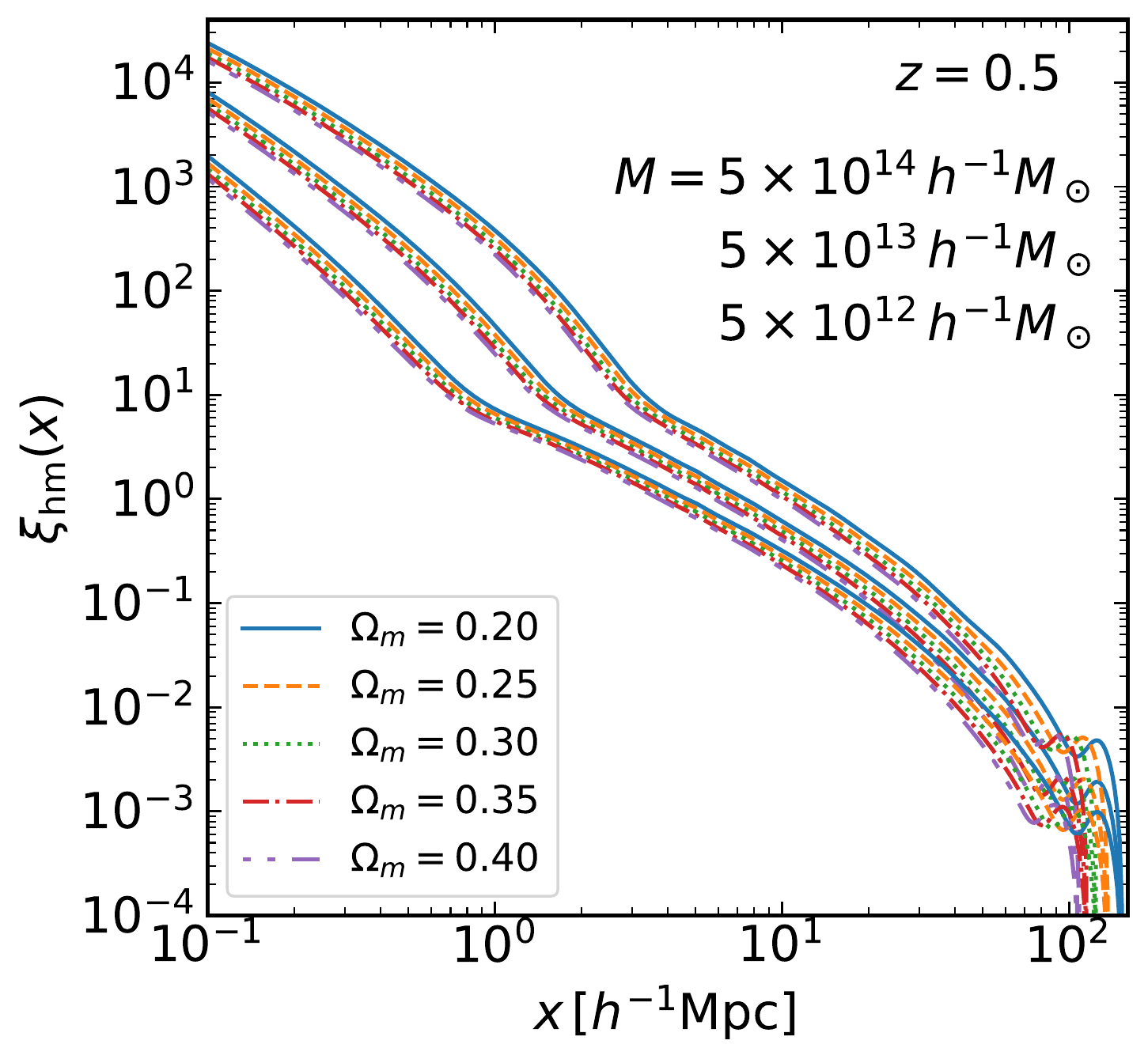}
\includegraphics[height=5cm,angle=0]{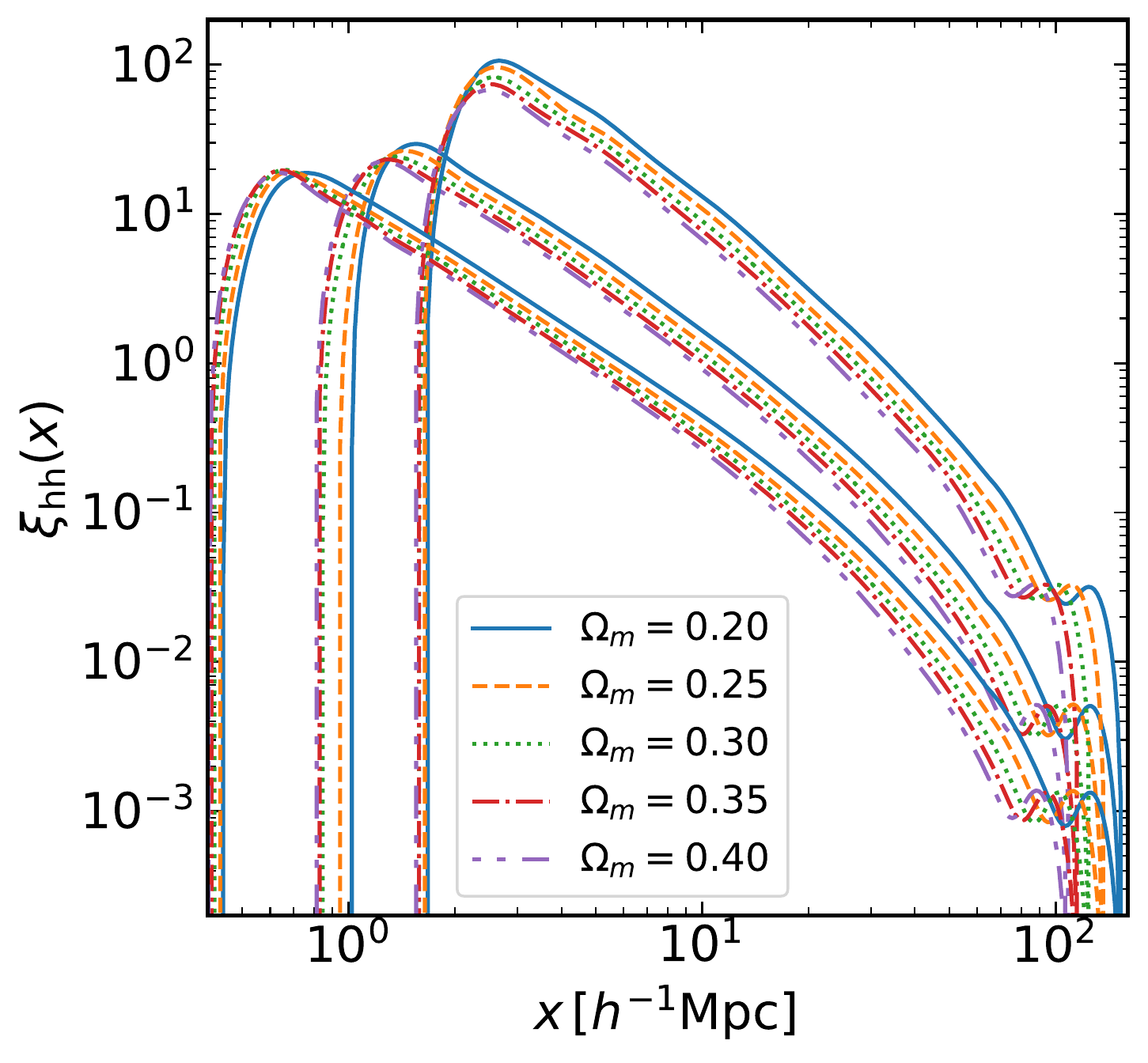}
\end{center}
\caption{A demonstration of our halo modules in \textsc{Dark Emulator} to predict halo-related statistical quantities as a function of cosmological parameters. We show how the halo mass function (left panel),
the halo-matter cross correlation function (middle) and the halo auto-correlation correlation function (right) vary with $\Omega_\mathrm{m}$ for a flat-geometry cosmology, but with $\sigma_8$ and other cosmological parameters being kept fixed to their fiducial values. Three sets of the lines in the left panel show the mass functions at redshifts
$z=0, 0.5$ and $1$, respectively, and the three sets in the middle and right panels show the correlation functions for halos at three masses as indicated in the figure legend, at redshift $z=0.5$.
Throughout this paper we use the spherical overdensity mass,
$M_{200}$, for the halo mass definition, where the mean mass overdensity within the halo boundary is 200 times $\bar{\rho}_{\rm m0}$.
\label{fig:demo1}
}
\end{figure*}
The halo mass function $\mathrm{d}n/\mathrm{d}M$
is defined as the comoving number density
of halos in the mass range $[M,M+\mathrm{d}M]$ and at redshift $z$,
for a given cosmological model denoted by its parameters $\bp$:
\begin{equation}
\frac{\mathrm{d}n}{\mathrm{d}M}(M,z,\bp) \,\mathrm{d}M.
\end{equation}
This is the first important quantity which we are going to calibrate with simulations.
We will build a module enabling the fast computation.

Two-point clustering properties of halos are characterized by the auto-correlation
functions of two halo samples and the cross-correlation functions of halos and matter,
which we denote as
\begin{equation}
\xi_{\rm hh}(x; M_1,M_2, z, \bp),
\end{equation}
and
\begin{equation}
\xi_{\rm hm}(x; M, z, \bp),
\end{equation}
respectively. Here we explicitly denote that the two halo samples in the halo-halo correlation function can have different masses, $M_1$ and $M_2$. We will omit the arguments $z$ and $\bp$ below for notational simplicity.

Fig.~\ref{fig:demo1} shows examples of quantities of our main interest, halo statistical quantities,
predicted by \textsc{Dark Emulator} that we develop in this paper.
The figure shows the halo mass functions, the halo-matter cross correlation functions, and the halo-halo auto correlation functions with varying density parameter $\Omega_{\rm m}$, but fixing other parameters to their fiducial values. In doing so, we keep the spatial flatness, the overall normalization $\sigma_8$ as well as the baryon fraction, and the change in $\Omega_{\rm m}$ is compensated by the density of dark energy $\Omega_\mathrm{de}$ as well as the Hubble parameter $H_0$.
We show in the left panel the prediction at three different redshifts, while in the middle and right panels, we consider three halo masses at $z=0.5$. Each of these quantities can be computed very quickly by the emulator, in
$\sim 100$~milliseconds
on a typical modern lap-top computer.

Once these statistical quantities of halos are given, we can compute galaxy observables, such as
those shown in
Eqs.~(\ref{eq:DeltaSigma}) and (\ref{eq:w_gg}), based on an empirical HOD model for the mean number of galaxies within a halo with mass $M$, $\avrg{N(M)}$. We here employ the functional form originally proposed by \citet{2005ApJ...633..791Z} and then slightly generalized by \citet{2015ApJ...806....2M}. The explicit formulae as well as the derivation of the resultant galaxy statistics can be found in Appendix~\ref{sec:HOD}.

On large scales, the galaxy statistics are computed as the weighted average of the corresponding halo statistics.
Specifically, using the halo occupation distribution that gives the average number of galaxies in halos of mass $M$,
$\avrg{N(M)}$, we can compute
\begin{equation}
\xi_{\rm gm}(x) = \frac{1}{\bar{n}_{\rm g}}\int\!\mathrm{d}M\frac{\mathrm{d}n}{\mathrm{d}M}(M)
\avrg{N\!(M)}\,\xi_{\rm hm}(x; M),\label{eq:xi_gm}
\end{equation}
for the cross and
\begin{eqnarray}
&&\xi_{\rm gg}(x) = \frac{1}{\bar{n}_{\rm g}^2}\int\!\mathrm{d}M_1\frac{\mathrm{d}n}{\mathrm{d}M}(M_1)\avrg{N\!(M_1)}\nonumber\\
&&\hspace{1cm}\times \int\!\mathrm{d}M_2\frac{\mathrm{d}n}{\mathrm{d}M}(M_2)\avrg{N\!(M_2)}
\xi_{\rm hh}(x; M_1, M_2),\label{eq:xi_gg}
\end{eqnarray}
for the auto correlation functions, where
the mean galaxy number density, $\bar{n}_\mathrm{g}$, is given by
\begin{equation}
\bar{n}_{\rm g} = \int\!\mathrm{d}M\frac{\mathrm{d}n}{\mathrm{d}M}
\avrg{N\!(M)}.\label{eq:ng}
\end{equation}
We now introduce mass-dependent halo bias functions
\begin{eqnarray}
&&\xi_\mathrm{hm}(x;M) = b_\mathrm{hm}(x; M)\,\xi_\mathrm{mm}(x),\\ &&\xi_\mathrm{hh}(x;M_1,M_2) = b^{(2)}_\mathrm{hh}(x; M_1,M_2)\,\xi_\mathrm{mm}(x),
\end{eqnarray}
for the cross and the auto correlation functions, respectively.
The galaxy correlation functions, Eqs.~(\ref{eq:xi_gm}) and (\ref{eq:xi_gg}), can be rewritten as
\begin{eqnarray}
\xi_{\rm gm}(x) &=& b_\mathrm{g}^\mathrm{cross}(x)\,\xi_{\rm mm}(x),\\
\xi_{\rm gg}(x) &=& \left[b_\mathrm{g}^\mathrm{auto}(x)\right]^2\,\xi_{\rm mm}(x),
\end{eqnarray}
where the corresponding galaxy bias functions are computed as
\begin{eqnarray}
b_\mathrm{g}^\mathrm{cross}(x) &=& \frac{1}{\bar{n}_{\rm g}^2}\int\!\mathrm{d}M\frac{\mathrm{d}n}{\mathrm{d}M}(M)\avrg{N\!(M)}b_\mathrm{hm}(x; M),\\
\left[b_{\rm g}^{\rm auto}(x)\right]^2 &=& \frac{1}{\bar{n}_{\rm g}^2}\int\!\mathrm{d}M_1\frac{\mathrm{d}n}{\mathrm{d}M}(M_1)\avrg{N\!(M_1)}\nonumber\\
&& \times \int\!\mathrm{d}M_2\frac{\mathrm{d}n}{\mathrm{d}M}(M_2)\avrg{N\!(M_2)}
b_{\rm hh}^{(2)}(x; M_1, M_2). \nonumber\\
\end{eqnarray}
Now the cross correlation coefficient for galaxy and matter fields, Eq.~(\ref{eq:cross_correl_coeff}), reads
\begin{equation}
r_{\rm gm}(x) = \frac{b_\mathrm{g}^\mathrm{cross}(x)}{b_\mathrm{g}^\mathrm{auto}(x)}.
\end{equation}
The condition, $r_\mathrm{gm} = 1$, is trivially satisfied when the halo auto bias function is written as a product of the two cross bias functions:
\begin{equation}
b_\mathrm{hh}^{(2)}(r; M_1, M_2) = b_\mathrm{hm}(r; M_1)\,b_\mathrm{hm}(r; M_2),\label{eq:bias_separable}
\end{equation}
for any halo mass $M_1$ and $M_2$. This relation would hold at sufficiently large separations\footnote{In Fourier space, however, a residual contribution is known to persist even on the large scale limit, and this behaves as a non-Poissonian shot noise term, $k^0$
\citep{seljak09,hamaus10,2013PhRvD..88h3507B}. Hence, it would not contribute significantly in configuration space.}.

On mildly nonlinear scales, one should take into account a violation of the relation~(\ref{eq:bias_separable}).
In particular, understanding the scale dependent function $r_\mathrm{gm}(x)$, and thus $b_\mathrm{hm}(x; M)$ and
$b_\mathrm{hh}^{(2)}(x; M_1, M_2)$, is crucial to recover the underlying matter correlation function
out of galaxy observables.
On strongly nonlinear scales, the one-halo term, namely clustering contribution due to pairs of
galaxy-galaxy or galaxy-matter within the same halo, dominates the signal over the two-halo term discussed so far.
Especially, the galaxy-galaxy lensing signal defined in Eq.~(\ref{eq:DeltaSigma}) in this regime
gives an information on the average mass of halos in which lensing galaxies reside. For
a simplistic scenario where HOD is a delta function at mass $M$, this halo mass information
tells us how the bias function should behave on large scales, breaking the degeneracy between bias and the underlying clustering signal. In more realistic settings, the small-scale information inferred from the galaxy-galaxy lensing helps to determine the HOD parameters and we can perform a quasi bias-free analysis
using the large-scale clustering signal.

All these analyses cannot be realized unless we have a good control of the model predictions for the
halo clustering signals including their dependence on mass as well as on cosmological parameters.
Therefore, the three quantities, $\mathrm{d}n/\mathrm{d}M$, $\xi_\mathrm{hh}$, and $\xi_\mathrm{hm}$,
are of our central interest in this paper (see Fig.~\ref{fig:demo1} for example plots of these quantities
varying $\Omega_\mathrm{m}$). Our emulator models these quantities at the core, and
predicts the galaxy statistics by combining them with a HOD prescription in an analytical manner.
In doing so, we pay attention to the evaluation speed of the statistics such that it is feasible to perform a
Markov-Chain Monte Carlo analysis of parameter inference in a high-dimensional parameter space,
e.g., a space including cosmological parameters as well as HOD parameters.

Two-dimensional projected clustering statistics can also be computed analytically, based on the three-dimensional
clustering signals predicted by our emulator. Alternatively, one might be tempted to project the matter particles,
halos or mock galaxies in a simulation box along an chosen one axis or direction, and then measure
the correlation signals in two dimensions to model the projected signals directly.
One could further increase statistics by combining the results from multiple projection directions.
In contrast to this conventional approach,
we would like to emphasize that our procedure, which first measures the correlation functions in three dimensions and then perform projection by the numerical integration along the line-of-sight,
is more advantageous in the sense that we automatically access the information in all the possible two dimensional
maps obtained by projection along all the possible different directions.
We also note that in our approach the projection width can be chosen as desired once the full three-dimensional
information is available.
Since redshift-space distortion can impact the projected statistics when the width is small, we implement a simple model to account this effect in the module that computes $w_\mathrm{gg}$.

\section{Simulation ensemble}
\label{sec:sims}

We summarize here basic features of the \textsc{Dark Quest} simulation suite.
All the simulations presented in this paper are listed in Table~\ref{tab:simulation_list}. More detailed explanations on each of the simulation suite will be given in the subsequent subsections.
We also describe details of postprocessing analyses.
\begin{deluxetable*}{l|llllll}
\tablecolumns{7}
\tablewidth{0pt}
\tabletypesize{\footnotesize}
\tablecaption{%
Summary of our simulation suites. We show the number of particles ($N_\mathrm{part}$), comoving box size ($L_\mathrm{box}$ in $h^{-1}$Mpc), cosmological model, random number seeds used in initial conditions (IC), the number of realizations per model or parameter set ($N_\mathrm{real}$) and the purpose of the simulations; calibration of either the halo mass function (HMF), halo-matter cross correlation function (HMCCF), halo auto correlation function (HACF) or the halo propagator (PROP), or other testing purposes such as the initial condition of simulations (IC).
}
\tablehead{\colhead{Class} &
\colhead{$N_\mathrm{part}$}
&\colhead{$L_\mathrm{box}$}
&\colhead{cosmology}
&\colhead{IC}
&\colhead{$N_\mathrm{real}$}
&\colhead{purpose}
}
\startdata
HR & $2048^3$ & $1,000$ & fiducial & random & $28$ & assessment of variance (HMF, HMCCF)\\
 & & & 20 models in Slice~1 & fixed$^{a}$ & $1$ & test of ICs\\
 & & & 100 models in Slice~1--5 & random & $1$ & Emulator (HMF, HMCCF; Slice~1--4 for training, Slice~5 for validation)\\
\hline
LR & $2048^3$ & $2,000$ & fiducial & random & $14$ & assessment of variance (HACF, PROP)\\
 & & & 100 models in Slice~1--5 & random & $1$ & Emulator (HACF, PROP; Slice~1--4 for training, Slice~5 for validation)\\
 & & & & & & (Slice~5 also for validation of HMF)\\
\hline
test & $256^3$ & $250$ & fiducial & fixed$^{b,c}$ & $1$ & convergence study (same resolution as LR)\\
& $512^3$  & & fiducial & fixed$^{b,c}$ & $1$ & convergence study (same resolution as HR)\\
& $1024^3$  & & fiducial & fixed$^{b}$ & $1$ & convergence study \\
& $2048^3$  & & fiducial & fixed$^{b}$ & $1$ & convergence study
\enddata
\tablenotetext{a}{Initial phases taken to be the same as one of the fiducial HR realization.}
\tablenotetext{b}{Exactly the same initial phases are employed for the six test simulations.}
\tablenotetext{c}{Initial conditions based on the Zel'dovich approximation are generated in addition to 2LPT. Also five initial redshifts $1+z_\mathrm{in} = 15, 30, 60, 120$ and $240$ are tested.}
\label{tab:simulation_list}
\end{deluxetable*}

\subsection{Simulation design}
\label{subsec:LHC}

One of the key elements for an efficient emulator is the sampling scheme of the models in a high-dimensional
input parameter space. It should be designed such that the hypervolume of interest is sampled as homogeneously
as possible. Indeed, Latin Hypercube Designs (LHDs) have been employed in previous studies to show a good
performance to construct the training data for emulators \cite[e.g.,][]{Coyote2}.
An LHD is a design achieved by first selecting a hyperrectangle, then dividing it into a regular lattice and
selecting only one sample in every lattice interval when projected into any one dimension.

LHDs are one of useful techniques employed in the literature of experimental design (see \citealt{DesignReview} for a recent review).
Imposing certain conditions, an LHD can have desirable space-filling and projection properties. Because of
these, they are often employed in black-box experiments, where the dependence of the outcome on input variables are completely unknown. While our situation is slightly different (i.e., the relation between inputs and outputs can be approximately modeled using fitting formulae in cosmology), LHDs have been a standard tool for the development of emulators in cosmological settings. In many cases of cosmology, one wishes to emulate a considerably large number of outputs. An experimental design highly optimized to one output can sometimes give a significantly inferior performance on other outputs. An LHD is expected to give, albeit non-optimal, a reasonable set of samples for all the outputs similarly to black-box experiments.

We here employ a variant of LHD, called
maximin-distance ``sliced'' LHD (\textsc{SLHD}) developed in \citet{SLHD}.
This is a technique to realize a hierarchy of maximin distance (i.e., the minimum distance between different
sampling points is maximized) LHDs: the whole samples are located to
construct an LHD, and they are classified into subgroups called ``slices'' with the same number of samples,
each of which independently satisfies the conditions for an LHD. In practice, a good space filling property
(i.e., a near maximin design) is ensured by minimizing the following quantity:
\begin{eqnarray}
\Phi(\mathbf{X}_N) = \frac{1}{2}\left(\phi_{\mathrm{all}}+\frac{1}{m}\sum_{t=1}^m \phi_t\right),\label{eq:SLHD}
\end{eqnarray}
where we denote by $\mathbf{X}_N = \{\mathbf{x}^{(i)} \,|\, i = 1,\dots, N, \,\mathbf{x}^{(i)}\in \mathbb{R}^n\}$
the locations of the whole $N$ samples in the $n$-dimensional input parameter space, and $\phi_\mathrm{all}$ and
$\phi_t$, respectively, stand for the cost function for the total and the $t$-th slice:
\begin{eqnarray}
&&\phi_\mathrm{all}(\mathbf{X}_N) = \left(\frac{2}{N(N-1)}\sum_{i, j\in\mathbf{X}_N}\frac{1}{d^r(\mathbf{x}^{(i)},\mathbf{x}^{(j)})}\right)^{1/r},\label{eq:SLHD1}\\
&&\phi_t(\mathbf{X}_{t}) = \left(\frac{2}{M(M-1)}\sum_{i, j\in\mathbf{X}_t}\frac{1}{d^r(\mathbf{x}^{(i)},\mathbf{x}^{(j)})}\right)^{1/r},\label{eq:SLHD2}
\end{eqnarray}
where $\mathbf{X}_t = \{\mathbf{x}^{(i)} \,|\, i = (t-1)M+1,\dots, tM, \,\mathbf{x}^{(i)}\in \mathbb{R}^n\}$ is
the samples in the $t$-th slice with $M=N/m$ members, the quantity $d(\mathbf{x}^{(i)},\mathbf{x}^{(j)})$ is
the distance between two samples, $\mathbf{x}^{(i)}$ and $\mathbf{x}^{(j)}$, and we use the standard Euclidean
distance, $\sqrt{|\mathbf{x}^{(i)}-\mathbf{x}^{(j)}|^2}$, for simplicity. This is equivalent to putting a uniform prior when the input parameter space is sampled. Here, the minimization of $\phi_\mathrm{all}$ or $\phi_t$ at the limit of $r\to \infty$
is equivalent to the maximization of the minimum distance among the design points, as the name ``maximin''
suggests.
An optimal \textsc{SLHD} is achieved by minimizing the mixture of $\phi_\mathrm{tot}$ and $\phi_t$ with the former
upweights
according to the ratio of the number of sampling points in the whole and sub samples
(i.e., Eq.~\ref{eq:SLHD}).
We use the parameter, $r=15$, which is the default value in the \textsc{SLHD} code.

\begin{figure*}[ht!]
\begin{center}
\includegraphics[width=16cm,angle=0]{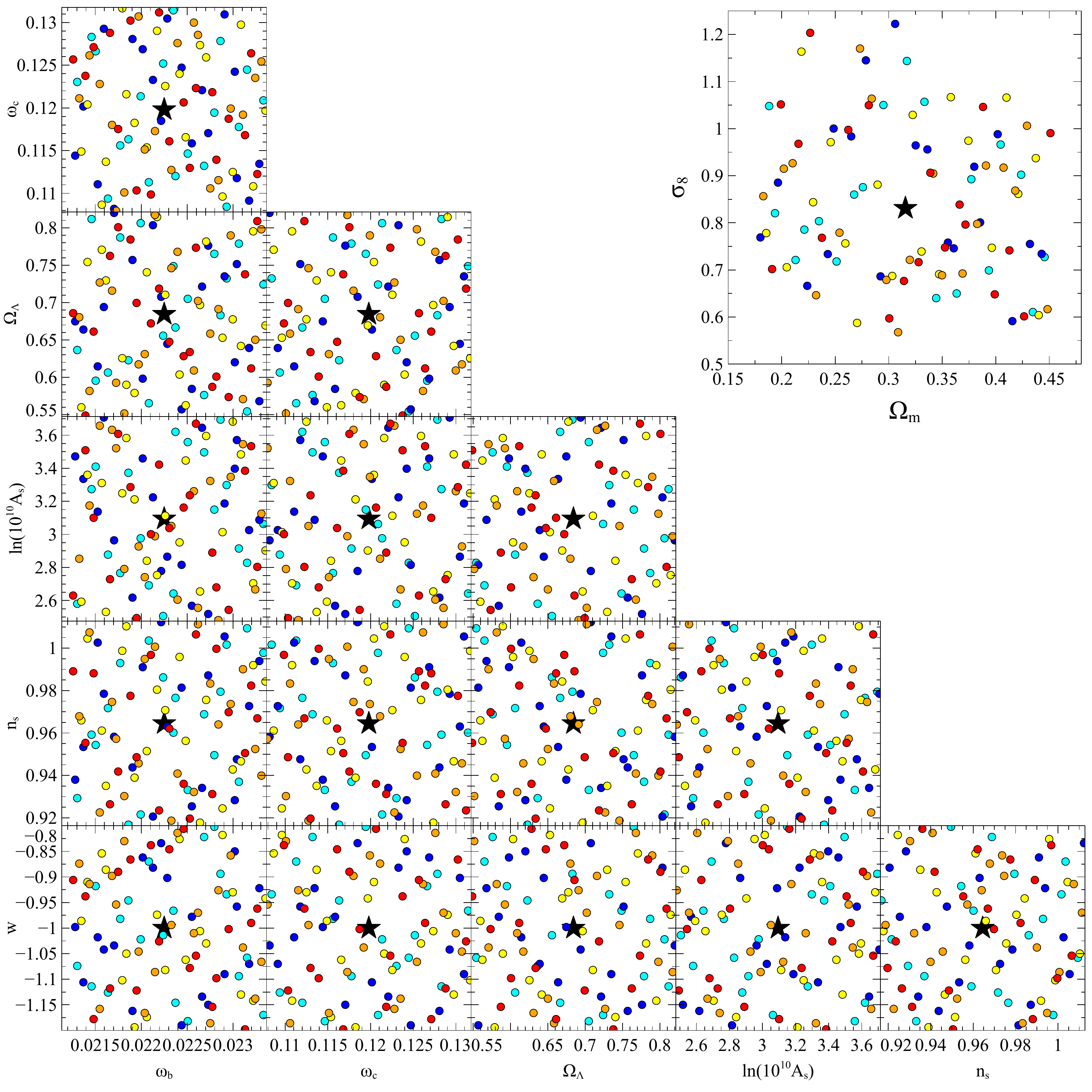}
\end{center}
\caption{\textsc{SLHD} sampling scheme with $5$ slices in the $6$ dimensional
cosmological
parameter space within the flat $w$CDM framework around the fiducial {\it Planck}
cosmology (star symbols).
The samples from the same slice are shown by the circles in the same color. In addition to the $6$ varied parameters in the $w$CDM model, we show the projection of samples
to the two-dimensional planes of derived parameters $\Omega_\mathrm{m}$ and $\sigma_8$ in the top-right panel.
\label{fig:SLHD}
}
\end{figure*}

This method allows a rather flexible design of samples unlike standard single slice LHDs, for which splitting the samples into a training and a validation set can ruin the desirable space-filling or projection properties before splitting.
For instance, training set and validation set are chosen from different slices in our case, both covering
the parameter space homogeneously, and the sample points in the two sets are guaranteed to be reasonably far
(i.e., no sample in the training set is very close from any sample in the validation set).
This is crucial for a stringent validation test because the accuracy of emulation
can be more objectively tested by such validation samples.
We sample $N=100$ cosmological models in total with $m=5$ slices
each of which is composed of $M=20$ samples in a $n=6$ dimensional parameter space. We consider the $w$CDM cosmology in the parameter range of
\begin{eqnarray}
&&0.0211375 < \omega_\mathrm{b} < 0.0233625,\nonumber\\
&&0.10782 < \omega_\mathrm{c} < 0.13178,\nonumber\\
&&0.54752 < \Omega_\mathrm{de} < 0.82128,\nonumber\\
&&2.4752 < \ln(10^{10}A_\mathrm{s}) < 3.7128,\nonumber\\
&&0.916275 < n_\mathrm{s} < 1.012725,\nonumber\\
&&-1.2 < w < -0.8,
\label{eq:params}
\end{eqnarray}
where $\omega_\mathrm{b} \equiv \Omega_\mathrm{b}h^2$ and $\omega_\mathrm{c} \equiv \Omega_\mathrm{c}h^2$ are
the physical density parameters of baryon and cold dark matter (CDM) with
$h=H_0/(100 \, \mathrm{km} \, \mathrm{s}^{-1} \, \mathrm{Mpc}^{-1})$ being the Hubble parameter,
$\Omega_\mathrm{de} \equiv 1 - (\omega_\mathrm{b} + \omega_\mathrm{c} + \omega_\mathrm{\nu})/h^2$
is the dark energy density parameter assuming a flat geometry of the universe,
$A_\mathrm{s}$ and $n_\mathrm{s}$ are the amplitude and tilt of the primordial curvature power spectrum
normalized at $0.05\,\mathrm{Mpc}^{-1}$ and $w$ is the equation of state parameter of dark energy.
As for the neutrino density $\omega_\nu \equiv \Omega_\nu h^2$, we fix to $0.00064$, corresponding to
$0.06\,\mathrm{eV}$ for the total mass of the three mass eigenstates (see Appendix~\ref{sec:neutrino}
for our approximate treatment of massive neutrinos). When computing the distance in Eqs.~(\ref{eq:SLHD1}--\ref{eq:SLHD2}), we linearly rescale the range of the six cosmological parameters in Eq.~(\ref{eq:params}) to $[0,1)$.
Our $100$ samples are shown in Fig.~\ref{fig:SLHD}, where the samples from the same slice are depicted
by the same color.

The parameter range above is centered at the best-fit cosmological parameters to the {\it Planck} CMB data
\citep{planck-collaboration:2015fj}:
$(\omega_{\rm b}, \omega_\mathrm{c}, \Omega_\mathrm{de}, \ln(10^{10}A_\mathrm{s}), n_\mathrm{s}, w) = (0.02225, 0.1198, 0.6844, 3.094, 0.9645, -1)$.
The fiducial {\it Planck} cosmology gives, as derived parameters, $\Omega_{\rm m}=0.3156$
(the present-day matter density parameter) and $\sigma_8=0.831$
(the rms linear mass density fluctuations within a top-hat sphere of radius $8~h^{-1}$Mpc).
We should note that the range of each cosmological parameter covered by our \textsc{SLHD} is sufficiently
broad such that the simulations can cover a range of cosmological models
that ongoing large-scale structure surveys can probe. The parameter range shown in Eq.~(\ref{eq:params})
corresponds to a change of $\pm 5\%$ for $\omega_\mathrm{b}$ and $n_\mathrm{s}$, $\pm 10\%$ for
$\omega_\mathrm{c}$, and $\pm 20\%$ for $\Omega_\mathrm{de}$, $\ln(10^{10}A_\mathrm{s})$ and $w$ from
their central values, which are much larger than the constraints by \citet{planck-collaboration:2015fj}.
As most of large-scale structure probes are sensitive to a combination of $\sigma_8$ and $\Omega_\mathrm{m}$,
we show in the top right panel of Fig.~\ref{fig:SLHD} the range of \textsc{SLHD} models in this
projected parameter space.
Note that the current-generation galaxy surveys have put constraints on the combination of $\sigma_8$ and $\Omega_\mathrm{m}$ at a precision of its 95\% CL region comparable with or smaller than supported range of our emulator. However, if the best-fit model inferred from the galaxy survey is away from the fiducial {\it Planck} cosmology, the posterior region might be outside the supported region \citep[e.g., see][for such an example]{Hikage19}. In such cases, one needs to supply an alternative model or perform additional simulations so that the support range of our emulator can cover the range inferred from the actual data. Or one could use an empirical approach to extrapolate the prediction outside the support range by using the halo model or other analytical method. This is beyond the scope of this paper, and will be explored, if needed, in a separate paper.

\subsection{Box size and resolution}
\label{subsec:box}
The simulations presented here are performed with $2048^3$ particles in comoving cubes with side length of either
$1\,h^{-1}\mathrm{Gpc}$ (hereafter, high-resolution runs: HR) or $2\,h^{-1}\mathrm{Gpc}$ (low-resolution runs: LR).
The mass of the simulation particle in HR (LR) simulations is $1.020\times10^{10}$ ($8.158\times10^{10}$)
$\,h^{-1}M_\odot$ for the fiducial {\it Planck} cosmological model, and varies with the value of
$\Omega_\mathrm{m}$ for different cosmological models.
We perform one LR and HR simulation at every 100 \textsc{SLHD} sampling points.
In addition we have performed $28$ ($14$) random realizations for the fiducial {\it Planck} cosmology under
the HR (LR) setting.
The total volume of $28$ or $112~(h^{-1}\mathrm{Gpc})^3$ for the HR or LR runs at the fiducial {\it Planck}
cosmology is sufficiently large compared to the SDSS volume, which is $\sim 4~(h^{-1}{\rm Gpc})^3$
corresponding to the comoving volume up to $z\sim 0.6$ over the solid angle of about 10,000 sq. degrees.
In the following we refer to each SLHD slice simply as ``Slice'', e.g. ``Slice~1''.
We will use the $20$ simulations in Slice~5 for a cross validation of the emulator and use only the $80$
simulations in Slice~1 to 4 for the training.
In addition, we run simulations with a smaller box size, $250\,h^{-1}\mathrm{Mpc}$
for the fiducial {\it Planck} cosmology, with several different numbers of particles,
$256^3$, $512^3$, $1024^3$ and $2048^3$,
to assess a numerical convergence of our results.
Note that the spatial resolution of the simulations with $512^3$ or $256^3$ particles in these small boxes is
equivalent to that of the main HR or LR simulation, respectively.

As we will show later, the mass resolution of our HR simulations is sufficient to accurately estimate the halo mass
function and the halo-matter cross correlation function in each halo mass bin down to the minimum mass of
$\sim 10^{12}\,h^{-1}M_\odot$, smaller than a typical host halo mass of CMASS or LOWZ galaxies
\citep[e.g. see Fig.~4 in][]{2015ApJ...806....2M}, where the LOW-Z galaxies roughly correspond to the
SDSS Luminous Red Galaxies (LRG) in their figure.
These simulations have already been used in \citet{2018ApJ...854..120M} to calibrate the mass-richness relation
of the redMaPPer clusters by comparing the model predictions of stacked lensing and abundance
with their measurements.
In addition the splashback features of halo edges traced by subhalos in the density and velocity space were
investigated by \citet{Okumura18} using these simulations.

On the other hand, the LR simulations are mainly used to calibrate the halo-halo auto correlation function,
which is noisier than the halo-matter cross correlation due to the larger shot noise, and thus
the precise calibration requires bigger-box simulations.
These simulations allow us to investigate large scale phenomena:
the alignment between the orientation of massive clusters and the large-scale structure surrounding them
were studied in \citet{Osato18} and \citet{Okumura17} using these simulations.
We will show below in more detail how different statistical quantities are evaluated from these HR and
LR simulations.

\subsection{Initial conditions}
\label{subsec:IC}
We generate initial conditions of individual $N$-body simulations using the
second-order Lagrangian perturbation theory \citep[2LPT]{scoccimarro98,crocce06a} implemented by
\citet{nishimichi09} and then parallelized in \citet{Valageas11a}.
We use the linear matter power spectrum computed by \textsc{CAMB} \citep{camb}, and generate Gaussian random fields
from this spectrum.
We compute displacements and velocities by 2LPT, for each particle located on the regular lattice.
The initial redshift is determined such that the rms displacement (at the linear order) is $25\%$ of
the mean inter-particle distance in one dimension, and this depends on the box size and cosmological parameters.
For the fiducial cosmological model, this condition roughly corresponds to $z=59$ and $29$ for the HR and LR runs,
respectively.
In Appendix~\ref{sec:IC}, we study how the results vary with the initial redshift as well as how the results are
altered if the Zel'dovich approximation \citep{zeldovich70} is used, instead of 2LPT,
to set up the initial conditions.

When we generate initial conditions for different cosmological models, we could adopt two ways regarding the
randomness of the realization.
The first possibility is to use the same random seed for different models as that for the fiducial
{\it Planck} cosmology.
This might be advantageous in the sense that the simulated large-scale structure shares the same randomness
and thus one can estimate how each Fourier mode grows in a different way depending on cosmological models
by reducing the sample variance, i.e. the dependence of structure growth on cosmological models.
Motivated by this, we perform a set of $20$ simulations with a fixed random number seed for the HR simulations
in Slice~1. This random number seed is the same as one of the $28$ realizations
of the fiducial {\it Planck} model.
However, a fixed random number seed across different cosmological models does not guarantee to give a converged
result in the final emulator in the sense that every simulation is affected by the same sample variance error
which never goes away by sampling many cosmological models.
By selecting a different random seed for each simulation, we hope that the sample variance should be reduced
in the final results to which the error in all the simulations propagates in a Bayesian manner.
We thus adopt varied random number seeds for the rest of our simulations.
We will see in Appendix~\ref{sec:seed} how the emulation results can change against these
difference choices of the initial seeds. The results shown in what follows are all based on the varied seed
simulations except for Appendix~\ref{sec:seed}.

\subsection{Time integration}
\label{subsec:time}
Once we generate a random realization following the method described in the previous subsection, we
simulate the distribution of particles using
the parallel Tree-Particle Mesh code \textsc{Gadget2} \citep[][]{gadget2}. We set the softening length to $5\%$ of
the mean inter-particle distance in one dimension. We employ the number of fast Fourier transform (FFT)
meshes twice larger than the number of particles in one dimension.
Other configuration parameters were previously calibrated \citep[e.g.][]{nishimichi09,Valageas11a,Takahashi12}.
The relevant parameters are: \textsc{ErrTolIntAccuracy} $=0.05$ for the time-integral accuracy,
\textsc{MaxSizeTimestep} $=0.03$ for the time stepping criterion, \textsc{MaxRMSDisplacementFac} $=0.25$ for
an additional limiter for the Particle-Mesh time step based on the rms particle displacement,
and \textsc{ErrTolTheta} $=0.5$ and \textsc{ErrTolForceAcc} $=0.001$ for the tree opening criterion that
controls the force accuracy.
In the references above, the convergence of the matter power spectrum was intensively tested
to confirm that the accuracy is better than one percent level.
Using $N$-body simulations with these carefully-tuned parameters, \citet{Takahashi12} provided
revised parameters for the halofit formula \citep{smith03}. As will be shown below, the convergence of
clustering signal of halos would be better once we adopt the number density-matching scheme for simulations with
different spatial resolution.
We thus believe that the parameters chosen to give a good accuracy on the matter power spectrum are already adequate
for a calibration of halo clustering quantities without further modification.

We store outputs of each $N$-body realization in 21 redshift bins in the range of $0\leq z \le 1.48$,
equally stepped by the linear growth rate for the fiducial {\it Planck} model. They are $1.48$, $1.35$, $1.23$,
$1.12$, $1.03$, $0.932$, $0.846$, $0.765$, $0.689$, $0.617$, $0.549$, $0.484$, $0.422$, $0.363$, $0.306$, $0.251$,
$0.198$, $0.147$, $0.097$, $0.048$ and $0$.
We use the same redshifts to dump snapshots for other cosmological models.
Since the time evolution of the statistics relevant for our purpose is slow and monotonic, these $21$ snapshots
are sufficient to be interpolated to make a prediction at an arbitrary redshift in between.

\subsection{Halo catalogs}
\label{subsec:halo_catalogs}

Since the aim of this paper is to accurately characterize halo clustering quantities,
the identification of halos in each $N$-body simulation output
is of crucial importance. There have already been comparison studies of different halo finders
\citep[e.g.,][]{Knebe11}. While halo properties appear to be relatively
robust, the ability of finding substructures can differ significantly depending on which algorithm to be used,
especially near the center of halos \citep{pujol:2014qy}. We thus simply select probable
host halos in which galaxies of interest reside, and discard subhalos from our primary
halo catalog when building an emulator of halo clustering quantities.

To identify dark matter halos in each simulation output, we, as our default choice,
employ \textsc{Rockstar} \citep{Behroozi:2013} that identifies dark matter halos and subhalos without distinction
based on the clustering of $N$-body particles in phase space. We supplementarily use
\textsc{Subfind} \citep{subfind} to study a dependence of the halo statistics on the finder
(see Appendix~\ref{app:halo}).
Throughout this paper we adopt $M \equiv M_{200\mathrm{m}}= 4\pi/3 (R_{200{\rm m}})^3 (200 \bar{\rho}_{\rm m0})$
for the halo mass definition, where $R_{200\mathrm{m}}$ is the spherical halo boundary radius within which the mean
mass density is $200$ times $\bar{\rho}_{\rm m0}$.
Again note that
$\bar{\rho}_{\rm m0}$ in the above equation is due to our use of the comoving coordinate,
and therefore $R_{200\mathrm{m}}$ is in the comoving length unit. We follow the default setting of the
\textsc{Rockstar} finder, and define the center of each halo from the center-of-mass location of a subset of
member particles in the inner part of halo, which is selected to minimize the uncertainty caused by the Poisson
noise and the positional dispersion of individual particles which is larger at the outskirt.
Our definition of halo mass includes all the $N$-body particles within the boundary $R_{200\mathrm{m}}$ around
the halo center (i.e., including particles even if those are not gravitationally bound by the halo).
The halo mass defined in this way is more relevant for weak lensing observables, which measure the total
enclosed mass within a given aperture.

After we identified halo candidates either by \textsc{Rockstar} or \textsc{Subfind}, we
determine whether they are central or satellite halos. When the separation of two different halos
(between their centers) is closer than $R_{200\mathrm{m}}$ of the more massive one,
we mark the less massive one as a satellite halo.
We keep halos with mass above $10^{12}h^{-1}M_\odot$ in the final halo catalog.
The dependences of the halo clustering on the halo finder, mass definition and the central-satellite split criterion are presented in Appendix~\ref{app:halo}.

\subsection{Hybrid Fourier-direct method to measure the correlation signal}
\label{subsec:measurement}

After simulations are done and halos are identified, we measure the clustering quantities. While
correlation functions can be accurately estimated by a direct pair counting, such a method can be computationally
expensive due to its $\mathcal{O}(N^2)$ scaling to the number of particles $N$.
We here develop a hybrid method that combines the direct pair counting method with a grid based method that
makes use of FFT.
The former is used to measure the clustering signal on small scales and the latter is on large scales. The FFT method suffers from inaccuracy near the grid spacing, but is robust for scales much larger than the grid size.

We measure the auto- and cross-correlation functions for halo-halo, halo-matter, and matter-matter pairs.
We employ $1024^3$ FFT grids for the large-scale signal and use the direct pair-counting method at scales below
$5\,h^{-1}\mathrm{Mpc}$ or $10\,h^{-1}\mathrm{Mpc}$ for the HR or LR runs that have 1 or 2~$h^{-1}{\rm Gpc}$ for
the box size, respectively.
These switching scales roughly correspond to five times the FFT grid spacing, which is chosen so that
the FFT method provides a good accuracy at scales greater than the switching scale.

\begin{figure}[t!]
\begin{center}
\includegraphics[width=8cm,angle=0]{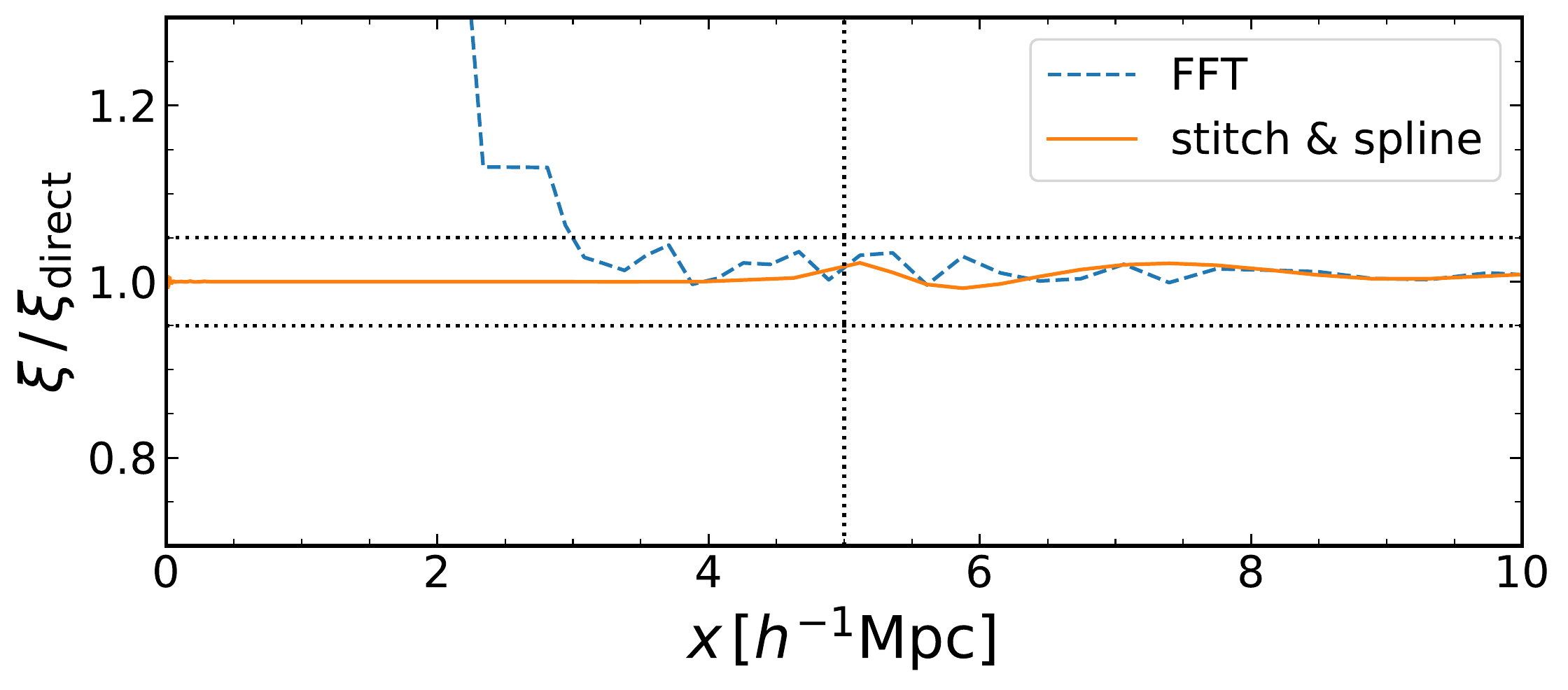}
\end{center}
\caption{Accuracy of our hybrid Fourier-direct method. We plot by the dashed line the ratio of the halo-matter cross correlation function measured with the FFT-based method to that from the direct pair counting, which should give the most accurate result. Compared to the reference result based on the direct pair counting, the FFT-based method shows overestimation at small pair separation. Also, it shows a noisy pattern at intermediate separations. Our final method, which combines the FFT with direct pair counting at $x=5\,h^{-1}\mathrm{Mpc}$ (vertical dotted line) and smoothed by cubic spline (see later discussion), is shown by the solid line. While a small residual can be seen near the switching scale of $5\,h^{-1}\mathrm{Mpc}$, the overall behavior is within our target accuracy.
\label{fig:hybrid}
}
\end{figure}

Figure~\ref{fig:hybrid} shows an example of our measurements of
the halo-matter correlation function for a halo sample with mass larger than $10^{13}\,h^{-1}M_\odot$
at $z=0$. For this exercise, we take one HR simulation for the fiducial {\it Planck} cosmology, and measure the correlation function with the direct pair counting up to $10\,h^{-1}\mathrm{Mpc}$ as a reference. Compared to this measurement we show the result of the FFT-based method (dashed line). Two features can be found from the ratio. First, the FFT-based method starts to deviate from the reference rather quickly as decreasing the separation below $\sim 2\,h^{-1}\mathrm{Mpc}$. This scale corresponds to about twice the grid size, and thus simply reflecting the resolution limit of FFT. Second, a noisy feature with can be observed on intermediate scales up to $x\sim 8\,h^{-1}\mathrm{Mpc}$. This is due to the discrete sampling of the pair separations (we can take only an integer vector in units of the grid spacing) together with the subtlety in the choice of the bin center, which we take as the geometric mean of the bin edges for simplicity. Note that this pattern appears to be almost the same for different random realizations and for different halo samples, supporting our interpretation above. To avoid the large error due to the first effect, we conservatively choose the switching scale to be $5\,h^{-1}\mathrm{Mpc}$ as indicated by the vertical dotted line. Furthermore, the stitched result is smoothed with a cubic spline function (see the next section for details) to reduce the second effect on intermediate scales, while keeping the time-consuming pair counting part only to a limited range of pair separations. Our default result that we will use in the emulator building is shown by the solid curve.

In our initial implementation, we accelerate the pair counting method by first sorting both particles and halos
in a coarse grid with $200^3$ cells.
We count pairs only in the same or the adjacent cells from which pairs can be closer than the matching scale.
The code is then updated to employ a more sophisticated sort-tile-recursive (STR) R-tree scheme for a more
efficient spatial indexing \citep{Mitsuhashi16}. Note that these different versions give identical results,
and the difference lies only in the speed to perform the exact pair counting.
Even with the help of these methods, the measurement of the matter-matter auto correlation function is
computationally expensive so that we cannot measure it from all the snapshots for all the models with
$2048^3$ particles. For this, we randomly select only $1/64$ of the simulation particles in the measurement.
This random selection increases the Poisson noise
to the measured signal, but the typical error caused by this is not important over scales of interest, roughly
larger than $0.1\,h^{-1}\mathrm{Mpc}$. While the main product of our emulator is the halo clustering quantities, we
also provide the matter auto correlation function for comparison, e.g. which can be used to estimate
the effective bias function of halos or galaxies under consideration.

In the FFT based measurement, we use the Cloud-in-Cells scheme \citep{hockney81} to assign matter particles or
halos to each grid density estimate and then perform the FFT.
We compute the product, $\delta_{1,\mathbf{k}} \delta^*_{2,\mathbf{k}}$, of two fields
(where $1$ and $2$ denote halos and/or matter) at each wavenumber vector $\mathbf{k}$ and then
Fourier-transform-back it to the configuration space. We take an average of the field product in each spherical
shell to estimate the correlation function at the radial bin.

\section{Emulation}
\label{sec:emu}
The \textsc{Dark Emulator} is constructed based on the \textsc{Dark Quest} simulation suite and the analysis
pipeline that was explained in the previous section. In this section we discuss how we construct different modules,
each of which predicts a statistical quantity of halos, and combine them to form \textsc{Dark Emulator}.
\subsection{Overall design}
\label{subsec:overall_design}

\begin{figure*}[ht!]
\begin{center}
\includegraphics[width=16cm,angle=0]{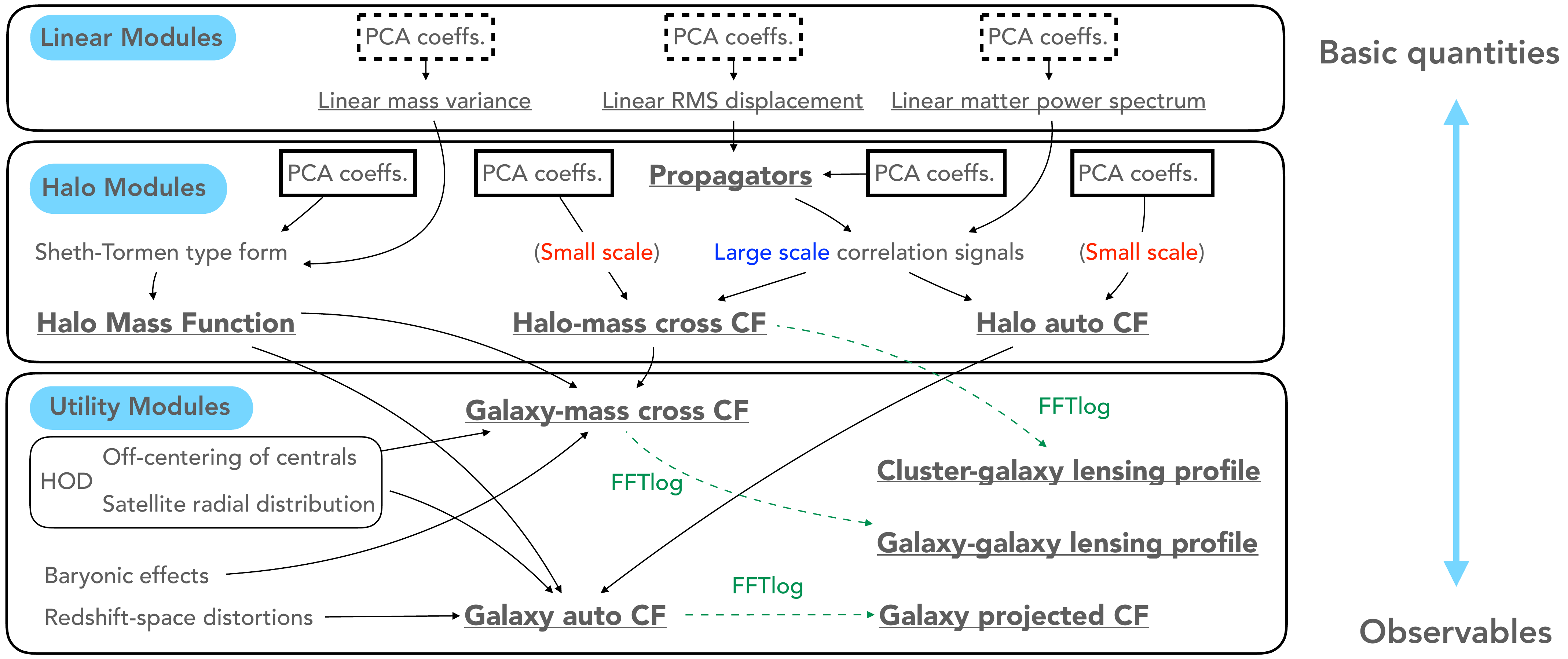}
\end{center}
\caption{
The layout of different modules of \textsc{Dark Emulator}. The cosmology dependence of the quantities in a square (i.e., ``PCA coeffs.'') are modeled by Gaussian Process, and those underlined are physical quantities evaluated in each of the module. The whole \textsc{Dark Emulator} code is made up of three groups of modules enclosed by a rounded rectangle box. The first group of modules, shown at the top of the figure, are for linear-theory quantities (\textsc{Linear Modules}). The second group
shows the modules for the abundance and clustering properties of halos (\textsc{Halo Modules}). These are calibrated with a suite of $N$-body simulations, and the core pieces of \textsc{Dark Emulator}. The other modules at the bottom work on the outputs of the \textsc{Halo Modules} and transform them into observable quantities (\textsc{Utility Modules}). These mainly connects halos to galaxies using an analytical prescription and project the three-dimensional quantities onto the two-dimensional sky.
\label{fig:DE_architecture}
}
\end{figure*}
The final goal of our work is to build an $N$-body simulation calibrated emulator that provides
an accurate prediction of galaxy clustering quantities as a function of cosmological parameters and parameters
needed to connect halos and galaxies for a given cosmological model.
There are various ways to do this. One way is to adopt {\it a priori} parametric prescription to connect halos and galaxies such as HOD, make mock catalogs of galaxies in each $N$-body simulation realization based on the assumed prescription,
measure galaxy clustering quantities from the mocks, and then build an emulator of galaxy quantities from the tabulated database with the model parameters in the prescription treated as the input variables in addition to the cosmological parameters.
This approach was, for example, employed in \citet{2015ApJ...810...35K}
\citep[also see][]{2018arXiv180405867Z}.
However, an emulator built based on such a method may produce inaccurate results with uncertainties
associated with galaxy-halo connection.
For instance, there is no guarantee that
a restricted HOD functional form assumed in the emulator can accurately
describe clustering properties for a sample of galaxies in a given survey.
In addition the radial profile of satellite galaxies in a given host halo has not yet been well constrained.
Furthermore, some of central galaxies might be off-centered from the true center.
Therefore, variations in galaxy clustering properties
cannot be incorporated in an emulator that employs the restricted model of the halo-galaxy connection.
Put another way, in this approach it is very difficult to modify or change an emulator after its construction to
include these variations and add flexibilities in the model predictions.

For this reason, we employ an alternative approach in this paper.
The core function of our emulator is to predict several basic halo clustering
quantities that are given as a function of cosmological parameters, halo mass, separation scale, and redshift.
We will combine the ``modules'' analytically at equation level, instead of using the mock catalogs,
by employing a halo-galaxy connection prescription (e.g. HOD)
to compute predictions of galaxy clustering quantities. The design of our emulator is illustrated in
Fig.~\ref{fig:DE_architecture}.
It is composed of three groups of modules surrounded by the rounded rectangular boxes, each of which has a number of functionalities as denoted by the text.
The first group are \textsc{Linear Modules}
which predict statistical quantities of the linear matter perturbations (see Appendix~\ref{sec:linear} for details). The second group is the core part, and predicts various statistical properties of dark matter halos (\textsc{Halo Modules}). Finally, at the bottom of the figure, we have \textsc{Utility Modules}, which combine the upper-level modules to compute observable quantities.
The key ingredient in this group is the prescription to connect halos and galaxies (see the items in the inset). Another functionality implemented here is to compute the projected clustering quantities such as galaxy-galaxy weak lensing correlation function by directly projecting the three-dimensional correlation function along the line-of-sight direction by numerical integration.
We also provide options to include possible baryonic corrections to the mass profile near the halo center, as well as redshift-space distortions (these effects will be presented in a separate paper).
Although we assume a specific HOD prescription as a working example of halo-galaxy connection, an user can change it and adopt another prescription to have the galaxy clustering quantities from \textsc{Halo Modules}.
Thus our method allows a flexible modification of the halo-galaxy connection, without the need for additional training based on numerical realizations of mock galaxy catalogs.

\subsection{Resolution study and matching scheme}
\label{subsec:matching}
Because of limited numerical resources such as memory and executive CPU time,
we can run only a finite number of $N$-body simulation realizations, where the size of each simulation is mainly determined by the number of $N$-body particles.
Even for a fixed number of particles, there is a trade-off between the resolution and the box size.
While the former is responsible for the minimum length scale and the minimum mass of halos down to which the simulation results are accurate, the latter defines the number of Fourier modes available in each simulation and thus controls the statistical precision.
The usual way to cover a wider dynamic range of the predictions is to combine simulations performed in different
box sizes and then stitch their results over separations or wavenumbers between neighboring box-size simulations.
Indeed, such a method was used in previous works such as
\citet{Coyote3}, \citet{Valageas11a}, and \citet{Takahashi12}, where the main goal was to calibrate the matter power spectrum. An analytical model based on perturbative calculation was further combined at the large scale limit in \citet{Coyote3} to suppress uncertainties due to the large sample variance near the wavenumber corresponding to the box size.

In this subsection we examine the numerical convergence of halo quantities using a set of
simulations with different resolutions. We then discuss a strategy
to combine the results of different simulations
to predict the clustering signals over wider ranges of halo masses and length scales.

\begin{figure}[h]
\begin{center}
\includegraphics[width=8cm,angle=0]{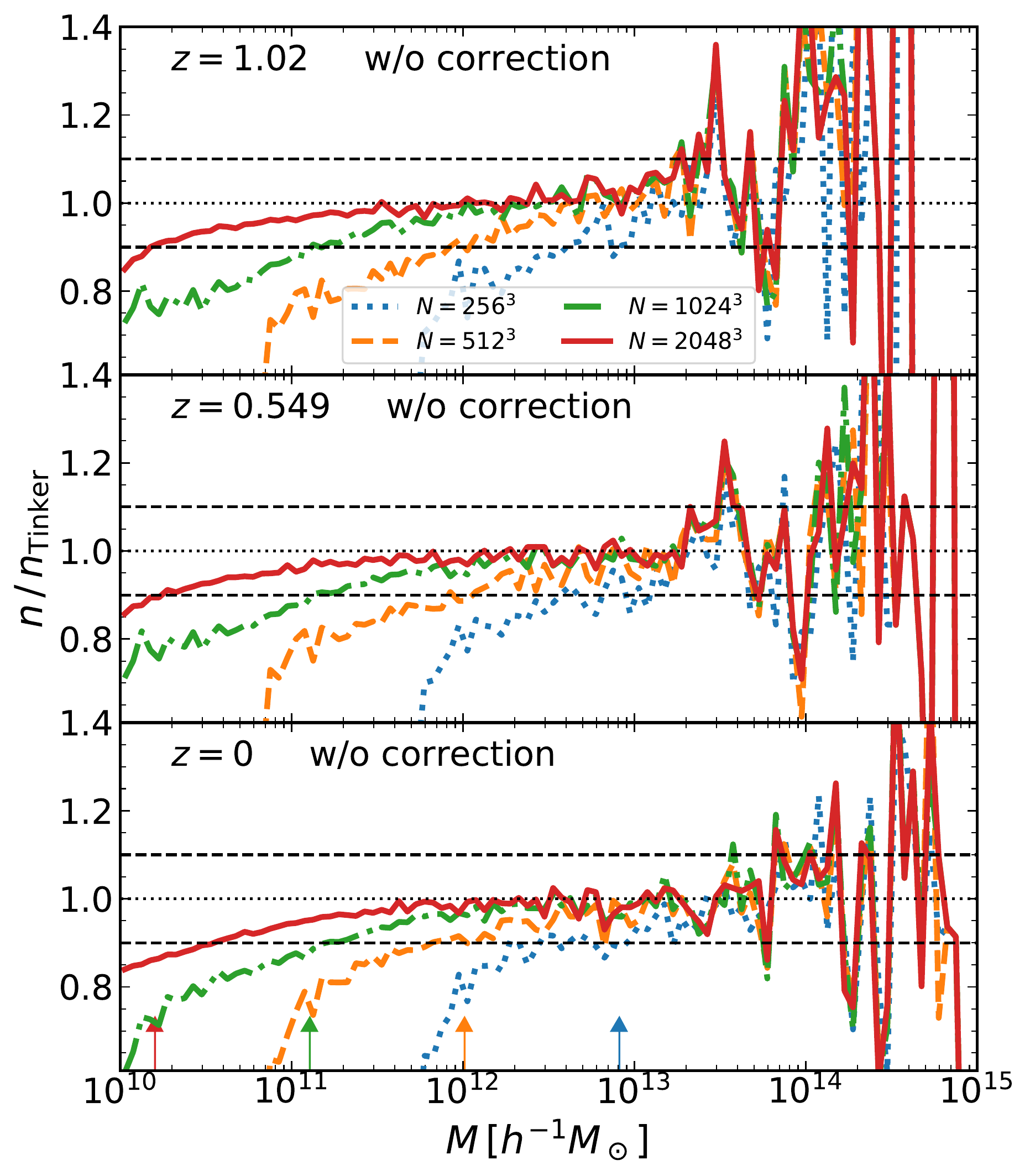}
\end{center}
\caption{Resolution study for the halo mass function. Here we fix simulation box size to $250\,h^{-1}{\rm Mpc}$ and compare
the mass functions measured from simulations with different mass resolutions.
We use simulations of $1\,h^{-1}{\rm Gpc}$ box and $2048^3$ particles (HR simulations)
to build an emulator of halo mass function, which are equivalent, in terms of the resolution,
to the $512^3$ simulation in this plot.
Plotted here is the ratio of simulation result to the fitting formula in \citet{Tinker08} at $z=0$. The arrows in the horizontal
axis denote halo mass which corresponds to 100 particles for the halo mass definition of 200 times the cosmic mean density.
The two horizontal dashed lines denote $\pm 10\%$ fractional difference from the \citet{Tinker08} mass function. The three panels show the ratio at $z=1.02, 0.549$ and $0$ (from the top to the bottom).
\label{fig:MF_resolution1}
}
\end{figure}

\subsubsection{Halo mass function}
\label{subsubsec:hmf}
In Fig.~\ref{fig:MF_resolution1}, we first examine the halo mass function (HMF) for
the fiducial {\it Planck} cosmology at three redshifts, $z=1.02, 0.549$ and $0$
, using four $N$-body simulations
with different numerical resolutions.
Note that the simulation with the lowest resolution among the four, which has $256^3$ particles,
has the same resolution as our main \textsc{LR} suite,
whereas the second from the worst, with $512^3$ particles, corresponds to the resolution of the \textsc{HR} suite. For reference, HMF in Fig.~\ref{fig:MF_resolution1} is normalized by the fitting formula of \citet{Tinker08}, with the mass definition of $200$ times the cosmic mean density. To have a fair comparison,
we integrated the \citet{Tinker08} HMF (hereafter Tinker HMF)
over halo masses in each mass bin, which is used when we measure the HMF from simulations.

We can see that the measured HMF better matches
the Tinker HMF down to lower masses as increasing the simulation resolution.
The four simulations agree with each other at the high mass end,
although the curves are noisy due to the Poisson noise. Note that these simulations are done in a comoving
box with a side length of $250\,h^{-1}\mathrm{Mpc}$,
which is smaller than our main simulations of $1$ or $2\,h^{-1}\mathrm{Gpc}$ used for the emulator development.
This suggests that our main simulations
have much lower Poisson noise at such high mass bins. The vertical arrows, from right to left for
higher resolution, denote the halo mass corresponding to 100 $N$-body particles.
The figure indicates that the simulation HMF at this mass scale is underestimated by about 10\% fairly
independently of redshift.
Thus, one needs at least several hundreds of particles to
estimate HMF to a percent accuracy.

We here propose a way to empirically correct for a systematic error in the estimated HMF due to numerical resolution. Our method is motivated by the method in \citet{Warren06},
which was developed for halos that are identified by the Friends-of-Friends (FoF) method.
They proposed that the FoF mass of each halo is calibrated as
\begin{eqnarray}
\tilde{M} = \left(1-N_\mathrm{p}^{-0.6}\right)\,M,
\label{eq:mass_correction1}
\end{eqnarray}
where $N_\mathrm{p}=M/m_{\mathrm{p}}$ is the number of member particles,
$m_{\mathrm{p}}$ is the $N$-body particle mass, and $\tilde{M}$ is the corrected mass.
Since the FoF algorithm tends to link physically unbound particles near the halo boundary
when the mass resolution is poor, an FoF halo mass tends to be \textit{overestimated} compared to the true mass.
Hence the FoF based HMF tends to be \textit{overestimated} for low halo masses that are affected by numerical resolution.
The correction factor is applied to each FoF halo in such a way that
the FoF mass is \textit{reduced} to correct for the overestimation in HMF.
This procedure was further confirmed in \citet{crocce10}, where the method was applied to FoF halos in
\textsc{MICE} simulations. On the other hand, our result in Fig.~\ref{fig:MF_resolution1} displays
a rather opposite trend: our HMF is {\it underestimated} in low-resolution simulations, implying
that a mass of each low-mass halo tends to be \textit{underestimated}.
Since we use the spherical-overdensity (SO) mass, the SO based HMF is affected by the matter density field around
the halo region, which tends to be underestimated in low-resolution simulations. This is different from the FoF
finder, and thus the opposite trend is understandable.

\begin{figure}[h]
\begin{center}
\includegraphics[width=8cm,angle=0]{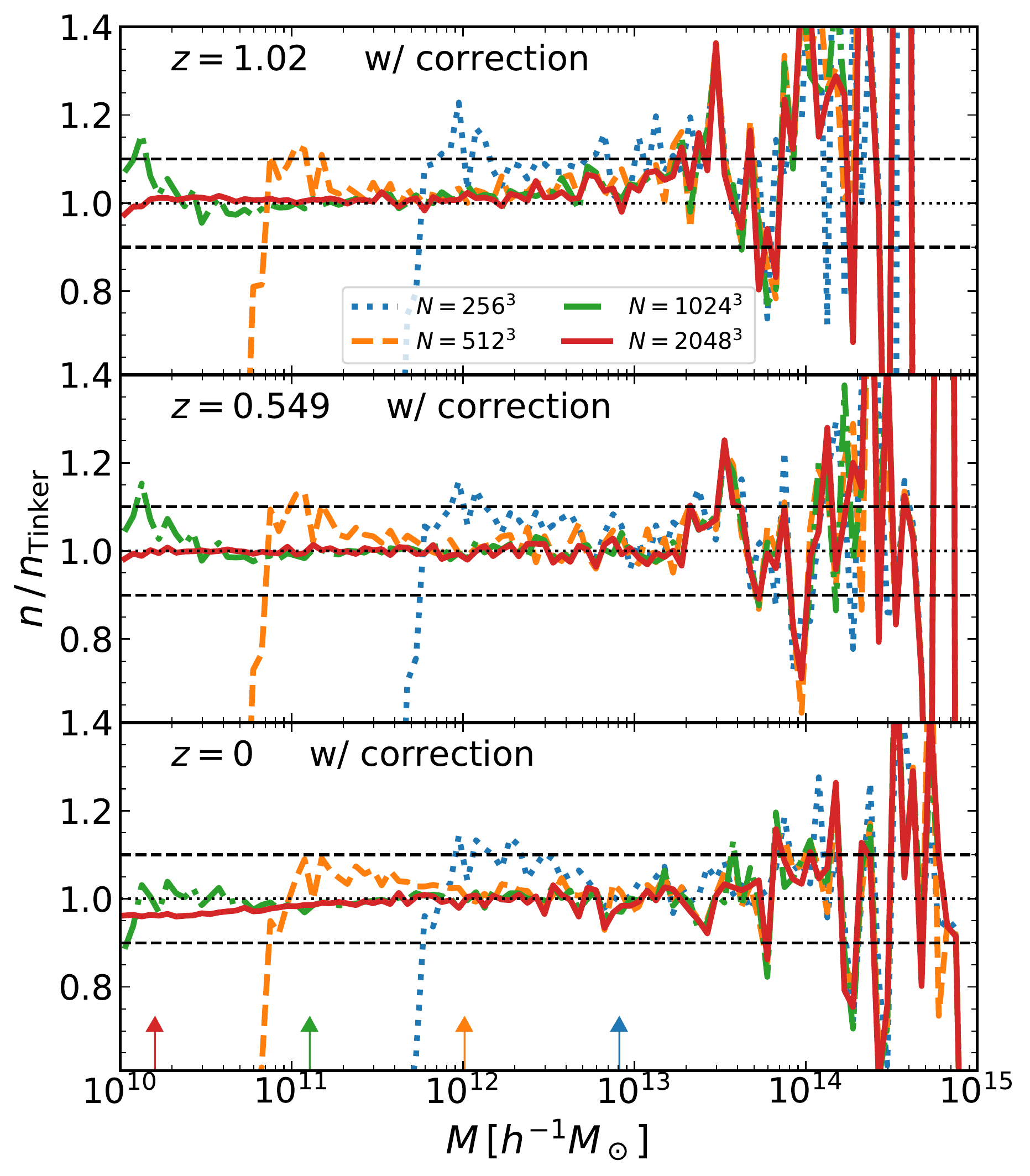}
\end{center}
\caption{Similar to
Fig.~\ref{fig:MF_resolution1}, but the plot shows the mass functions when inaccuracies
in individual halo masses due to limited mass resolution are corrected for
according to Eq.~(\ref{eq:mass_correction2}).
\label{fig:MF_resolution2}
}
\end{figure}

We thus use the following equation to correct for the SO mass:
\begin{eqnarray}
\tilde{M} = \left(1+N_\mathrm{p}^{-0.55}\right)\,M,
\label{eq:mass_correction2}
\end{eqnarray}
where $N_\mathrm{p}$ the number of particles within $R_{200\mathrm{m}}$ in the SO mass definition.
We employ a slightly different power of $N_\mathrm{p}$ from Eq.~(\ref{eq:mass_correction1}) to have a better calibration.
After this correction, the HMFs from different resolution simulations
better agree with each other down to the resolution limit denoted by the vertical arrows,
as shown in Fig.~\ref{fig:MF_resolution2}.
Below the mass limit, the SO halo mass becomes over-corrected, yielding an overestimation in HMF.
These trends after the correction appear to be very similar at different redshifts.
While a further refinement of the empirical function given by Eq.~(\ref{eq:mass_correction2})
would be possible in principle, we do not use halos with less than $200$ particles when we calibrate HMF.
Note that we here employed a slightly conservative threshold of 200 particles compared to the case of 100
particles as discussed in Fig.~\ref{fig:MF_resolution2}.
Furthermore, we use only the \textsc{HR} simulations (resolution equivalent to the one with $512^3$ particles
here), for development of the HMF emulator.
We can determine HMF accurately down to $\sim 10^{12}\,h^{-1}M_\odot$ and this number varies depending on the
cosmological model as we will discuss below.

\subsubsection{Correlation functions}
\label{subsubsec:CF}

\begin{figure*}[ht!]
\begin{center}
\includegraphics[width=8cm,angle=0]{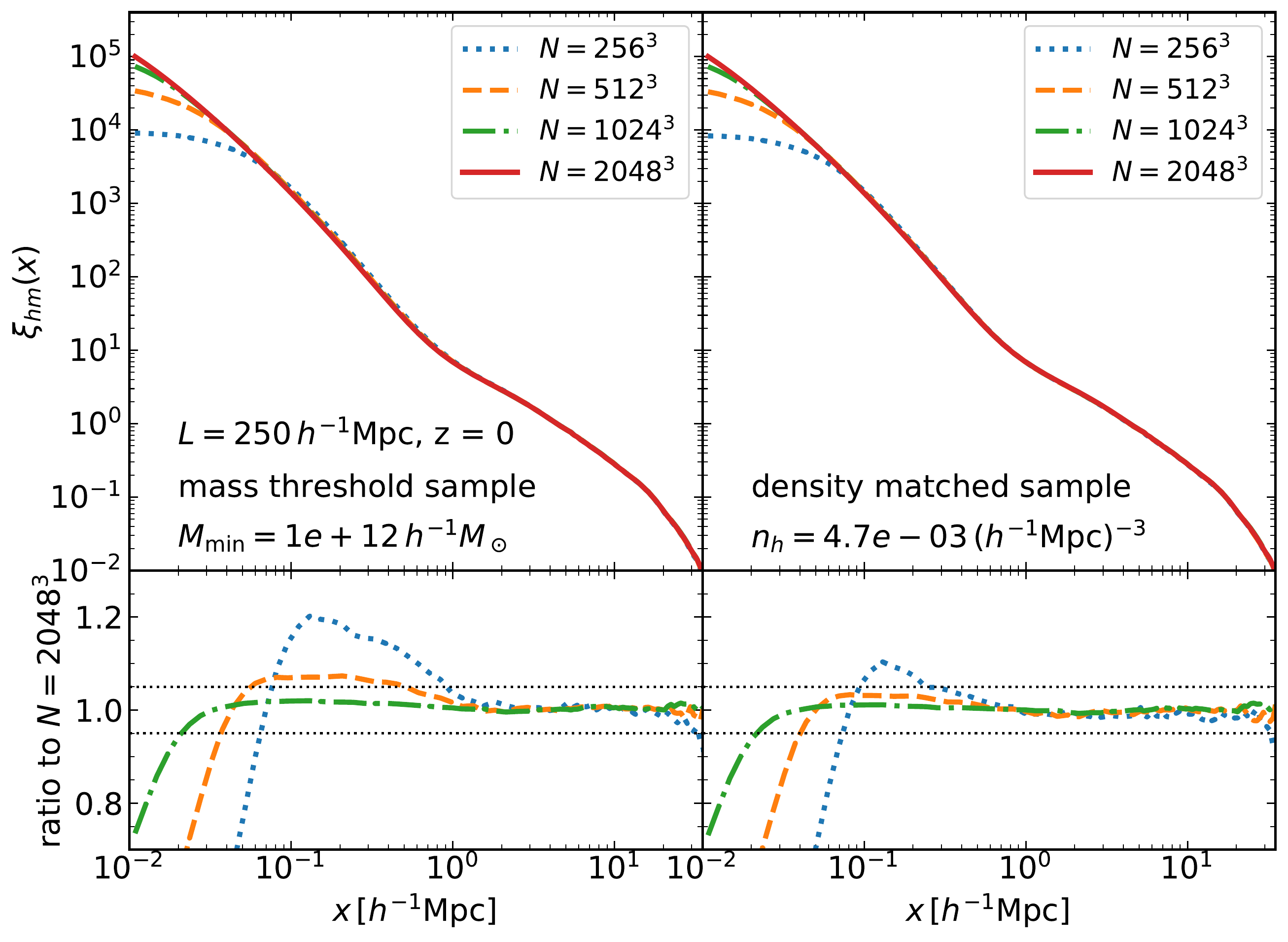}
\includegraphics[width=8cm,angle=0]{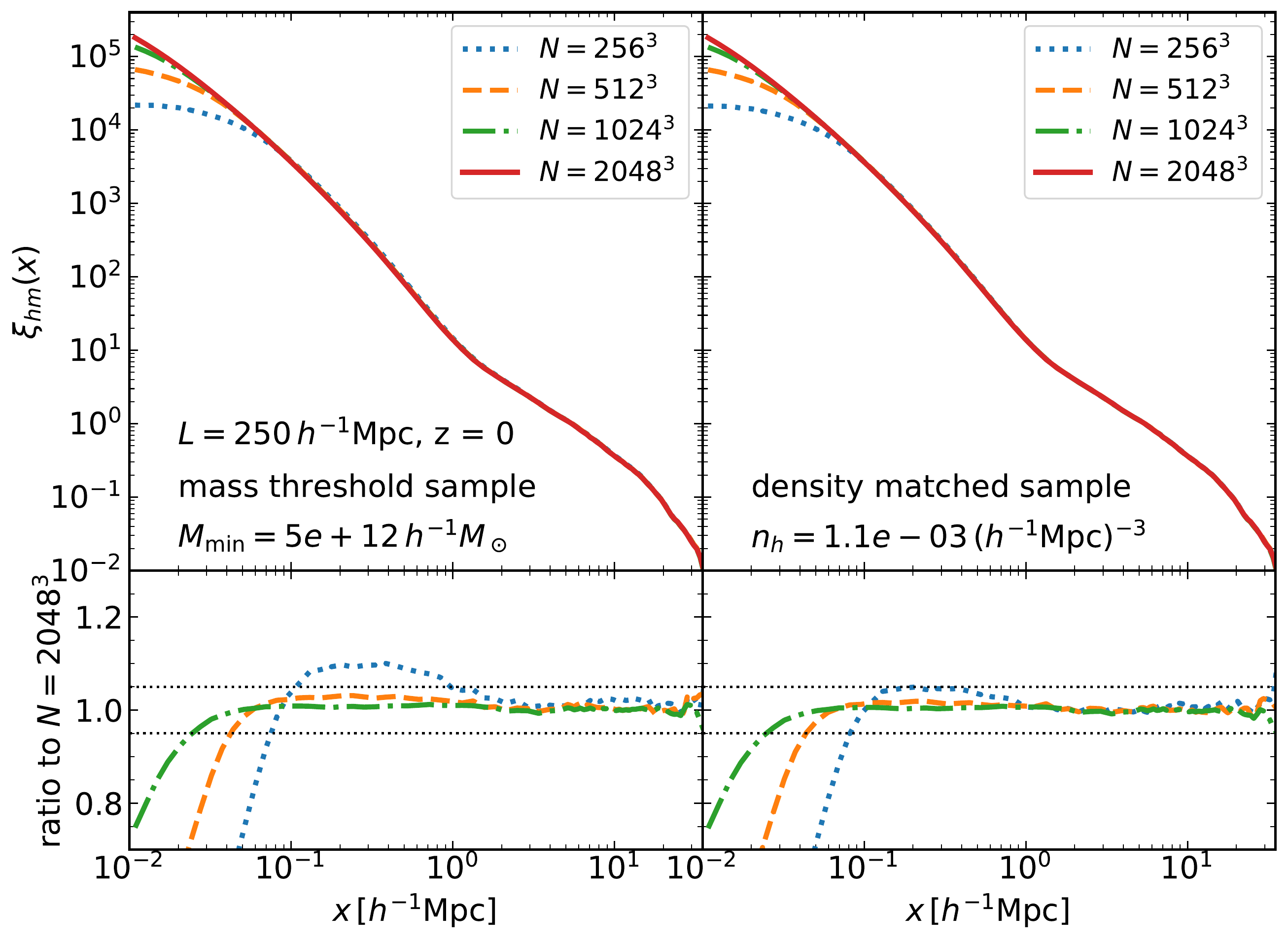}
\end{center}
\caption{Resolution study for the halo-matter cross correlation function using
simulations with different mass/spatial resolutions in a small box ($250\,h^{-1}\mathrm{Mpc}$ a side) as in Fig.~\ref{fig:MF_resolution1}.
For the module of the halo-matter cross correlation, we use the HR simulations
($1\,h^{-1}{\rm Gpc}$ box size and $2048^3$ particles) which are equivalent to the $512^3$ simulations in this plot. We show the results for the mass threshold samples in the two left panels, while in the right panels the halo samples are chosen so as to have an equal number density above the mass threshold. We show in the bottom panels the ratio to the simulation with $2048^3$ particles as a reference.
\label{fig:match_cross1}
}
\end{figure*}

\begin{figure*}[ht!]
\begin{center}
\includegraphics[width=8cm,angle=0]{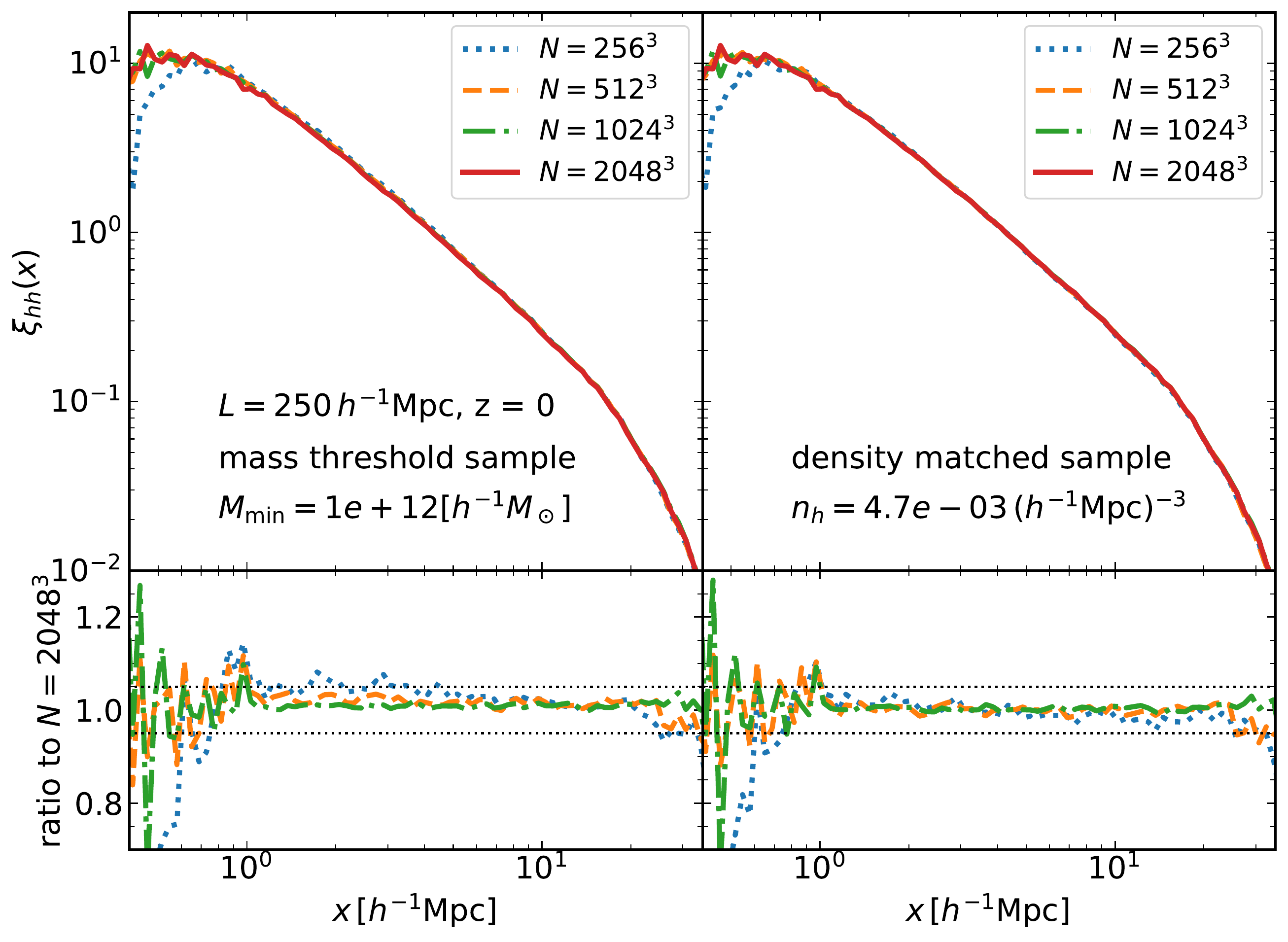}
\includegraphics[width=8cm,angle=0]{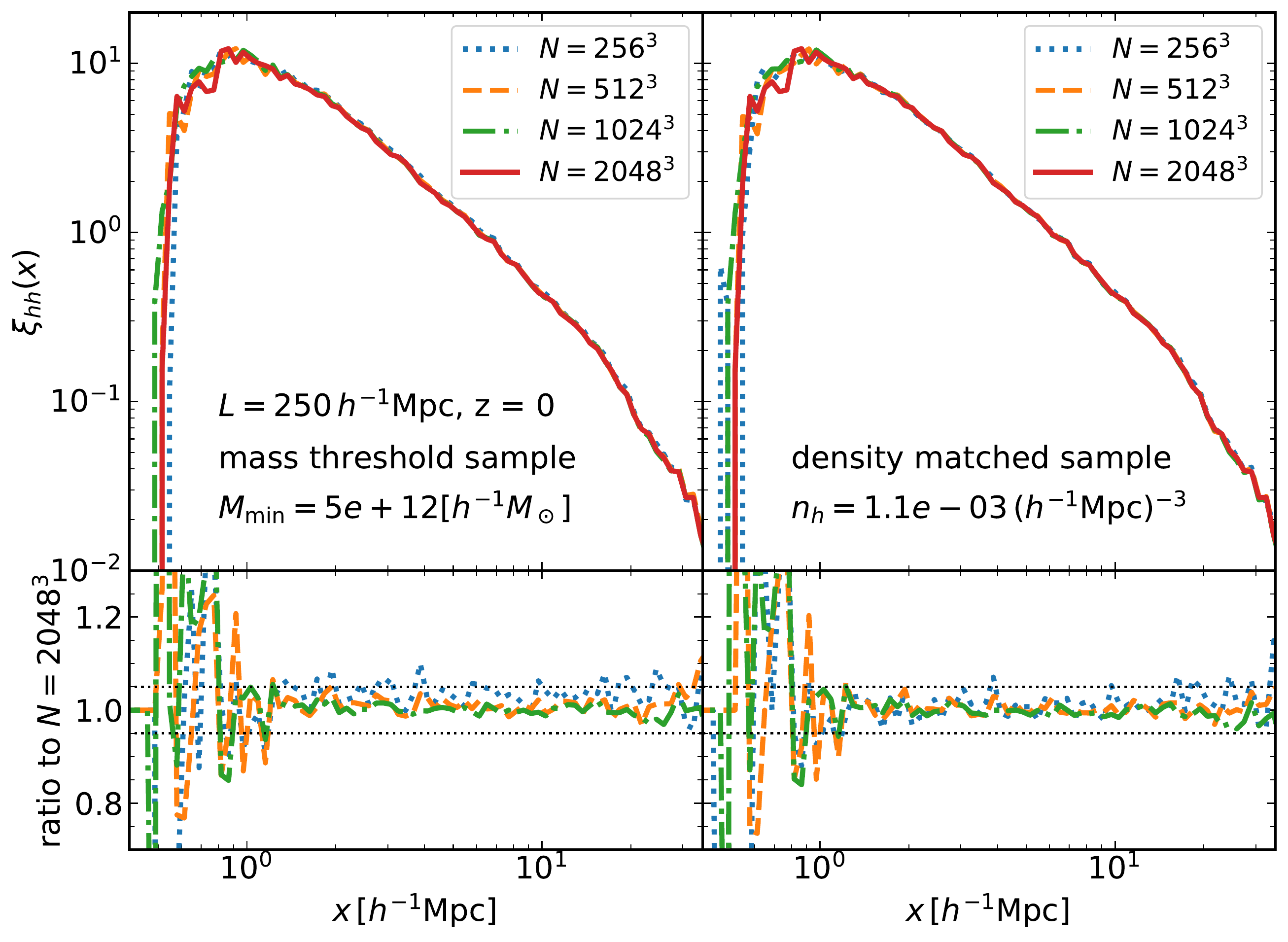}
\end{center}
\caption{Resolution study for the halo-halo auto correlation function from simulations with different mass/spatial resolutions as in the previous figure.
For the module that computes the halo-halo correlation, we use simulations of $2\,h^{-1}{\rm Gpc}$ box size and $2048^3$ particles (hereafter
LR simulations) that are equivalent to the $256^3$ simulations in this plot.
\label{fig:match_auto1}
}
\end{figure*}

Next we check clustering correlation functions of halos. In doing so,
we need to consider subsamples of halos divided by
halo discriminators such as
halo mass, and then consider the clustering correlations as a function of different subsamples.
In the left plot of each panel in Figs.~\ref{fig:match_cross1} and \ref{fig:match_auto1},
we show the halo-matter cross- and halo-halo auto-correlation functions for a mass threshold sample of halos with
$M\ge 10^{12}~h^{-1}M_\odot$ or $5\times 10^{12}~h^{-1}M_\odot$,
measured from the four simulations as in Fig.~\ref{fig:MF_resolution1}.
Here the threshold mass $10^{12}~h^{-1}M_\odot$ corresponds to halos with more than
100 member particles for simulations with more than $512^3$ particles,
whereas it corresponds to halos with only $\sim 10$ particles for $256^3$ simulation.
Note that we did not apply the correction from Eq.~(\ref{eq:mass_correction2}) for halo masses in these figures.
While all the correlation functions agree with each other at large separations, the smaller scales clearly show
the effect of numerical resolution; the measurements from a lower-resolution simulation start to deviate from
those from a higher-resolution simulation on scales smaller than $\sim 1\,h^{-1}\mathrm{Mpc}$.
The comparison of Figs.~\ref{fig:match_cross1} and \ref{fig:match_auto1} reveals that the deviation is larger
for a halo sample of smaller mass threshold.
The inaccuracy is ascribed to several facts.
In a lower resolution simulation, halo masses around the mass threshold are not determined accurately on individual
halo basis due to the lack of numerical resolution as discussed in Fig.~\ref{fig:MF_resolution1}.
Thus the halo sample of a given mass threshold becomes different from that of higher-resolution simulation.
Moreover the mass distribution around each halo in a lower resolution simulation is simulated less accurately.

In this paper we employ a slightly different sample of halos to develop the emulator. Rather than using the mass as the primary proxy of the different clustering strength of halos
we consider
mass threshold samples and label each sample in terms of the number density of halos above the threshold. We expect two advantages from this
conversion:
first, the cosmology dependence of the noise level of various statistics are weaker compared to the samples labeled by the mass. Indeed, we know that the mass of the heaviest halos available in each simulation can be quite different among different cosmological models and at different redshifts. Second, as
quite obvious from Fig.~\ref{fig:MF_resolution1}, the masses inferred from simulations are quite sensitive to the resolution especially at the low mass end. To see this more qualitatively, we show in
right plot of each panel in Figs.~\ref{fig:match_cross1} and \ref{fig:match_auto1}
the clustering signals for the mass threshold samples with a fixed number density
in each simulation. Here the number density is the same as that of the mass threshold sample for the highest-resolution simulation ($2048^3$)
in the left plot,
and the mass threshold for other simulations are determined to match the target number density.
Now an agreement between different resolution simulations is better than in the left panel, reflecting the fact that
the number of halos in the sample is less affected by numerical resolution compared to the mass-threshold sample.
Nevertheless, the lowest resolution simulation still exhibits a
relatively large deviation for the halo-matter cross correlation at small scales,
especially for the sample corresponding to $10^{12}\,h^{-1}M_\odot$, because the matter distribution
in high density regions is less accurately simulated in such a low resolution simulation.
To avoid this inaccuracy, we thus use only the \textsc{HR} simulations to
estimate the halo-matter cross correlation function for different cosmological models.

On the other hand, Fig.~\ref{fig:match_auto1} shows a slightly better agreement among the four simulations with different resolutions,
implying that the halo-halo auto correlation is relatively robust against the numerical resolution.
Note that larger scatters in the ratio on
small scales ($\lesssim 1\,h^{-1}\mathrm{Mpc}$)
are due to the halo exclusion effect, which states that the correlation signal is sharply suppressed on small scales
due to the fact that no halo pair can exist below $R_{\rm 200}$ radius of the larger one by construction in our halo sample.
Thus a slight misestimation of the halo radii due to numerical resolution can lead to a large error in the correlation signal at a fixed scale around the typical $R_{\rm 200}$ of the sample.
Since our final product is the \textit{galaxy} correlation function and the one-halo term gives a dominant contribution around these scales,
the scatter seen here does not largely affect the predictions of galaxy auto correlation function as we will show later.
Based on these results, we use the \textsc{LR} simulations to estimate the auto correlation functions for different
cosmological models.

In summary we use the correlation functions of halos measured for halo samples with different number densities.
When this is combined with the HMF module that gives the halo number density as a function of mass,
one can compute the halo correlation function for a given mass threshold instead of the number density.
Furthermore we can compute the correlation function of halos in an infinitesimally narrow mass bin
by taking the numerical derivative of the correlation functions for a mass threshold halo sample with respect to the threshold mass.

\subsubsection{Large-scale limit}
\label{subsubsec:largescale}

\begin{figure}[h!]
\begin{center}
\includegraphics[width=8cm,angle=0]{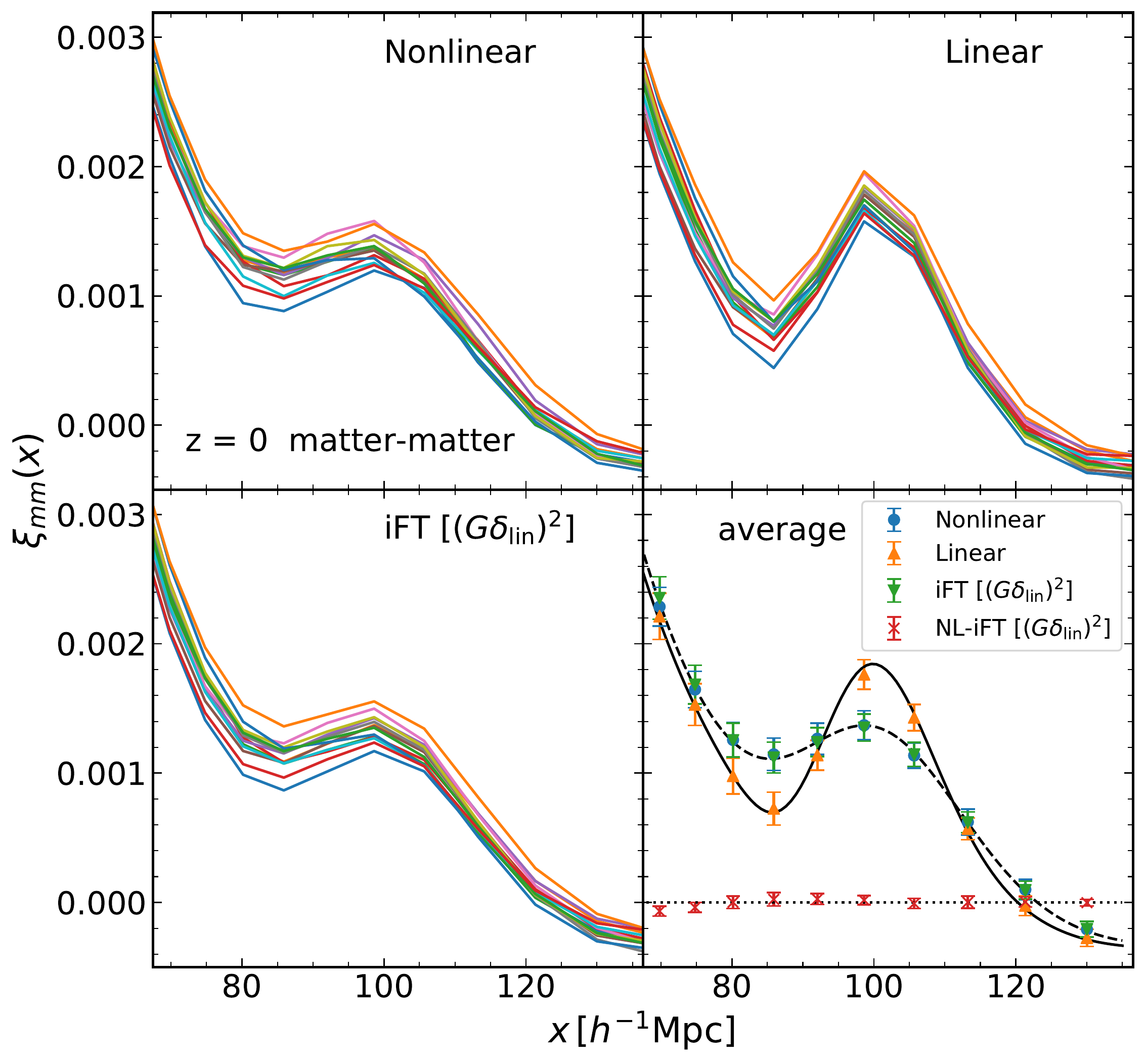}
\end{center}
\caption{Matter auto correlation function around the BAO scale.
We show in the upper-left panel the correlation function measured from the $14$ LR simulations for the fiducial
{\it Planck} cosmology at $z=0$. The upper-right panel shows the linear correlation function for the Gaussian random realizations which correspond to the initial conditions used in the simulations of
the upper-left panel. The lower-left panel shows the results computed by taking the inverse Fourier transform of the product of the propagator and the linear power spectrum, referred to as ${\rm iFT}[G_\mathrm{m}^2(k)P_\mathrm{lin}(k)]$ here,
for the same random realizations (see text for details).
Finally, the lower-right panel shows the average of curves in the other panels over the $14$ realizations.
The cross symbols with errorbars show the difference between the full nonlinear curves and the propagator-based
model for the random realizations considered here. The solid and the dashed curves denote analytical calculations
for the linear theory and the propagator model, respectively.
\label{fig:BAO_ximm}
}
\end{figure}

\begin{figure*}[t!]
\begin{center}
\includegraphics[height=8cm,angle=0]{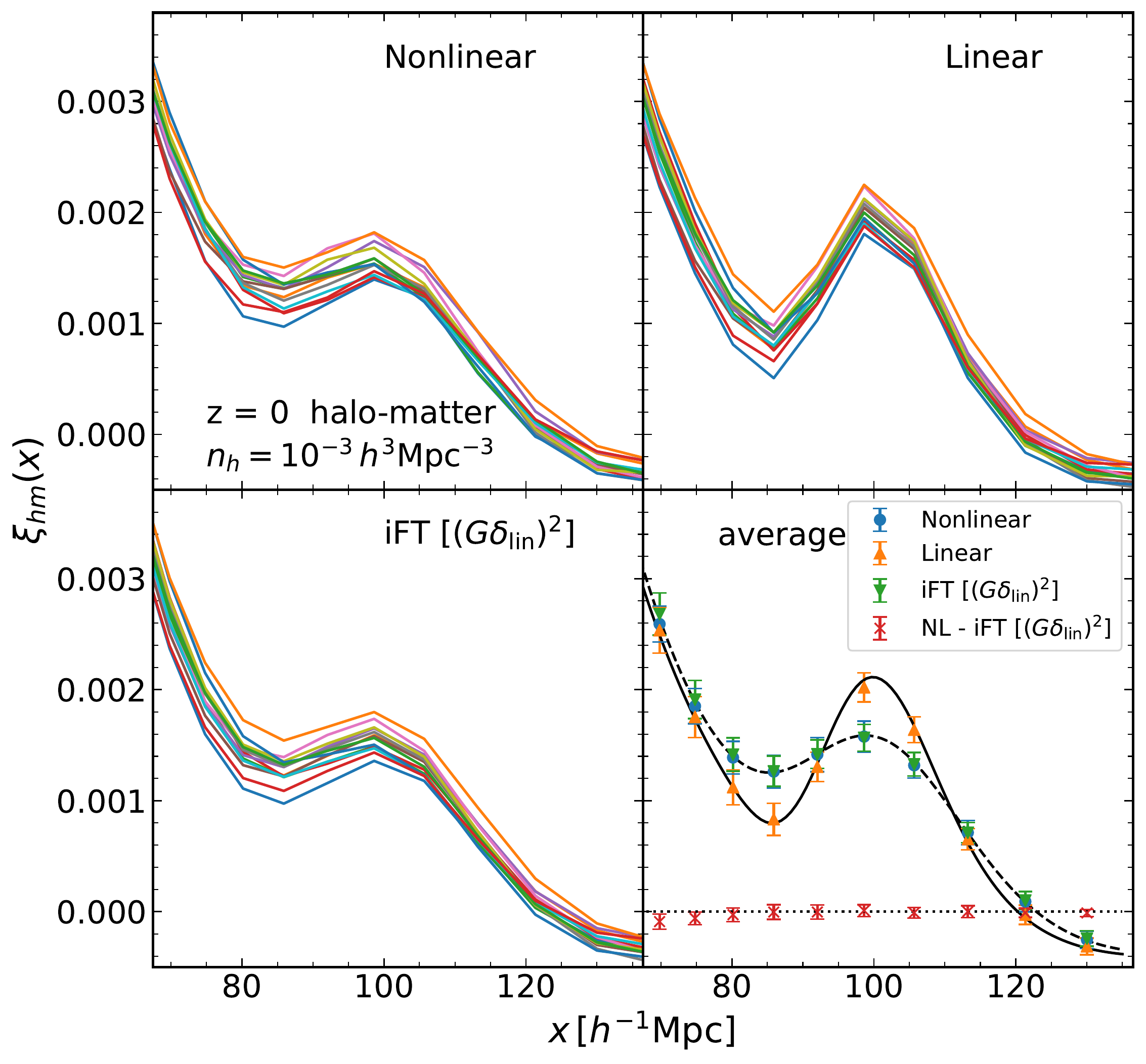}
\includegraphics[height=8cm,angle=0]{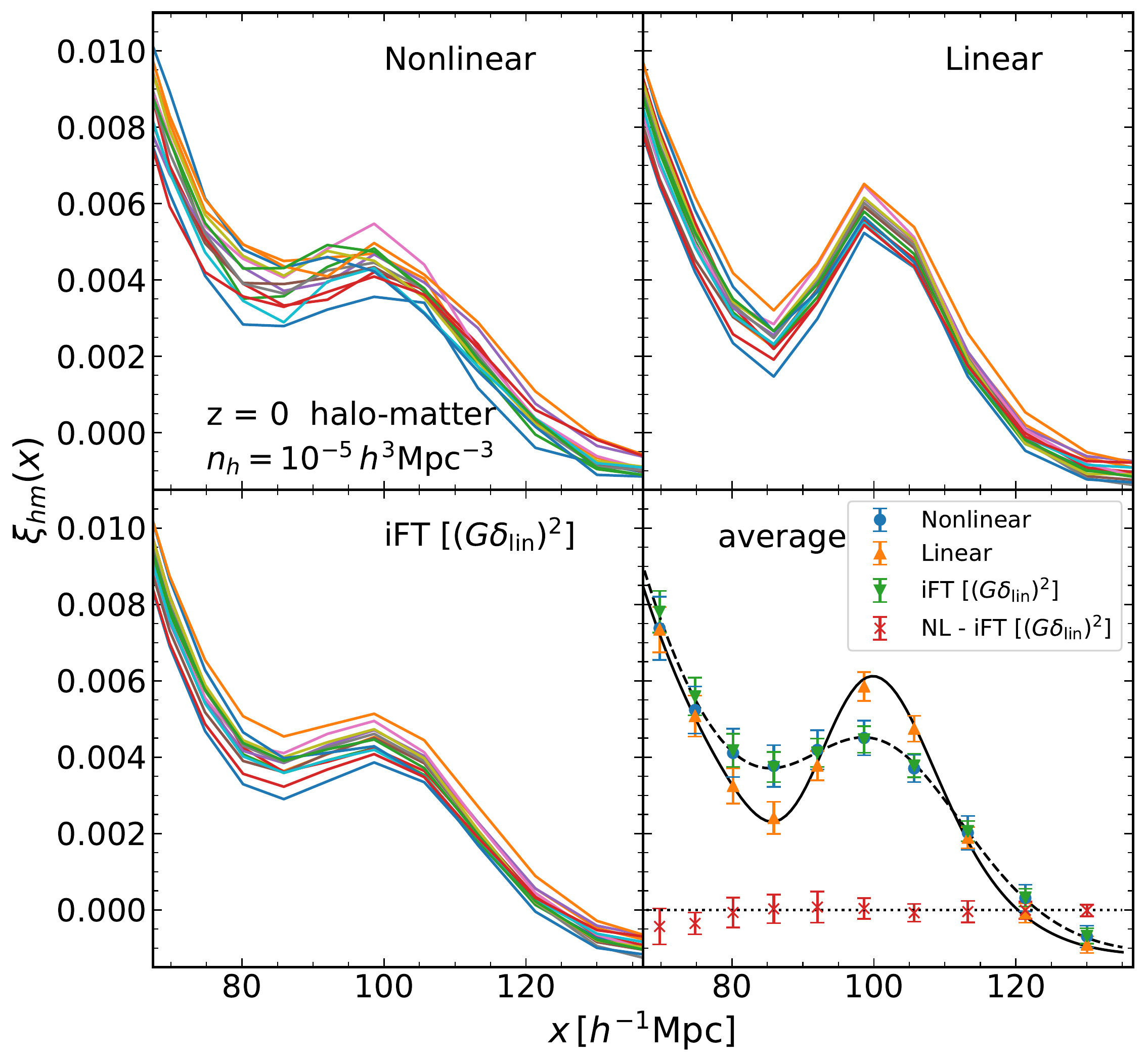}
\end{center}
\caption{
Similar to
Fig.~\ref{fig:BAO_ximm}, but for the halo-matter cross correlation functions. Here we
show the results for halo samples with number density of $10^{-3}$
and $10^{-5}~(h^{-1}{\rm Mpc})^{-3}$ in the left and right panel, respectively.
\label{fig:BAO_xihm}
}
\end{figure*}

\begin{figure*}[t!]
\begin{center}
\includegraphics[height=8cm,angle=0]{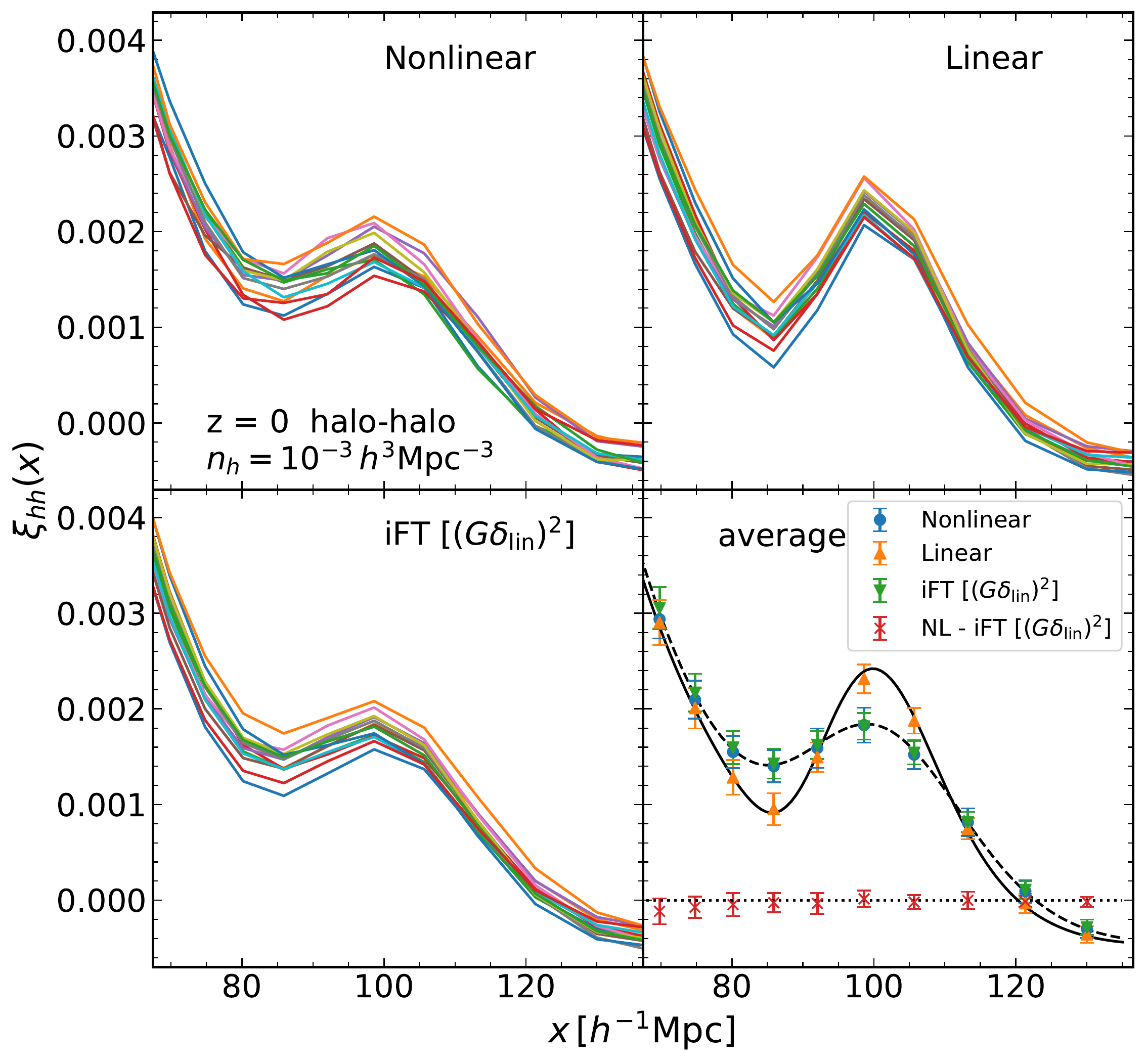}
\includegraphics[height=8cm,angle=0]{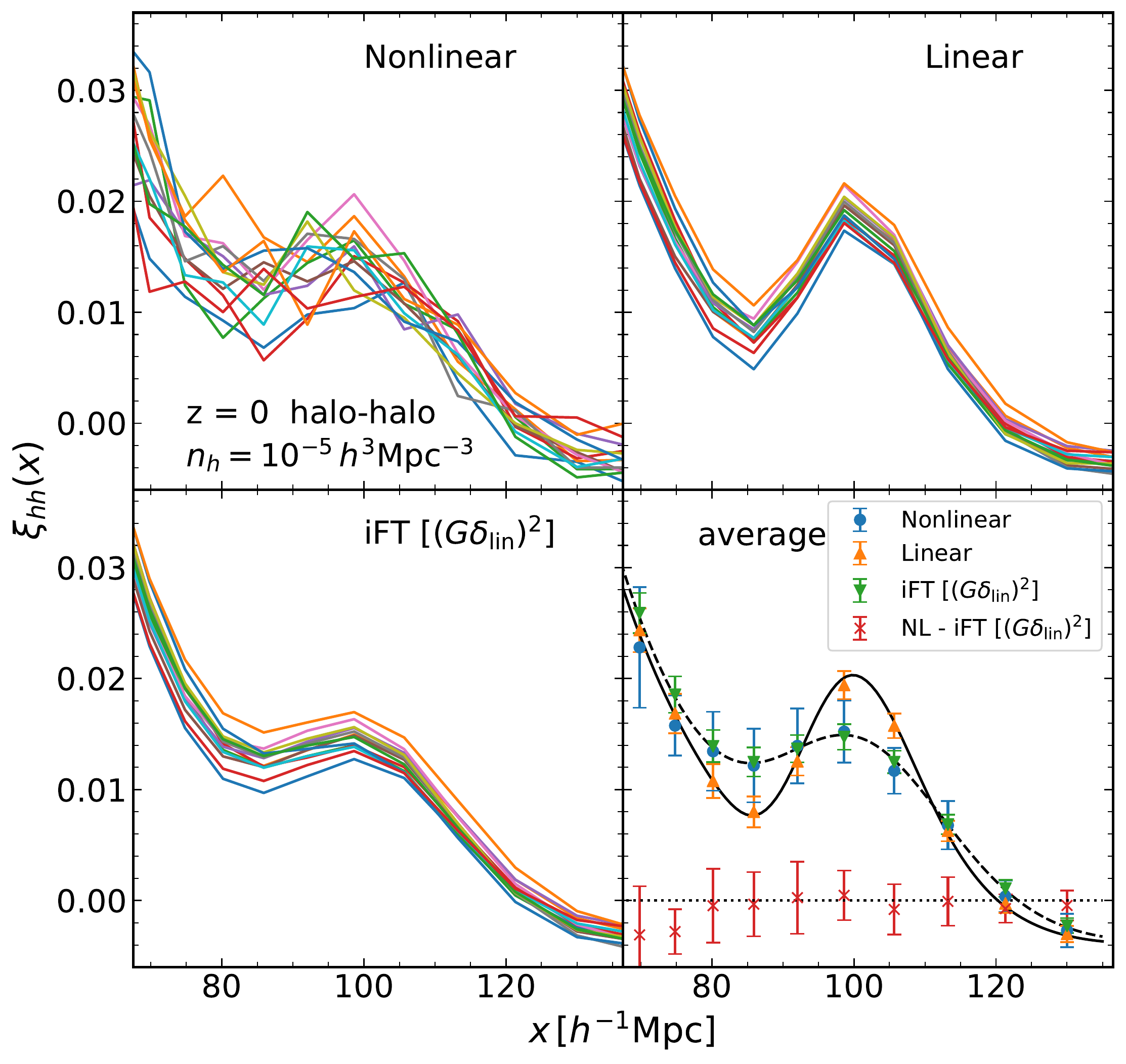}
\end{center}
\caption{Similar to
Fig.~\ref{fig:BAO_xihm}, but for the halo-halo auto correlation functions for the two number-density selected samples as before.
\label{fig:BAO_xihh}
}
\end{figure*}
The clustering correlation functions of halos measured from each of the simulations become considerably noisy on very large scales, around the baryon acoustic oscillation (BAO) scale due to the significant
sample variance due to the finite simulation volume even for our LR simulations of $2~h^{-1}{\rm Gpc}$ size.
To overcome this obstacle, we employ a semi-analytical approach based on the \textit{propagator}~\citep[e.g.,][]{crocce:2006uq},
which captures most of the expected linear and nonlinear effects around the BAO scale.
We then stitch the prediction with the direct simulation results to obtain model predictions over a wide
range of scales, as described below.

Figs.~\ref{fig:BAO_ximm}, \ref{fig:BAO_xihm}, and \ref{fig:BAO_xihh} show
the matter auto, halo-matter cross, and halo auto correlation functions on large scales, respectively.
The solid curves in the upper-left panel of each figure depict the correlation functions measured
from each of the $14$ LR realizations for the fiducial {\it Planck} cosmology.
Clearly, the realization-to-realization scatter is large.
For comparison, the upper-right panels show the linear-theory predictions which we computed using the same Gaussian random realizations as in the initial conditions of each simulation
in order to properly take into account the sample variance effect
(see Sec.~\ref{subsubsec:gamma1} for our method to determine the linear bias parameter from the halo correlation functions).
The scatter among the realizations
seen in the linear predictions are comparable to the corresponding nonlinear counterparts, except for the halo-halo auto correlation function with low number density
of $10^{-5}~(h^{-1}{\rm Mpc})^{-3}$ (i.e., the right panels of Fig.~\ref{fig:BAO_xihh}).
This suggests that the primary source of the scatters is indeed the sample variance in the initial conditions, and the shot noise adds only a moderate scatter for low-density samples of halos.

Since our varied cosmology simulation suite is in principle performed only once at each model, the large scatters in the measured correlation function make it difficult to construct an accurate emulator. Unlike the matter power spectrum, we cannot switch to a parameter-free perturbative calculation on these large scales because we have to know the halo bias that is not accurately described by a simple, analytical prescription which often ignores
the dependence on scale and cosmology.
We need an appropriate method where the sample variance is
sufficiently reduced and at the same time the large-scale bias of halos under consideration is properly taken into account.

Another important effect on the BAO scale, in addition to the large-scale bias, is the damping of the BAO feature. This is clearly visible from comparison of the upper-left and -right panels in Figs.~\ref{fig:BAO_ximm}, \ref{fig:BAO_xihm} and \ref{fig:BAO_xihh}.
It is known that this effect is to a large extent due to the large-scale bulk motion of the cosmic fluid which can be accurately modeled by the \textit{propagator} \citep{crocce:2006uq}. In their paper the propagator for the matter field is defined by
\begin{eqnarray}
\left\langle\frac{\partial \Phi_{a,\mathbf{k}}}{\partial\delta_{\mathrm{lin},\mathbf{k}'}}\right\rangle \equiv \delta_\mathrm{D}^3(\mathbf{k}-\mathbf{k}')G_a(k)\, ,
\label{eq:def_propagator}
\end{eqnarray}
where $\Phi_a$ can be either the density or the velocity divergence field of matter. Note here and in what follows that the linear density field $\delta_\mathrm{lin}$ and its power spectrum $P_\mathrm{lin}$ are always scaled by the linear growth factor to the same redshift as other quantities such as $\Phi_a$ or $G_a$. The function $G_a(k)$ is called the (two-point) propagator, which shows a damping form very close to a Gaussian shape toward high $k$. This function can be interpreted to describe how much memory of the initial density field
($\delta_{\mathrm{lin}}$) persists in the final (nonlinear) fields ($\Phi_a$).
One can analytically show that this function is exactly a Gaussian with its variance equal to the inverse square of the rms displacement field in case of the Zel'dovich dynamics for a Gaussian initial condition.

In most of
resummed perturbation theories, the leading order contribution to the mixed power spectrum of two fields $\delta_a$ and $\delta_b$ is expressed as $G_a(k)G_b(k)P_\mathrm{lin}(k)$, where the subscripts $a$ and $b$ can be the density or the velocity divergence of matter or any tracers \citep[e.g.,][]{crocce06b,crocce08,bernardeau:2008lr} and $P_\mathrm{lin}(k)$ is the linear matter
power spectrum.
Inverse Fourier Transform (iFT)
of this combination gives a reasonable prescription on the two-point correlation function around BAO scale:
\begin{eqnarray}
\xi_{a,b,\mathrm{tree}}(r) = \mathrm{iFT}\left[G_a(k)G_b(k)P_\mathrm{lin}(k)\right],\label{eq:xi_tree}
\end{eqnarray}
where we put the subscript \textit{tree} to indicate that this quantity is the tree-level result (i.e., the leading-order diagrams) of the resummed perturbation theories.
Indeed Eq.~(\ref{eq:xi_tree}) with a simple Gaussian approximation of the propagator can already explain the damping of BAO peak in the matter correlation function very accurately \citep[e.g.,][]{matsubara08a}. By going into higher orders, a sub-percent-level shift in the BAO peak location to a smaller separation scale can be realized \citep{crocce08}. This would be important in interpreting the BAO-related distance
measurements
from actual observations.

\begin{figure*}[t!]
\begin{center}
\includegraphics[width=15cm,angle=0]{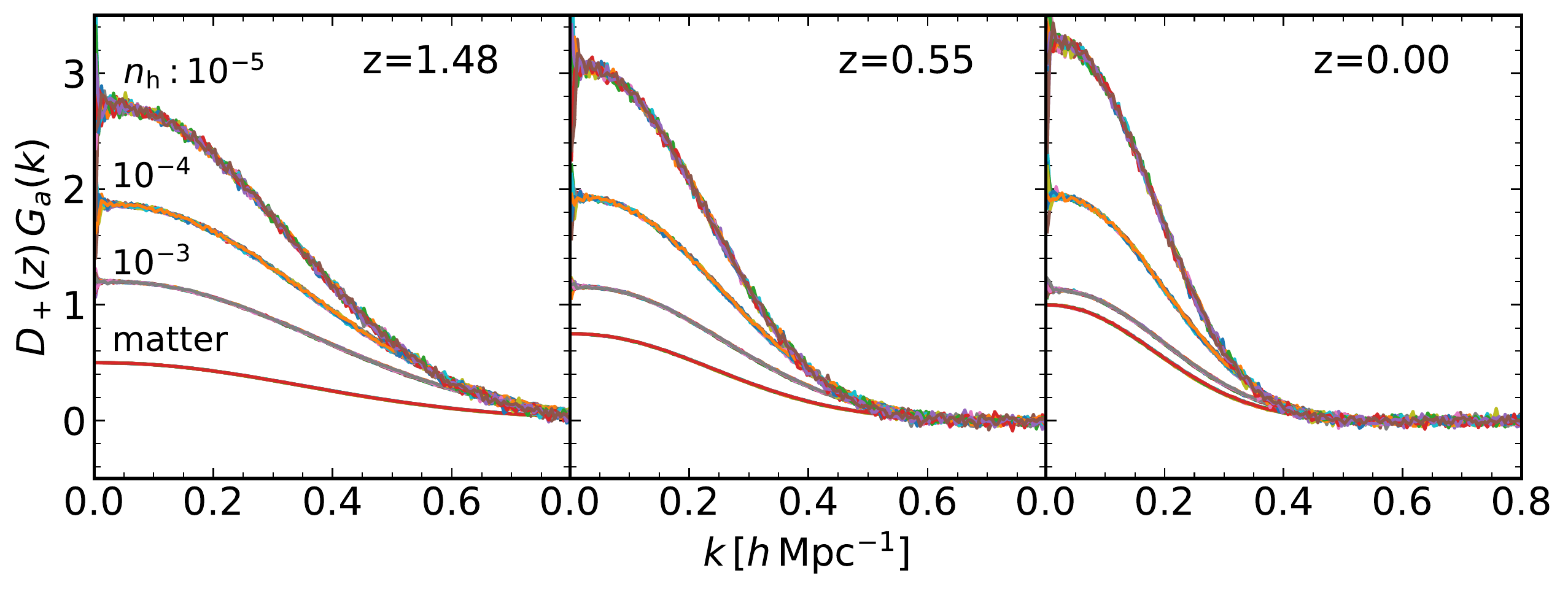}
\end{center}
\caption{Propagator for the matter and halo density fields (see Eq.~\ref{eq:def_propagator} for definition of the propagator).
We here consider the halo samples with number densities $n_{\rm h}=10^{-5}$, $10^{-4}$, and $10^{-3}\,(h^{-1}{\rm Mpc})^{-3}$, and show the results at redshifts $z=1.48$,
$0.55$ and $0$ in the left, middle, and right panels, respectively.
We multiply the linear growth factor $D_+(z)$ to reduce the dynamic range.
\label{fig:propagator}
}
\end{figure*}

We now consider the propagator for halos. One can define the propagator by simply replacing $\Phi_a$ with the
density field of halos in a given sample. In what follows we denote by $G_a(k)$ with the subscript $a$
either the matter or the halo density field. In case of halos the low-$k$ limit
of the function corresponds to the linear bias factor. A damping behavior at high $k$ should be
similar to that of the matter field, and this damping
is responsible for the smearing of BAO peak measured through the clustering of halos.
We show the functions for matter and halos with different number densities and at different redshifts in Fig.~\ref{fig:propagator}, which are measured
from the $14$ LR realizations
of the fiducial {\it Planck} cosmology. We estimate the function by taking
\begin{eqnarray}
G_a(k) = \frac{P_{a,\mathrm{lin}}(k)}{P_{\mathrm{lin}}(k)},
\label{eq:prop_estimator}
\end{eqnarray}
where $P_{a,\mathrm{lin}}(k)$ is the cross power spectrum of tracers
``$a$'' and the linear density field.
In this estimator, we use the linear power spectrum $P_\mathrm{lin}(k)$ measured from the linear density field used for the initial condition instead of the theoretical smooth function, and the sample variance is largely suppressed by taking this ratio.
Indeed, the scatter among
the $14$
realizations seen in the figure is rather small, especially for low number density halo samples compared to the scatter in the corresponding auto correlation function in Fig.~\ref{fig:BAO_xihh}.
The overall trend of this function looks very simple, as already
discussed; it appears to be
a Gaussian-like damping function with a linear bias factor at the low-$k$ limit which depends on the halo number density.
In addition
we can see that the damping starts at smaller $k$ at lower redshifts, reflecting the fact that the information in the initial density field remains more on larger scales and at higher redshifts.

To summarize, our strategy to describe the large-scale limit of matter or halo correlation functions is to emulate the function $G_a(k)$ for both matter and halo fields, and substitute
it into Eq.~(\ref{eq:xi_tree}). Likewise, we take the same combination for the random fields $\delta_\mathrm{lin}$ used in the initial conditions, which we schematically denote as $\mathrm{iFT}[(G\delta_\mathrm{lin})^2]$. We show this model in the lower-left panel of Figs.~\ref{fig:BAO_ximm}, \ref{fig:BAO_xihm}, and \ref{fig:BAO_xihh} for the random realizations corresponding to the $14$ simulations shown in the upper panels.
The curves obtained in this way appear to be very similar to the direct simulation results
in the top-left panels.

Finally, the average of these curves are shown by the downward triangles with errorbars in the lower-right panel.
They are almost indistinguishable from the circle symbols for the nonlinear correlation function directly measured from the nonlinear fields. Indeed, their difference shown by the crosses are consistent with zero.
A closer look at the scale dependence of this residual
indicates a small pattern that would cause a small shift on the BAO peak toward a smaller scale. It tends to be positive around the inflection point of the correlation function (at around $90\,h^{-1}\mathrm{Mpc}$), and negative at scales smaller than $80\,h^{-1}\mathrm{Mpc}$. Since in most cases these features are within the errorbars, which correspond to the scatter among realizations, we simply ignore this small residual in the following discussion.

We also show the continuous limit of the model, Eq.~(\ref{eq:xi_tree}), by the dashed line. This is the expectation value of the downward triangles in the limit of an infinite number of realizations. Our final model for the large-scale correlation function is this line. With this procedure, we can reduce the sample variance significantly since the prediction is based on the noiseless linear power spectrum $P_\mathrm{lin}$.
Our approach works well even in the case of halo auto correlation function for a halo sample with small number
density (see the upper-left panel in the right part of Fig.~\ref{fig:BAO_xihh}); the unaccounted shot-noise effect
only adds a random scatter and no systematic trend can be seen in the residual. We will explain how we switch from
the direct measurement of the correlation functions to the prescription based on the propagator explained here
in \ref{subsubsec:halos_implementation}.

\subsection{Implementation detail and performance}
\label{subsec:implementation}
We have so far described building blocks of our \textsc{Dark Emulators}.
Below we describe how to model their cosmological dependences.

Our basic strategy for emulator development is as follows.
First we build a data vector for each of the four main halo functions (halo mass function, halo-matter cross
correlation, halo-halo auto correlation, and the propagator) including their dependence on redshift, separation,
and the number density, which can be translated into the halo mass threshold, from simulation realizations of each cosmological model.
Second, we apply Principle Component Analysis (PCA) to the data vector, which allows
for a huge dimensionality reduction of the data vector by keeping only a handful of most significant principle
component (PC) coefficients \citep[also see][for the similar method for the matter power spectrum]{Coyote3}.
In doing so, we use a public PCA package, \textsc{empca}~\citep{empca}, which allows us to introduce a weighting
to the input data vector. An advantage of this weighting method is that we can put a zero weight to
missing data.
Third, we apply Gaussian Process (GP) regression to the significant PC coefficients for different cosmological
models in order to have a quick GP interpolation of the model prediction of each of the halo functions in
an arbitrary cosmological model.
As for the GP regression, we use a public code, \textsc{george}~\citep{george}.
We adopt a stationary kernel function with either \textsc{ExpSquared}, \textsc{Exp}, \textsc{Matern32}, or
\textsc{Matern52}, and pick one for each PC coefficients based on the likelihood to explain the data after
optimization.

In building the emulator, we use multiple realizations for the fiducial {\it Planck} cosmology to estimate errors
in the PC coefficients.
Assuming that the errors are independent of cosmology, we add the errors in square into the diagonal components of
the GP kernel function.
Unless otherwise stated, we use $80$ simulations in Slice~1 to 4 from either in
the \textsc{HR} or the \textsc{LR} suite.
The remaining $20$ models in Slice~5 as well as the fiducial {\it Planck} model are used for a cross validation of
the emulator outputs.

We describe details of the actual implementation of the four main halo modules in the following subsections. The
connection to the galaxy statistics will be explained in the subsequent section.

\subsubsection{Halo mass function}
In this section we describe how to build a module of the halo mass function.
As shown in the Figs.~\ref{fig:MF_resolution1} and \ref{fig:MF_resolution2}, the fitting formula by
\citet{Tinker08} works very well at least
for the fiducial cosmology at $z=0$. The fitting function we use in the following
is a modified version of the earlier model in \citet{press74} \citep[also see][]{sheth02}, given by
\begin{eqnarray}
&&\frac{\mathrm{d}n}{\mathrm{d}M} = f(\sigma_M)\frac{\bar{\rho}_\mathrm{m}}{M}\frac{\mathrm{d}\ln \sigma_M^{-1}}{\mathrm{d}M},
\end{eqnarray}
with
\begin{eqnarray}
&&f(\sigma_M) = A\left[\left(\frac{\sigma_M}{b}\right)^a + 1\right]\exp\left(-\frac{c}{\sigma_M^2}\right).
\label{eq:Tinker_HMF_form}
\end{eqnarray}
Here the mass variance $\sigma_M^2$ is given by
\begin{eqnarray}
\sigma_M^2 = \int\!\!\frac{k^2\mathrm{d}k}{2\pi^2} P_{\rm lin}(k;z)\left|\tilde{W}_R(k)\right|^2,
\end{eqnarray}
where $P_{\rm lin}(k;z)$ is the linear matter power spectrum at redshift $z$, and
$\tilde{W}_R(k)$ is the Fourier transform of a top-hat filter of radius $R$ that is specified by an input halo mass
$M$ via $R = (3M/4\pi\bar{\rho}_{\mathrm{m},0})^{1/3}$.
\citet{Tinker08} showed that HMF measured in simulations is well fitted by the above functional form with
time-dependent coefficients:
\begin{eqnarray}
&&A(z) = 0.186\,(1+z)^{-0.14},\label{eq:Tink_bigA}\\
&&a(z) = 1.47\,(1+z)^{-0.06},\label{eq:Tink_smalla}\\
&&b(z) = 2.57\,(1+z)^{-\alpha},\label{eq:Tink_b}\\
&&c(z) = 1.19,\label{eq:Tink_c}\\
&&\alpha = -\left(\frac{0.75}{\log_{10}(\Delta/75)}\right)^{1.2}.
\end{eqnarray}
The overdensity $\Delta$ is $200$ in our halo mass definition.

\begin{figure*}[t!]
\begin{center}
\includegraphics[width=18cm,angle=0]{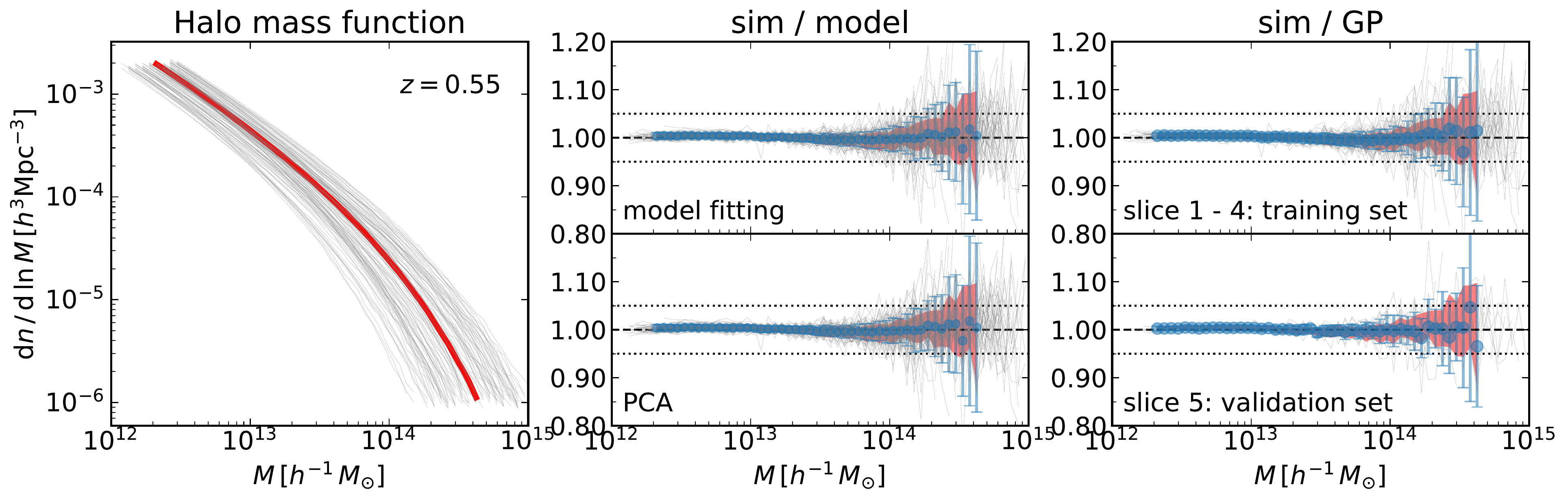}
\end{center}
\vspace*{-2em}
\caption{Modeling of the halo mass function (HMF).
{\it Left panel}: Variations in HMF at $z=0.55$, which are measured from each simulation of 100 cosmological
models in HR simulation suite (each simulation has a 1~$h^{-1}{\rm Gpc}$ on side).
The red curve shows HMF for the fiducial {\it Planck} cosmology.
{\it Middle-upper panel}: We model the HMF in each simulation by a functional form of \citet{Tinker08} (Eq.~\ref{eq:Tinker_HMF_form}), where we
estimated best-fitting parameters of $A$ and $a$ to the simulated HMF, but used the same $b$ and $c$ in Eqs.~(\ref{eq:Tink_b}) and (\ref{eq:Tink_c}). Each gray
curve is the ratio of the simulated HMF to the best-fit Tinker HMF for each of 100 cosmological models. The point and errorbar at each mass bin, in this and following
panels, denote the mean and scatter of the ratios at the mass bin.
The shaded region in this and other plots
denotes statistical uncertainties in HMF that are estimated from scatters of HMFs in the 28 realizations of {\it Planck} cosmology.
The horizontal dotted lines denote $\pm 5$ percent in the fractional difference.
{\it Middle-lower panel}: To model the redshift and mass dependence of HMF in each cosmological model,
we performed the principal component analysis (PCA) to the best-fitting Tinker parameters, $A$ and $a$, at each of 21 output redshifts over the range $0<z<1.48$;
hence 42 data points in each cosmological model (see text for details). The plot shows that keeping the six most significant PC coefficients
gives almost identical accuracy as compared to the results after the model fitting (the upper panel). The loss of accuracy induced in this procedure is less than one percent in all the cases.
 {\it Right-upper panel}: We performed the Gaussian process (GP) regression to the PC coefficients at
80 sampling points in Slice~1 to 4 in 6-dimensional cosmological parameter space.
{\it Right-lower panel}: Validation test of the GP interpolation, i.e. our HMF emulator module, showing
how the GP interpolation can reproduce the simulated HMF in each of 20 cosmological models in Slice~5, which are not used in the GP regression.\label{fig:HMF_model}
}
\end{figure*}
\begin{figure}[t!]
\begin{center}
\includegraphics[width=8cm,angle=0]{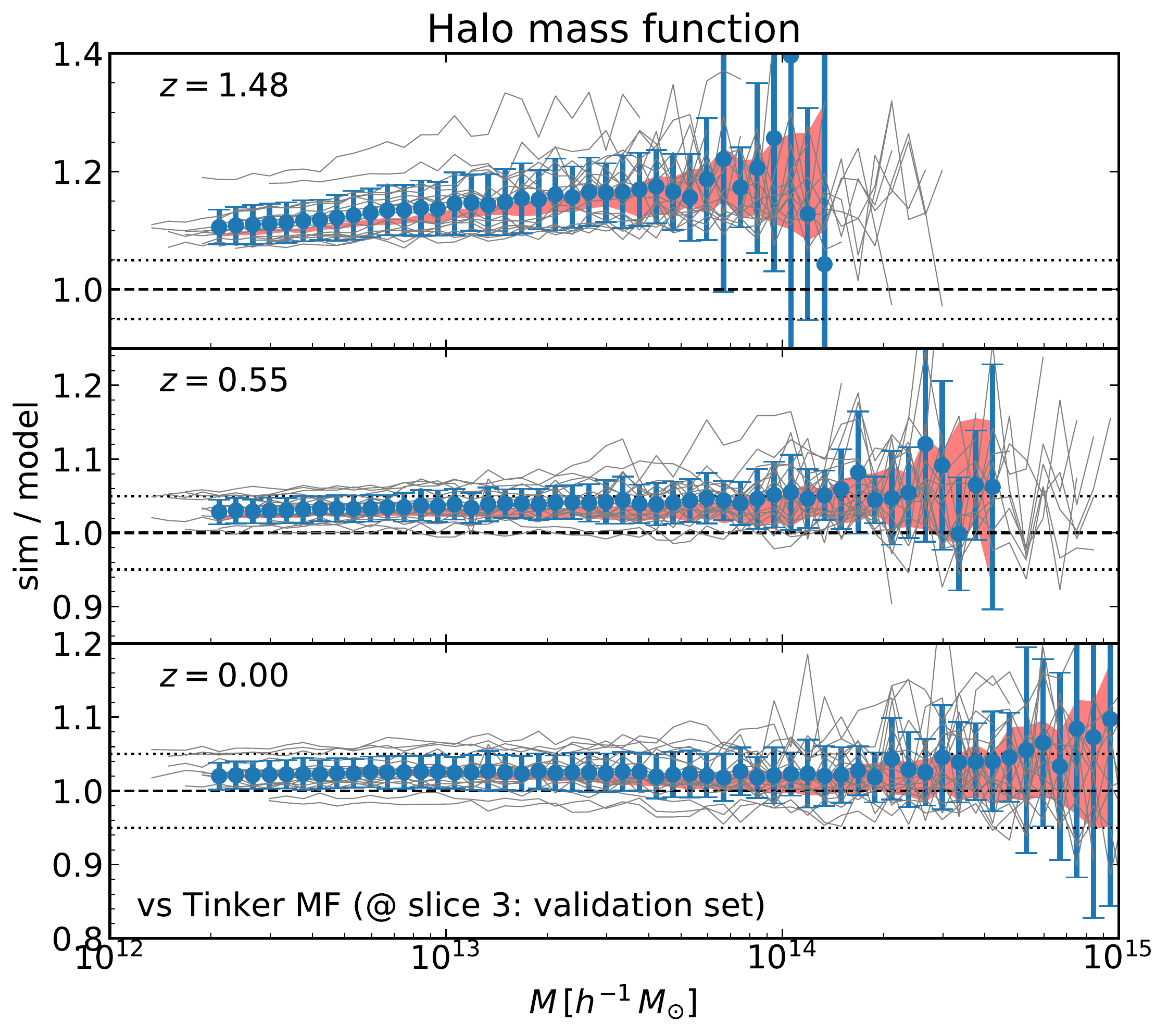}
\end{center}
\caption{Comparison of our simulation HMF with the original Tinker fitting formula \citep{Tinker08}.
\label{fig:HMF_Tinker}
}
\end{figure}

The variation in the HMF over our $100$ cosmological models can be found in the left panel of Fig.~\ref{fig:HMF_model} (gray curves). We employ the HR suite here and choose to show the HMF at $z=0.55$ as a typical redshift of the CMASS galaxies.
We also show by the red curve the HMF for the fiducial {\it Planck} cosmology (the mean of 28 realizations).
The 100 models are taken from the 5 \textsc{SLHD} slices (hereafter Slice~1, 2, \dots, 5) each of which consists of
20 different cosmological models as described in Section~\ref{subsec:LHC}. The variation in the HMF is quite large,
and it is not so obvious whether or not the universal form of Eq.~(\ref{eq:Tinker_HMF_form}) can explain it.

Before constructing an HMF module, we first test the accuracy of
Tinker HMF against our simulation suite (\textsc{HR} runs).
Fig.~\ref{fig:HMF_Tinker} compares the simulated HMF with the original Tinker HMF prediction (using the coefficients in Eqs.~\ref{eq:Tink_bigA}--\ref{eq:Tink_c})
for each of 20 cosmological models in Slice~5, at three redshift bins, $z=0, 0.55$, and $1.48$, respectively. The comparison indicates a larger deviation of Tinker formula from the simulation results as redshift increases. At $z=0$, the ratio of our HR simulation suite to the Tinker MF shows typically $5\%$ scatter from the Tinker MF, with the mean slightly biased from unity (by $\sim 2$ to $3\%$). The overall slope of the HMF is already very well captured by this formula, and the error is mostly on the amplitude.
At higher redshifts, both the bias and the scatter grow. At $z=1.48$, the slope of the ratio shows a clear mass
dependence with a larger bias toward the massive end.
We leave further discussion on the inaccuracy of the original Tinker HMF formula to Appendix~\ref{sec:IC}, where we
discuss that this discrepancy is mainly due to the fact that the Tinker
HMF was calibrated against simulations using initial conditions based on the Zel'dovich approximation at an initial
redshift that is not high enough.

From this exercise, we decide to update some of the parameters in the Tinker HMF. We drop the assumption of the HMF
universality and allow the parameters $A$ and $a$ to vary as a function of cosmological models. For the parameters
$b$ and $c$, we keep the values given by Eqs.~(\ref{eq:Tink_b}) and (\ref{eq:Tink_c}), where $b$ determines the
slope of HMF at low mass end and
$c$ determines the cutoff at high mass end. Our HR suite is most accurate over the intermediate range of halo
masses, where the overall amplitude and the slope are controlled by the parameters $A$ and $a$, respectively.
Therefore, we recalibrate these parameters for each of our simulations.

The middle-upper panel of Fig.~\ref{fig:HMF_model} addresses how the functional form
given by Eq.~(\ref{eq:Tinker_HMF_form}) can fit the simulated HMF for each of 100 cosmological models with the
updated parameters. Here we estimated the best-fitting parameters of $a$ and $A$ that reproduce the simulated HMF.
In the fitting, we consider two sources of errors. For high mass bins, we consider statistical uncertainties due to
the Poisson noise of the number counts.
We also include a phenomenological penalty term at low mass bins, where the correction
(Eq.~\ref{eq:mass_correction2}) plays a significant role.
That is, in the model fitting we include the following statistical errors in the number counts of halos at
each mass bin:
\begin{eqnarray}
\frac{\Delta N_\mathrm{h}}{N_\mathrm{h}} = \frac{1}{\sqrt{N_\mathrm{h}}} + \frac{1}{N_\mathrm{p}},
\end{eqnarray}
where $N_\mathrm{h}$ is the number of halos at each mass bin,
and $N_\mathrm{p}$ the number of $N$-body member particles, $M/m_\mathrm{p}$, at the logarithmic bin center.
Moreover, we do not include any mass bin of halos which are defined by $N_\mathrm{p}<200$.
The figure shows that our model HMF generally gives a very good fit to simulated HMF for each of the 100
cosmological models, where the ratio is very close to unity well within $\pm 5\%$ accuracy
denoted by the horizontal dotted lines. At high mass end, the simulation data points are dominated by Poisson noise
due to too small number of halos per bin.
The circle point and error bar at each mass bin are the average and scatter in the ratios of the best-fit HMF to the simulated HMF for the 100 cosmological models.
For comparison, the red-color shaded region denotes scatters among the 28 realizations of {\it Planck} cosmology,
giving an estimate of the sample variance for volume of 1~$(h^{-1}{\rm Gpc})^3$.
The typical accuracy of the model as indicated by the error bars are $\sim 1$ ($3$) $\%$ at $10^{13}$ ($10^{14}$) $\,h^{-1}M_\odot$.

We then compress the data vector, $\mathbf{d}=(A_0,a_0,\dots,A_{20},a_{20})$, which consists of the fitting
parameters $A$ and $a$ at each of 21 redshifts (therefore 42 data in total) for
each cosmological model, using PCA. Combining all data vector for 100 cosmological models as well as 28
realizations of the fiducial {\it Planck} cosmology (therefore $128\times 42=5376$ data points in total),
we decompose the data vector for the $i$-th simulation, $\mathbf{d}_i$, into the principal components (PCs) as
\begin{eqnarray}
\mathbf{d}_i = \sum_{j=1}^{n}~ \alpha^{\rm HMF}_{i,j} \mathbf{e}^{\rm HMF}_j,
\label{eq:hmf_pca}
\end{eqnarray}
where $\mathbf{e}^{\rm HMF}_j$ is the $j$-th eigenvector with $42$ components, which is independent of cosmology or
simulation realization, and $\alpha^{\rm HMF}_{i,j}$ is the $j$-th PC coefficient for the $i$-th simulation.
After various checks we find that keeping the six most significant PC coefficients for each cosmological model,
corresponding to $n=6$ in Eq.~(\ref{eq:hmf_pca}),
is sufficient to keep the error induced in this step to a sub-percent level.
The accuracy level after applying the PCA method
is not degraded to an extent easily visible by eye when we compare the middle-lower and the middle-upper panels of Fig.~\ref{fig:HMF_model}.
In doing the PCA analysis, we downweight the components $A(z)$ by a factor of ten compared with $a(z)$ to
compensate their different dynamic ranges.

\begin{figure}[t!]
\begin{center}
\includegraphics[width=8cm,angle=0]{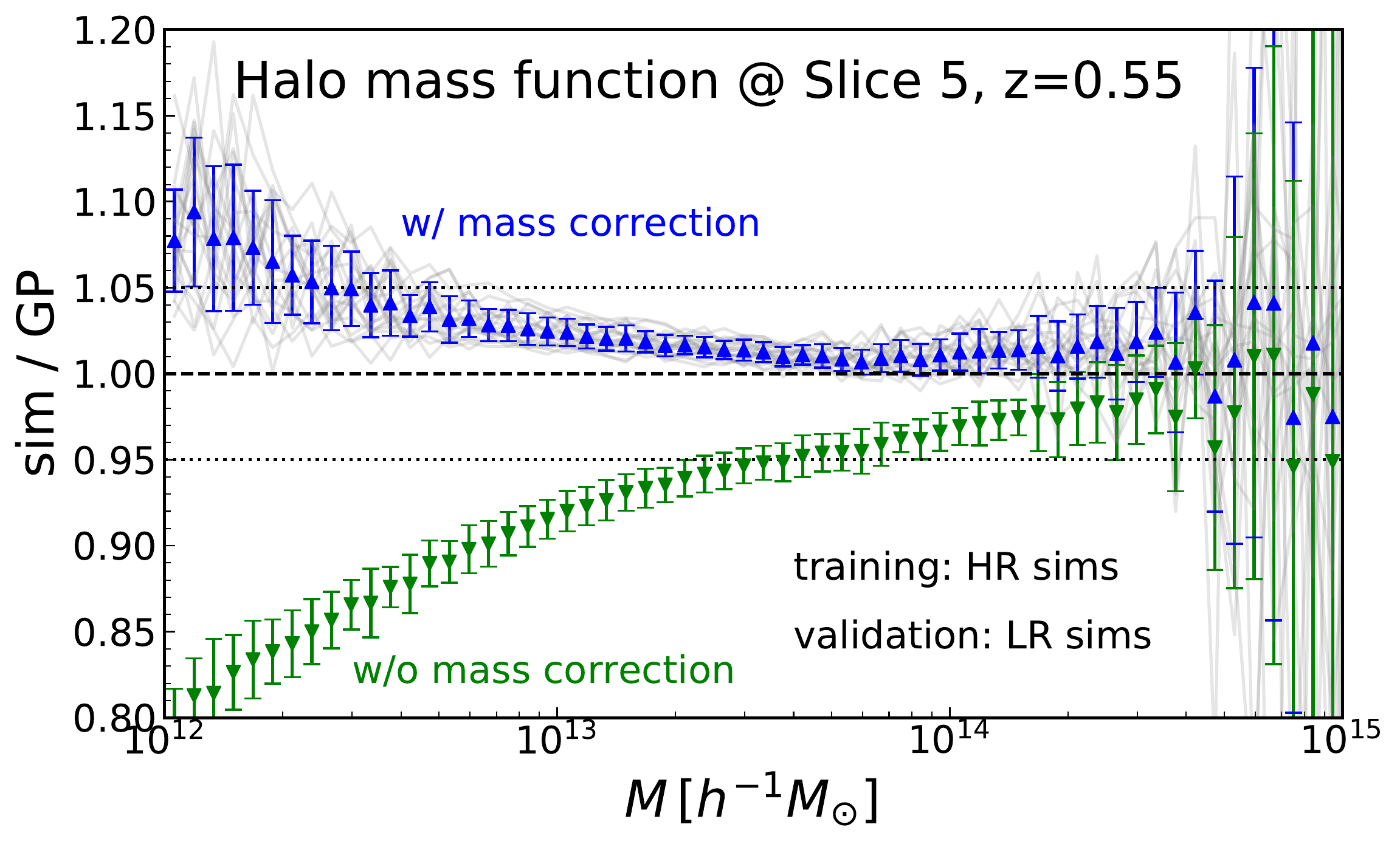}
\end{center}
\caption{Comparison of the emulator prediction against the simulations in LR suite. We consider the 20 cosmologies in Slice~5 at $z=0.55$, and show the ratio of the measurements from the LR
suite to the emulator predictions. We show the results both with (upper) and without (lower) the mass correction (\ref{eq:mass_correction2}). Similarly to Fig.~\ref{fig:HMF_model}, we show the mean and scatter among the models by the error bars and the individual cosmologies are shown by the gray solid curves.
\label{fig:massfunc_lowres}
}
\end{figure}
Our next task is to collect the six PC coefficients ($\alpha^{\rm HMF}_{i,j}$ in Eq.~\ref{eq:hmf_pca})
for different cosmological models, and then to perform GP regression to interpolate each of the six PC coefficients
between the sampled cosmological models each of which is located at a particular position in 6-dimensional
cosmological parameter space. To do this, we apply the GP regression to the PC coefficients
for the 80 cosmological models included in Slice~1 -- 4 as the training set, excluding the 20 cosmological models
in Slice~5 (see Section~\ref{subsec:LHC} for details). Note that we do not include the fiducial {\it Planck}
cosmology in this GP regression either.
The right-upper panel
of Fig.~\ref{fig:HMF_model}
compares the GP regression HMF with the simulated HMF for each of the 80 cosmological models in the training set.
The GP does not perfectly reproduce the results of PCA at each of sampled cosmological models because we take into account the statistical uncertainties
of the training data in the regression. Nevertheless the plot shows that after applying the GP
regression, the rms in the ratio among different cosmological models are kept below $\sim 1$ ($3$) $\%$ on $M \lesssim 10^{13}$ ($10^{14}$) $\,h^{-1}M_\odot$.

The right-lower panel of Fig.~\ref{fig:HMF_model} is the most important plot that
gives an assessment of the performance of HMF emulator for an arbitrary cosmological model.
The plot compares the GP interpolated
HMF with the simulated HMF for each of the 20 cosmological models in Slice~5 that are not used in the GP regression
and serve as a cross validation sample.
The HMF emulator achieves a great accuracy to predict the HMF, better than a few percent in the amplitude up to
halo masses of a few times $10^{14}~h^{-1}M_\odot$. The
performance is degraded for more massive halos, but the inaccuracy (averaged value denoted by the circle at each bin) is comparable with the statistical scatter.
The good performance suggests that our GP method does not suffer from an overfitting to the training set.

While the performance of the emulator can be assessed fairly precisely for halo masses up to $\sim 10^{14}\,h^{-1}M_\odot$, the scatter among the models appears to be large for cluster sized halos. While the large scatter could be simply due to the inaccuracy of the emulator, it can be partly due to the large Poisson noise which can affect significantly the measurements used as the reference due to the small number of available
cluster-scale halos in the simulations. As a final check of the accuracy of the emulator at the high-mass end, we compare the emulator prediction to the measurement from the LR simulations, which have bigger volume and thus less affected by the Poisson noise.

In Fig.~\ref{fig:massfunc_lowres}, we show the ratio of the mass function measured from these LR
simulations to the emulator
prediction, which is trained based on the HR simulation suite. We show the mean and the scatter among the 20 cosmological models in Slice~5, with individual cosmology result (gray solid curve).
We apply the correction of the mass based on the number of member particles (Eq.~\ref{eq:mass_correction2}) for the upper set of curves, while we do not apply this to the lower set of curves. Since the size of the correction is pretty large for mass less than several times $10^{13}\,h^{-1}M_\odot$ for these set of simulations, we cannot derive a clear conclusion for these masses given the empirical nature of the correction. For more massive halos, the simulation results after the correction are very close to the emulator prediction. The scatter among the models is much smaller than what we can see in Fig.~\ref{fig:HMF_model}, suggesting that the large scatter in the previous figure is indeed mainly due to the large Poisson noise in the reference simulations. However, a closer look at Fig.~\ref{fig:massfunc_lowres} reveals that the mean of the ratio among the cosmological models for cluster-size halos are systematically above unity by $\sim 1$ to $2\%$. This would be a slight inaccuracy of the emulator due to the use of the HR simulation suite with a smaller volume. Since the size of this systematic error is comparable to those discussed for less massive halos based on the HR suite, we do not further consider the possibility to combine the measurements from the LR and HR simulation suites, and instead stick to the emulator build based only on the HR suite for the halo mass function.

\subsubsection{Halo-matter cross-correlation function}
\label{subsubsec:xihm}
We now discuss the halo-matter cross correlation function. We first measure the cross correlations for
13 mass threshold halo samples with different number densities
at each of 21 redshifts from
each simulation run, where we define the halo samples in 13 logarithmically-spaced bins in the range of $n_\mathrm{h}=[10^{-8.5},10^{-2.5}]\,(h^{-1}\mathrm{Mpc})^{-3}$ (i.e., 2 bins per decade).
For each measurement, we have $140$ separation bins ($40$ logarithmic bins from $0.01$ to $5\,h^{-1}\mathrm{Mpc}$
for the direct pair counting method and $100$ linear bins from $5$ to $500\,h^{-1}\mathrm{Mpc}$ for the FFT method;
see Section~\ref{subsec:measurement} for details).
We thus have $13\times21\times140 = 38,220$ data points per simulation.

\begin{figure*}[t!]
\begin{center}
\includegraphics[width=18cm,angle=0]{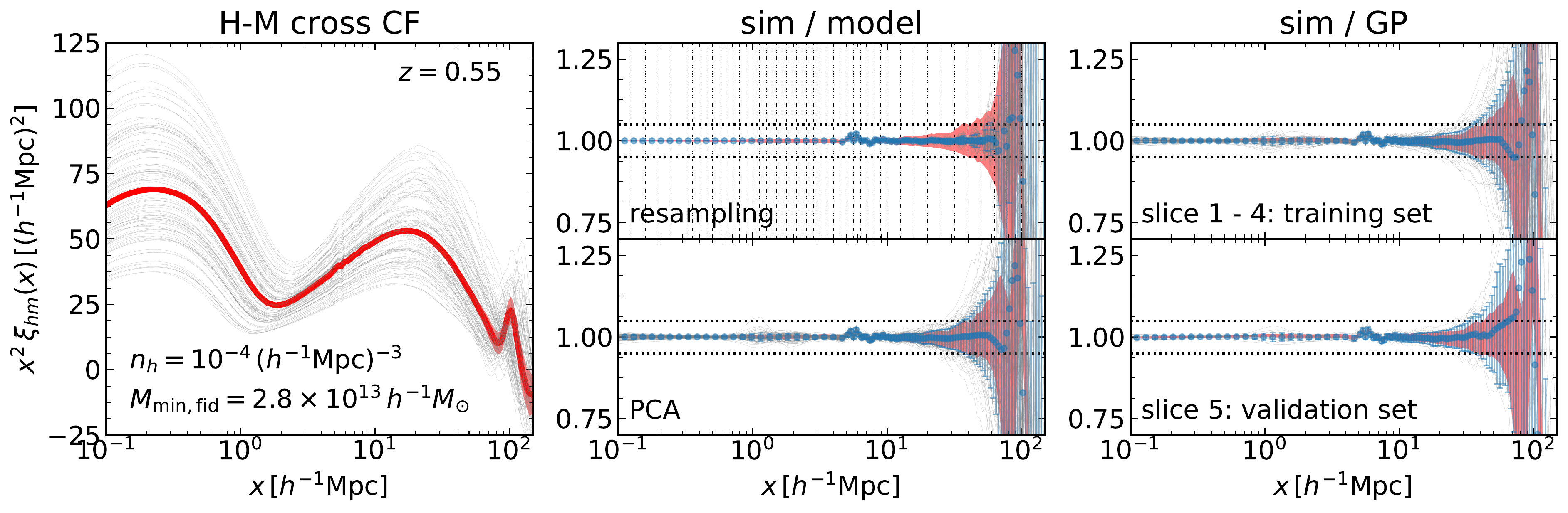}
\end{center}
\caption{Modeling of halo-matter cross correlation functions similarly to Fig.~\ref{fig:HMF_model}. Here we consider the halo sample at $z=0.55$ and with
number density $n_{\rm h}=10^{-4}\,(h^{-1}{\rm Mpc})^{-3}$, corresponding to
the mass threshold of $M\ge 2.8\times 10^{13}\,h^{-1}M_\odot$ for the {\it Planck} cosmology, as an example.
In the middle-upper panel we employ a resampling of separation bins, where
the resampling points are denoted by the vertical dashed lines, and then model the cross correlations by a cubic spline interpolation. The plot shows
the interpolation results compared to the correlations directly measured from 100 simulations (see text for details). The middle-lower panel shows the results when
we model the cross correlations using the PCA analysis for data vector including cross correlations in separation bins, halo sample bins and 21 redshift bins (18,018 data points). Here we show the results obtained by using the five most significant PC coefficients (therefore a huge dimension reduction from 18,018 to 5).
The right panels show the emulator predictions for the
training (upper) and validation (lower) cosmologies that are obtained
after applying the GP regression to the PC coefficients for the training simulations at 80 cosmological models. We show, by the red shaded region in the panels, statistical
uncertainties that are estimated from scatters in the 28 realizations of  the fiducial {\it Planck} cosmology.
\label{fig:cross_model}
}
\end{figure*}

The left panel of Fig.~\ref{fig:cross_model} shows variations in the halo-matter cross correlations for 101
different cosmological models in HR simulation suite,
for the mass threshold halo sample
with the number density of $10^{-4}~(h^{-1}{\rm Mpc})^{-3}$ and at $z=0.55$.
This halo sample is chosen to roughly mimic the number density and the large-scale bias of the CMASS galaxies.
The plot displays
rich cosmological dependences over scales ranging through the one-halo, two-halo terms to BAO scales.
We then reduce the dimensionality of the data vector by first re-sampling the separation bins and then applying
a PCA. The former is done using a cubic spline interpolation of the original data points up to
$100\,h^{-1}\mathrm{Mpc}$ with more data points around the one- and two-halo transition scale
(i.e., around a Mpc scale) as depicted by the vertical dotted line in the second panel of
Fig.~\ref{fig:cross_model}. We take $66$ data points after the resampling.
This procedure does not degrade the accuracy on small scales by no more than $3\%$ ($\lesssim40\,h^{-1}\mathrm{Mpc}$).
One might notice a small wiggly feature around $6\,h^{-1}\mathrm{Mpc}$. This is due to the grid effect of
the FFT method around the switching point to the direct pair-counting method as we described
in Fig.~\ref{fig:hybrid}.
Our spline function tries to remove this spurious feature to some extent by forcing the curve to be smooth. Since the raw simulation measurements employed here as the numerator still suffer from this artifact, the feature is still present in the ratio.

Our data vector still has $13\times21\times66= 18,018$ components per simulation, which is quite large.
We apply PCA to this data vector.
As in the case for HFM, we combine all the data vector from 128 simulations (100 for the varied cosmological models
plus the 28 {\it Planck} cosmology simulations), corresponding to
$128\times 18,018=2,306,304$ data points, and parameterize the halo-matter cross correlation into its PCs as
\begin{eqnarray}
\left.\xi_{\rm hm}(x,n_{\rm h},z)\right|_i=\sum_{a=1}^n~ \alpha^{\rm CCF}_{i,a} e_a^{\rm CCF}(x,n_{\rm h},z),
\end{eqnarray}
where $\left.\xi_{\rm hm}(x,n_{\rm h},z)\right|_i$ is the halo-matter cross correlation at separation $x$
for the halo sample with number density $n_{\rm h}$ and at redshift $z$ in
the $i$-th simulation, $e_a^{\rm CCF}$ is the $a$-th PC eigenvectors given as a function of separation $x$, $n_{\rm h}$ and $z$ (cosmology-independent), and $\alpha^{\rm CCF}_{i,a}$
is the $a$-th PC coefficient for the $i$-th simulation. Thus the eigenvectors $\{e^{\rm CCF}_{a}\}$
describe dependences of the halo-matter cross-correlation on separation, halo sample (halo number density), and redshift.
The different eigenvectors, $e^{\rm CCF}_a $ and $ e^{\rm CCF}_b$ with $a\ne b$, are orthogonal to each other.
The PC coefficients $\{\alpha^{\rm CCF}_{i,a}\}$ describe the dependences on different simulations, i.e. cosmological models.
In applying this PCA, we adopt the weight for each data point by $n_\mathrm{h} \, x\,[1-\exp(-x^2)]$ ($x$ is in
units of $h^{-1}{\rm Mpc}$), such that the data points containing the halo sample of higher number density are
upweighted, and data points at large separation ($x$) are relatively upweighted;
we empirically find the functional form of weight that satisfies these conditions.
As shown in the right-upper panel of Fig.~\ref{fig:cross_model}, we find that keeping the five significant
PC coefficients
reproduces the simulation results within a 5\% accuracy
on $x\lesssim 30\,h^{-1}\mathrm{Mpc}$ for each of 100 cosmological models, despite the huge
dimensionality reduction (from $18,018$ to $5$). While we can see a relatively large deviation from unity at around $1\,h^{-1}\mathrm{Mpc}$ for a few
cosmological models, the rms among the 100 models, as shown by the error bars, is typically below $2\%$ level on small scales up to $\sim 20\,h^{-1}\mathrm{Mpc}$ and reaches to $5\%$ at $\sim 40\,h^{-1}\mathrm{Mpc}$.
Since the halo-matter cross-correlation is a smooth function anyway, the PCA decomposition of the halo-matter cross
correlations works very well.

However, note that the PCA description can not well describe the correlation at very large separations.
The scatter of the curves around unity is
consistent with the statistical error of the simulation data due to a finite number of simulation realizations or equivalently a finite simulation volume as shown by the red shaded region, which is estimated from the $28$ simulations at the fiducial {\it Planck} cosmology.
We will instead use a different prescription, the propagator method, for the very large
scale to overcome both sample variance in the simulation data
and inaccurate modeling, as we will describe later.

We then perform GP regression to model the cosmology dependence of the PC coefficients for 80 cosmological models
in Slice~1 -- 4, which are our training set, similarly to Fig.~\ref{fig:HMF_model}.
Again note that we do not include the 28 {\it Planck} cosmology realizations for this GP regression.
The right-upper panel of Fig.~\ref{fig:cross_model} shows that the GP regression does not degrade the accuracy compared to the results after the PCA by more than 1\%
for the 80 cosmological models.
The right-lower panel gives a validation of our emulator; the GP interpolation
reproduces equally well the cross correlation function
for each of 20 validation cosmological models in Slice~5, which are not used in the GP regression.
The scatter is similar to the statistical error denoted by the shaded region (the sample variance of 28 {\it Planck}
cosmology simulations). Thus again, our GP regression interpolation performs well without significant overfitting.

In Appendix~\ref{sec:extra_dependence}, we show performance of the GP interpolation for different redshifts
as well as different sample of halos that are characterized by the different number density in each cosmological
model.

\subsubsection{Halo auto-correlation function}
\label{subsubsec:xihh}

\begin{figure*}[t!]
\begin{center}
\includegraphics[width=18cm,angle=0]{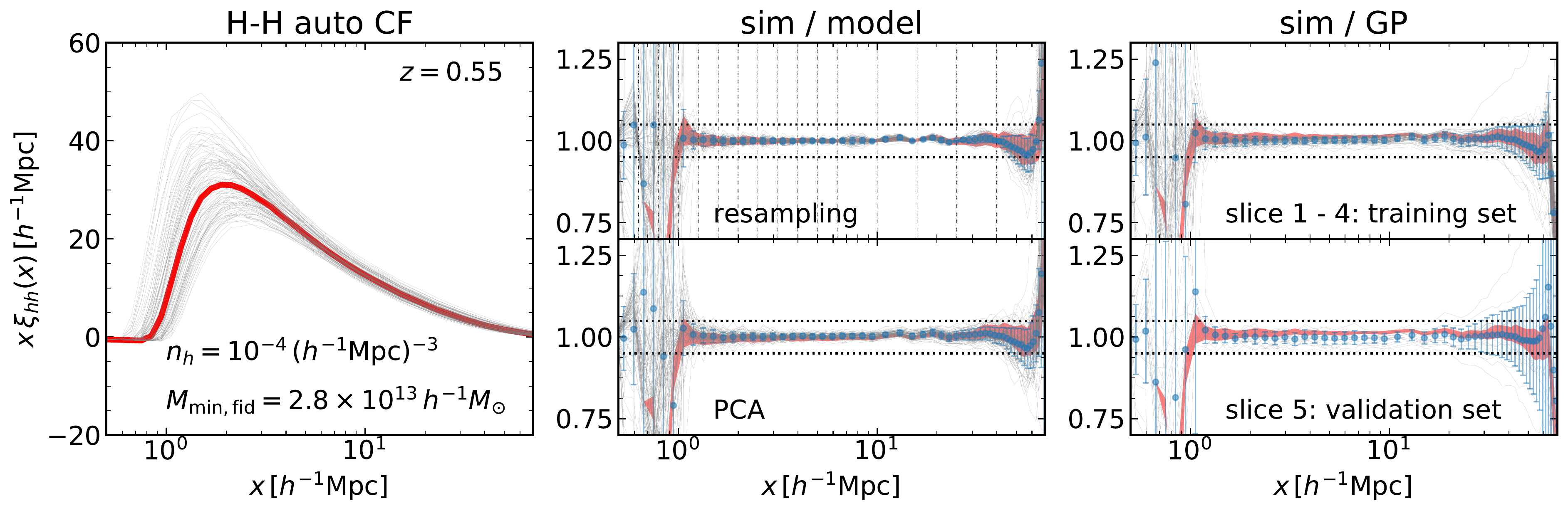}
\end{center}
\caption{Modeling of the halo-halo auto correlation functions, similarly to Fig.~\ref{fig:cross_model} for the
halo-matter cross-correlation functions.
As before, we consider halo samples with $n_{\rm h}=10^{-4}\,(h^{-1}{\rm Mpc})^{-3}$ at $z=0.55$.
Variations in the function over the 101 cosmological models (left), the accuracy of our modeling procedures
(resampling: middle-upper, PCA: middle-lower) and the performance after applying the GP regression (training set:
right-upper, validation set: right-lower) are shown, with the red shaded region indicating the scatters among the
14 simulations of the fiducial {\it Planck} cosmology.
\label{fig:auto_model}
}
\end{figure*}

We next discuss the halo auto correlation functions.
In this case we generally need to consider two mass threshold halo samples of different number densities, and thus
the dimension of input data vector is even larger than that for the halo-matter cross correlation functions.
However, we cannot obtain a meaningful signal for halo samples with number density lower than
$\sim10^{-6}\,(h^{-1}\mathrm{Mpc})^{-3}$ unlike the halo-matter cross correlation function due to the large
Poisson noise. Hence we consider here only $8$ bins with high number density
out of the $13$ bins that we considered for the cross correlation function.
We thus consider $36 \, (=8(8+1)/2)$ combinations for the two halo samples to form a matrix of auto correlation
functions at each of $21$ redshifts and at each separation bin, per simulation.

The left panel of Fig.~\ref{fig:auto_model} shows variations in the halo auto correlation functions
for 101 cosmological models for the mass threshold halo sample with number density
$n_{\rm h}=10^{-4}~(h^{-1}{\rm Mpc})^{-3}$ and at $z=0.55$, which are now computed from LR simulation suite.
The plot shows rich cosmological dependences in the halo auto correlations over the range of separation scales.

We then reduce dimensionality of the data vector by first re-sampling the separation bins.
We originally have $185$ bins for each of auto correlation functions, $40$ from the direct pair counting method ($0.1\,h^{-1}\mathrm{Mpc}<x<10\,h^{-1}\mathrm{Mpc}$), and $145$ for the FFT method (up to $300\,h^{-1}\mathrm{Mpc}$). Since the auto correlation function has a smooth shape at scales below
BAO scale, we can significantly reduce the number of sampling points.
As in the halo-matter cross correlation case,
we use a cubic spline interpolation to obtain the new data vector, and the vertical dotted lines in the
middle-upper plot denote the locations of new sampling points ($21$ points in total).
The error level after this procedure is around $3\%$ for the worst cases and is $1$--$2\%$ in terms of the rms among the models on scales $1\,\lesssim x / [h^{-1}\mathrm{Mpc}] \lesssim 40$. The larger error on smaller scales are due to the significant halo exclusion effect.

Even after the above resampling of separation bins we still have $21\times21\times36=15,876$ data points per
simulation. The next task is to reduce the dimensionality of the data vector using PCA.
As in the case for the halo-matter cross-correlations, we combine all the data vector from 114 simulations
(100 cosmological models plus 14 {\it Planck} cosmology simulations), corresponding to
$114\times 21\times 21\times 36=1,809,864$ data points,
and parameterize the halo auto correlation into its PCs as
\begin{eqnarray}
\left.\xi_{\rm hh}(x; n_1, n_2, z)\right|_i
=\sum_{a=1}^n~ \alpha_{i,a}^{\rm ACF}e_a^{\rm ACF}(x,n_1,n_2,z),
\end{eqnarray}
where $\left.\xi_{\rm hh}(x,n_1,n_2,z)\right|_i$ is the halo correlation function at separation $x$
for the two mass threshold halo samples with number densities $n_{\rm h} = n_1$ and $n_2$, respectively,
and at redshift $z$ in
the $i$-th simulation, $e_a^{\rm ACF}$ is the $a$-th PC eigenvector given as a function of separation $x$, $n_1$,
$n_2$, and $z$, and $\alpha^{\rm ACF}_{i,a}$
is the $a$-th PC coefficient for the $i$-th simulation. In applying the PCA, we adopt the weight, given as $n_1n_2x^{2}$, that is given as a function of
the two number densities $n_1$ and $n_2$ and the separation $x$.
However, since we use the LR suite here, there is a case that we cannot define a sample of halos with the highest number density, e.g.
$n_{\rm h}=10^{-2.5}\, (h^{-1} \mathrm{Mpc})^{-3}$, depending on redshifts and cosmological models.
In such cases we set the weight for PCA to zero.
After some experiments, we find that keeping the eight significant PC coefficients well reproduces the simulation results.
To be more quantitative, we can maintain the error level of a few $\%$ with a slight degradation toward the large scales ($\gtrsim 10\,h^{-1}\mathrm{Mpc}$)
as shown
in the middle-lower panel of Fig.~\ref{fig:auto_model}. Thus we made a huge dimensionality reduction of the data points (from $1.8\times 10^6$ to 8).
The error induced by this GP regression is negligibly small
except for very large scales, where we stitch to the propagator-based prescription as we will describe below.

Next we perform GP regression to model the eight PC coefficients for 80 cosmological models: 80 different
cosmological models in Slice~1 -- 4.
The right-upper plot of Fig.~\ref{fig:auto_model} shows that the GP interpolation reproduces
the simulation results for each of 80 cosmological models, which are the training set for the GP regression, within $1$ to $5$\% rms accuracy depending on the scale.
The right-lower panel gives a validation of our emulator; the GP interpolation
gives $2$--$3\%$ rms accuracy on $1\lesssim x / [h^{-1}\mathrm{Mpc}] \lesssim 20$ for the 20 cosmological models in Slice~5, which are the validation set.

It is worth noting that there are differences in the GP regression results between the halo-matter cross- and halo
auto-correlations. First, the small separation data points have large scatters, which is
not seen in the cross-correlation function. This is due to the fact that the auto-correlation function is
suppressed on these small scales due to the halo exclusion effect. We eventually expect a zero-crossing of the
auto-correlation function at the scale where the halo exclusion effect is dominant, and thus the ratio at the
zero-crossing scale becomes large and noisy.
Second, the accuracy of the emulator prediction is worse than that for the cross-correlation because of the larger
noise of auto-correlation measurements due to the Poisson noise originating from the discreteness of halos.

In Appendix~\ref{sec:extra_dependence}, we show the performance and validation for the halo auto-correlation
functions for halo samples of different number densities and
different redshifts as well as the cross-correlations between two halo samples
with different number densities.

\subsubsection{Propagator}
\label{subsubsec:gamma1}
The final piece of our halo modules is the propagator that describes the large-scale
correlation functions around BAO scales. Since the halo-matter cross correlation function involves
the propagators of halo and matter, we here study both. We use the \textsc{LR} suite for this purpose as
the damping behavior of the propagator is known to be mainly due to
the large-scale bulk motion. Since the estimation of the propagator is done by using
the cross correlation between the halo (or matter) density field to the linear density field used in setting up initial conditions of the simulations (i.e., Eq.~\ref{eq:prop_estimator}), the measured data does not show discreteness noise unlike what we see in the halo auto-correlation functions. Therefore, we can measure it to a
reasonably accurate precision for all the $13$ number-density bins for the mass threshold samples from $10^{-2.5}$ to $10^{-8.5}\,(h^{-1}\mathrm{Mpc})^{-3}$ used for the halo-matter cross correlation functions.

Since the shape of the propagator is roughly a Gaussian with its width being the rms displacement as
shown in Fig.~\ref{fig:propagator}, we parameterize it as
\begin{eqnarray}
G_a(k) = \left(g_0 + g_2 \, k^2 + g_4 \, k^4\right)\exp\left(-\frac{\sigma_{\mathrm{d},\mathrm{lin}}^2k^2}{2}\right),\label{eq:prop_model}
\end{eqnarray}
where $g_0$ can be interpreted as the linear bias at the large scale limit and $\sigma_{\mathrm{d},\mathrm{lin}}$
is the linear rms displacement for the matter field at the redshift under consideration, which is computed by the
linear module. The other coefficients, $g_2$ and $g_4$, are fitting parameters, introduced to capture a departure
from the simple Gaussian form.
In the above, the subscript $a$ in the left hand side stands for either $m$ (matter) or $h$ (halo), and the factor
$g_0$ is set to unity for the matter propagator.

\begin{figure*}[t!]
\begin{center}
\includegraphics[width=18cm,angle=0]{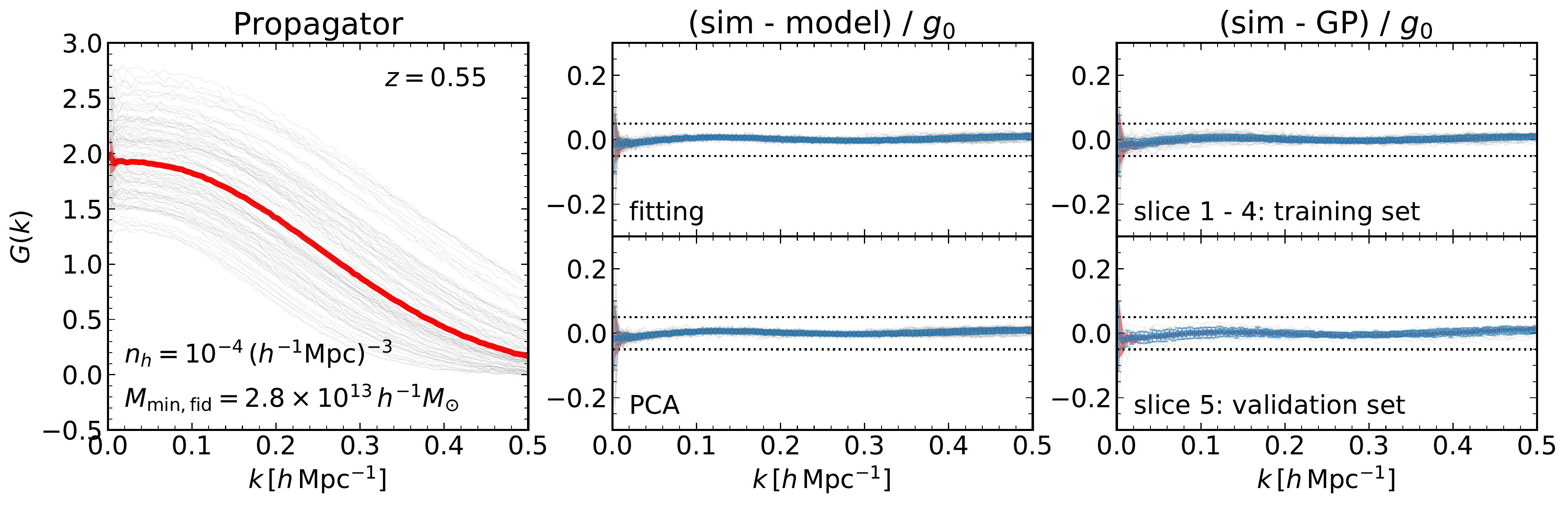}
\end{center}
\caption{
Modeling of the halo propagator, similarly to Figs.~\ref{fig:cross_model} and \ref{fig:auto_model} for correlation
functions. As before, we consider halo samples with $n_{\rm h}=10^{-4}\,(h^{-1}{\rm Mpc})^{-3}$
at $z=0.55$. Variations in the function over the 101 cosmological models (left), the accuracy of our modeling
procedures (model fitting: middle-upper, PCA: middle-lower) and the performance after applying the GP regression (
training set: right-upper, validation set: right-lower) are shown, with the red shaded region indicating the
scatter among the 14 simulations of the fiducial {\it Planck} cosmology.
\label{fig:gamma_model}
}
\end{figure*}

We show in Fig.~\ref{fig:gamma_model} our modeling detail of this function for halo samples with number density
$10^{-4}\,(h^{-1}\mathrm{Mpc})^{-3}$ at $z=0.55$. As before, we show variations in the function for the 101
cosmological models in the left panel, with the red shaded region showing the mean and the scatter of this function
for the fiducial {\it Planck} cosmology (the shaded region is also shown in the other panels, though they are
heavily overlapped with other symbols and thus difficult to see). We can see that the propagator is always a simple
decaying function of wavenumber, with a strong dependence on cosmology in the amplitude and the typical wavenumber
at which the curve is decaying.

We then fit the data using Eq.~(\ref{eq:prop_model}) and show the residual in the middle-upper panel. Here and also
in the other three panels, we normalize the residual by $g_0$, which is the low-$k$ limit of this function, to see
the importance of the residual relative to the overall amplitude of the function. While we see a small wiggly
pattern in the residual, the typical amplitude of this pattern is below a few
percent
level, which is sufficiently
small for our purpose.

At this point, we have three fitting parameters per halo sample, and thus $819\,(=3\times13\times21)$ data points
per simulation for the halo propagator. As before we reduce the dimensionality by applying the PCA.
The data points are approximated by
\begin{eqnarray}
\mathbf{d}^{\rm PRO}_i(n_h, z)
=\sum_{a=1}^n~ \alpha_{i,a}^{\rm PRO}\mathbf{e}_a^{\rm PRO}(n_h,z),
\end{eqnarray}
where $\mathbf{d}^{\rm PRO}_i(n_h, z)$ is the vector formed with the three fitting parameters for the
halo sample with number density $n_{\rm h}$ and at redshift $z$ in
the $i$-th simulation, $\mathbf{e}_a^{\rm PRO}$ is the $a$-th PC eigenvector given as a function of $n_h$ and $z$, and $\alpha^{\rm PRO}_{i,a}$
is the $a$-th PC coefficient for the $i$-th simulation. In applying the PCA analysis, we adopt the weight, simply given as $n_h$.
As before, since we use the LR runs here, there is a case that we can not define a sample of halos with the highest number density, e.g.
$n_{\rm h}=10^{-2.5}\, (h^{-1}\mathrm{Mpc})^{-3}$, depending on redshifts and cosmological models.
In such cases we set the weight for PCA to zero.
After some experiments, we find that keeping the four most significant PC coefficients well reproduces the simulation results, as shown in the middle-lower panel of Fig.~\ref{fig:gamma_model}. The extra error induced by the PCA is below 1\% level.

The remaining task is the same as before: train a GP regression using the $80$ cosmological models in Slice~1 to 4 and validate the results using the remaining $20$ models in Slice~5 (as well as the fiducial {\it Planck} cosmology). Although the variance of the residuals among the models shown in the right panels seems to be somewhat larger than that in the middle panels, the prediction of GP stays within the $\pm 5\%$ band shown by the horizontal dotted lines. More importantly, the accuracy of GP for the validation set is not degraded compared to that for the training set, implying that there is no problem of overfitting to the training data.

The validation tests for other halo samples as well as the matter propagator at various redshift can be found in Appendix~\ref{sec:extra_dependence}. In the current implementation, we model the matter propagator following exactly the same procedure for halos. The only difference is that we have one less free parameter ($g_0$ should always be unity for matter) and we need only two PCA components to ensure the accuracy.

\section{Usage of the emulator}
\label{sec:demo}
We have explained how each of the basic modules are modeled and tested in the previous section.
These correspond to the \textsc{Halo Modules} in Fig.~\ref{fig:DE_architecture}, which predict the statistical
properties of dark matter halos.
Now, in this section, we show several demonstrations of how to use the emulator to predict properties of dark
matter halos. Also we show
how to combine the predictions to compute clustering statistics of galaxies
(i.e., usage of the \textsc{Utility Modules} at the bottom in Fig.~\ref{fig:DE_architecture}).

\subsection{Halo properties}
\label{subsec:halos}

\subsubsection{Implementation detail}
\label{subsubsec:halos_implementation}

One application of our emulator is to make predictions for halos in the mass range of galaxies to clusters. Since our modules that compute halo clustering properties directly predict the signals as a function of the cumulative halo number density, which is discretely sampled every $0.5$ dex, it might not be so
practically
useful as it is.
To obtain the predictions of halo correlation functions at a given halo mass, we have to interpolate over the sampled number densities and convert it to the halo mass in a given cosmological model. Hence, if one wants to predict the clustering signals as a function of halo mass, an inaccuracy in the conversion from the mass to the number density can be a new source of error.

\begin{figure}[h]
\begin{center}
\includegraphics[width=8.8cm,angle=0]{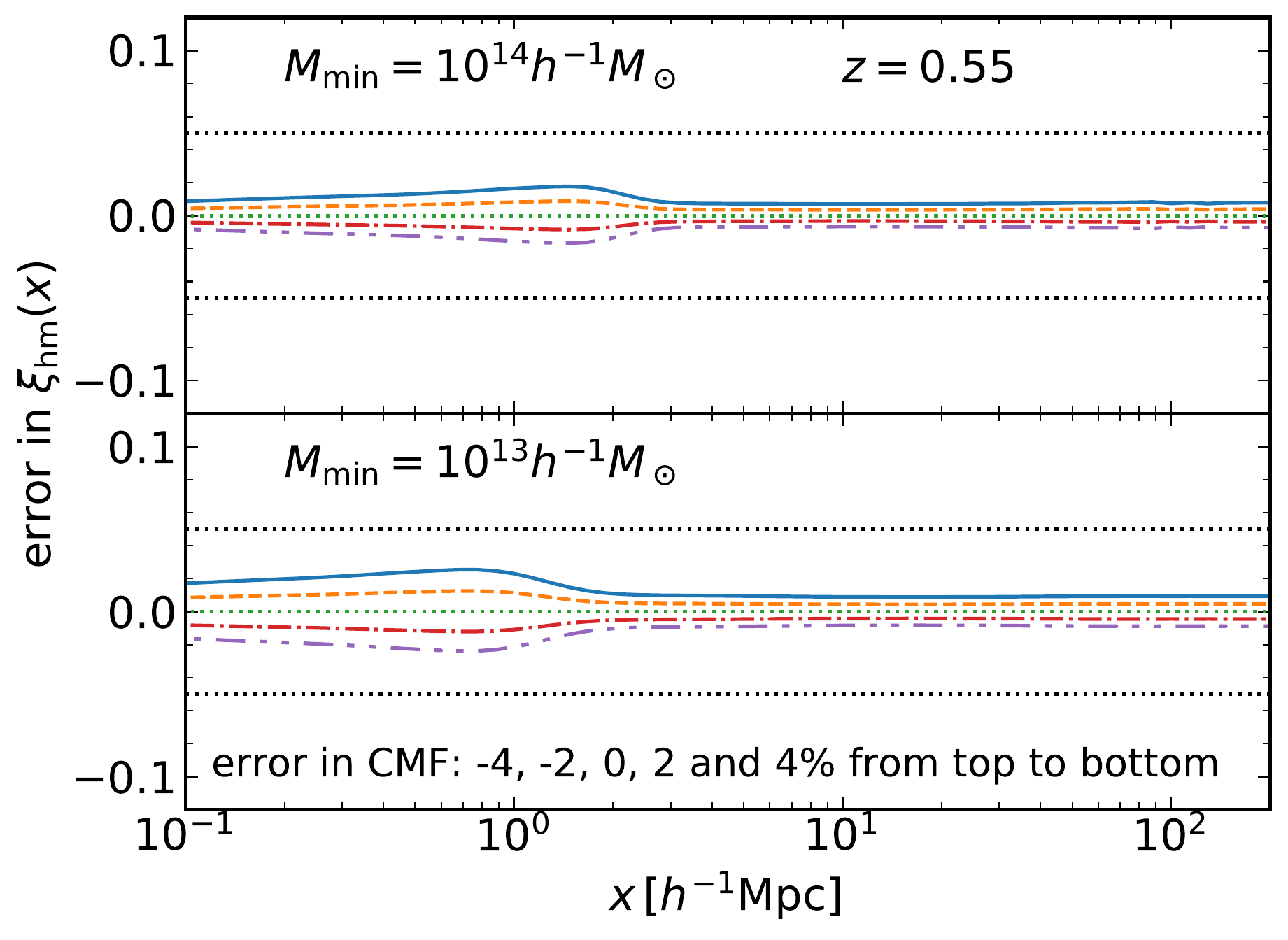}
\includegraphics[width=8.8cm,angle=0]{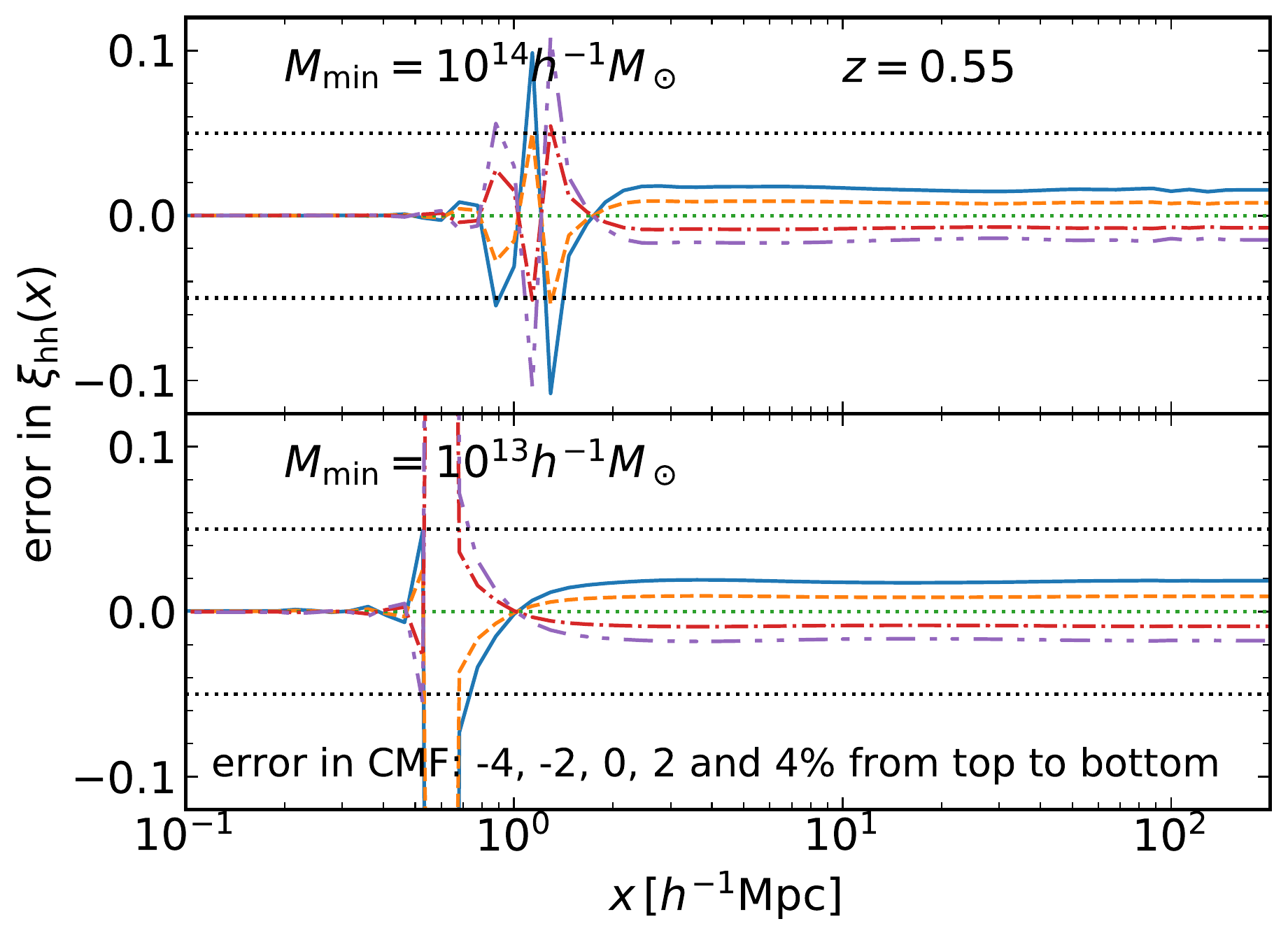}
\end{center}
\caption{The impact of a possible error in the conversion between the cumulative halo number density and the mass threshold on the emulator predictions of halo correlation functions. As a working example, we here consider the cross-correlation functions,
$\xi_{\rm hm}(x)$, in the upper panel, and the halo auto-correlation functions, $\xi_{\rm hh}(x)$, in the lower panel
for the two mass thresholds ($10^{14}$ and $10^{13}\,h^{-1}\mathrm{M}_\odot$) for the {\it Planck} cosmology
at $z=0.55$. Here we first use the halo mass function module to compute the cumulative
halo number density for the mass threshold, shift the number density by $-4, -2, 2$ or $4\%$, and then input the shifted number density into the emulator to obtain the shifted predictions of $\xi_{\rm hm}(x)$ and $\xi_{\rm hh}(r)$, respectively (see text for details). The figure shows the ratio of the shifted correlation function to the fiducial prediction. The significant features around
$x\simeq 1\,h^{-1}{\rm Mpc}$ in $\xi_{\rm hh}(x)$ are due to the halo exclusion effect that would not be present for the galaxy correlation function.
\label{fig:err_prop}}
\end{figure}

In Fig.~\ref{fig:err_prop} we study how the conversion to the cumulative halo number density to a target halo mass,
using the halo mass function module, causes a possible error in the predictions of halo correlation functions.
To do this, we consider the emulator outputs of halo correlation functions for two mass threshold samples with $M_\mathrm{min}=10^{13}$ and $10^{14}h^{-1}M_\odot$ for the fiducial {\it Planck} cosmology at $z=0.55$. Then we use the following method to
propagate a possible error in the
halo number density into an error in predicting the halo correlation functions as a function of the halo mass. i)
We first use the halo mass function module to compute the cumulative number density for the halo mass thresholds, $M_\mathrm{min}=10^{13}$ and $10^{14}h^{-1}M_\odot$. ii) We multiply the number density by a factor of 0.96, 0.98, 1.02 or 1.04,
respectively, which is intended to mimic a possible error in the number density calibration by
$-4, -2, 2$ or $4\%$, respectively. iii) We then obtain the emulator predictions of halo correlation functions by
inserting the shifted values of halo number density in the emulator.
Here $\pm 4\%$ error in the cumulative halo number density is considered as a rather pessimistic case because
Fig.~\ref{fig:HMF_model} shows that a typical error in the mass function is smaller in terms of the rms among the models ($\sim 1$ $(3)\%$ at $10^{13}$ $(10^{14})$ $h^{-1}M_\odot$).
In addition the error in the halo mass function seen in Fig.~\ref{fig:HMF_model} would be partly canceled when we consider the error on the \textit{cumulative} halo number density.
The upper panel shows the ratios of the shifted halo-matter cross-correlation functions, $\xi_{\rm hm}(x)$,
to the fiducial prediction. The figure shows
a constant shift in the two-halo regime, a slightly larger shift in the one-halo term, and a bump-like feature at transition scales between the two regimes.
These are caused by changes in
the linear bias and
the mass profile,
respectively. The size of the fractional shift in the cross-correlation function is smaller than that on the cumulative mass function, with a slight decreasing trend toward higher masses. In the lower panel, the auto correlation function, $\xi_{\rm hh}(x)$, shows a larger shift in the two-halo regime reflecting the fact that it scales as bias squared. A sharp feature can be found where the halo exclusion effect kicks in. Since the latter part  is dominated by one-halo term in case of the \textit{galaxy} correlation function, the final shift would be much smaller. Even with the pessimistic case of a $4\%$ error in the cumulative mass function, the induced shift in  the correlation functions are well within the $\pm 5\%$ band and mostly within $\pm 3\%$ level. When we consider a realistic error on the cumulative mass function (i.e., a few percent or below), the error on the correlation function arising from this is smaller than the typical error in the emulator in both cases.

\begin{figure}[h]
\begin{center}
\includegraphics[width=8.8cm,angle=0]{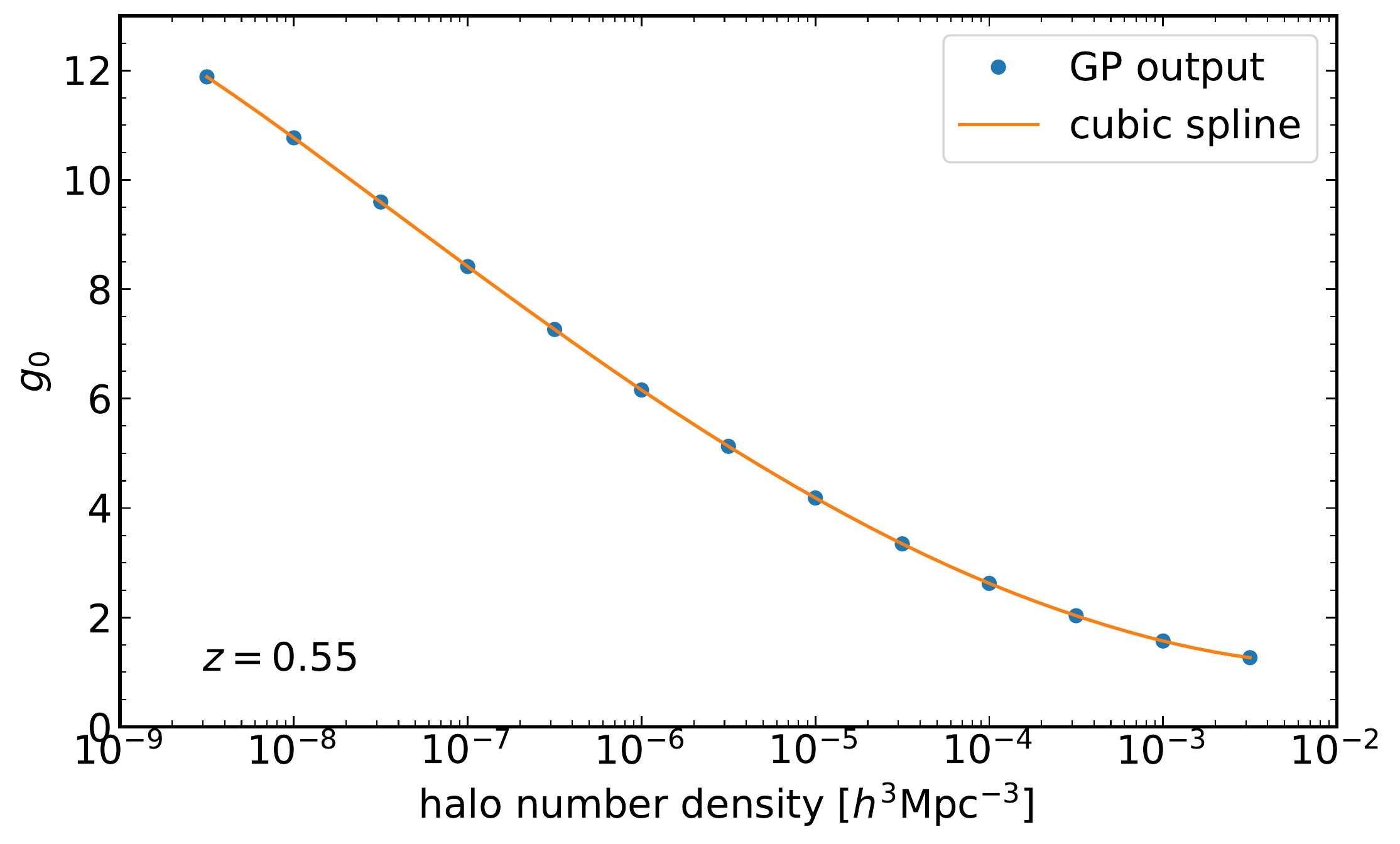}
\end{center}
\caption{Halo bias as a function of the halo number density for the fiducial cosmological model at $z=0.55$. The symbols are the direct output of our emulator and the solid curve is its interpolation using the cubic spline.
\label{fig:bias_ndens}
}
\end{figure}

As another example of the applications, we show in Fig.~\ref{fig:bias_ndens} the output of the emulator for the large-scale bias as a function of the halo number density (symbols). Here, the large-scale bias is defined as the fitted parameter $g_0$ in Eq.~(\ref{eq:prop_model}), which is the $k\rightarrow 0$ limit of the propagator (Eq.~\ref{eq:def_propagator}).
The plot shows the result for the fiducial {\it Planck} cosmology and at $z=0.55$.
We interpolate these data points using the cubic spline function to make a prediction at any halo number density in the range shown here. Note that we use the logarithm of the halo number density, instead of the raw values of the number density, for which our sampling is uniform.

\begin{figure}[h]
\begin{center}
\includegraphics[width=8.8cm,angle=0]{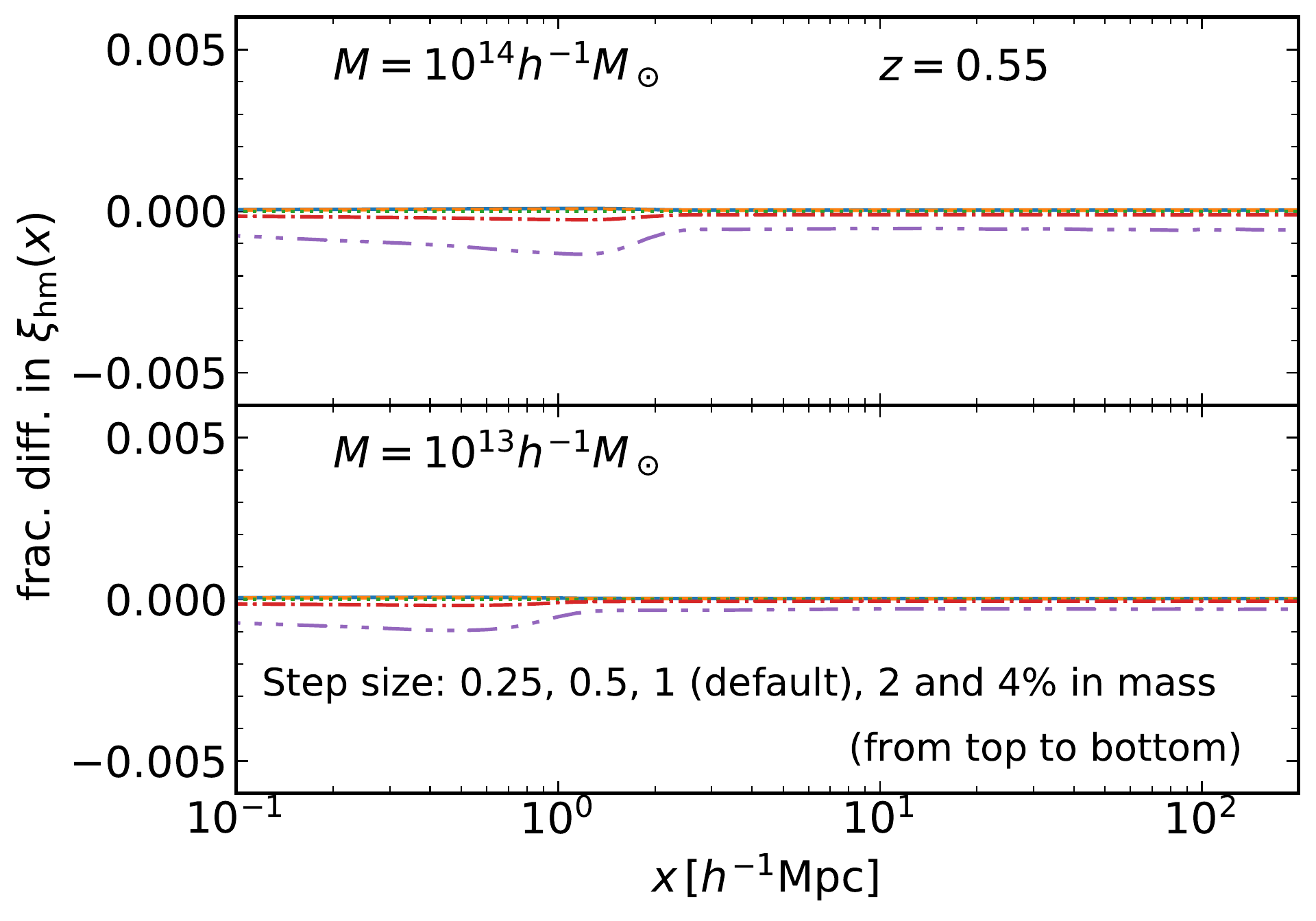}
\includegraphics[width=8.8cm,angle=0]{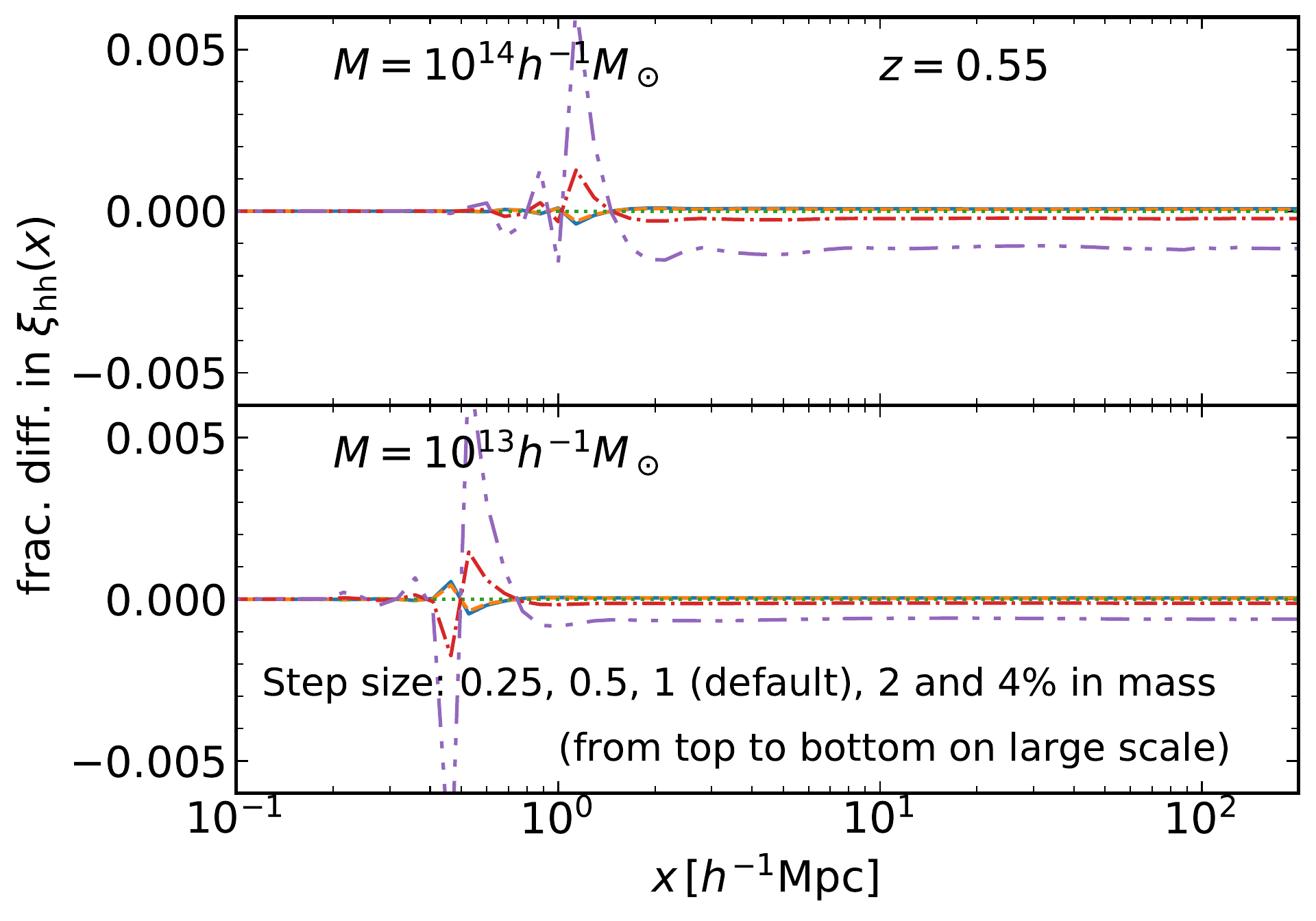}
\end{center}
\caption{Stability of the finite difference evaluation of the correlation functions at a given halo mass. We show the fractional change in the halo-matter cross (upper) and halo auto (lower) correlation function. We employ the default step size of $\pm 1\%$ in mass as the reference for this figure. Notice the rather narrow range of the vertical axis.
\label{fig:err_stepsize}
}
\end{figure}

\begin{figure}[h]
\begin{center}
\includegraphics[width=8.8cm,angle=0]{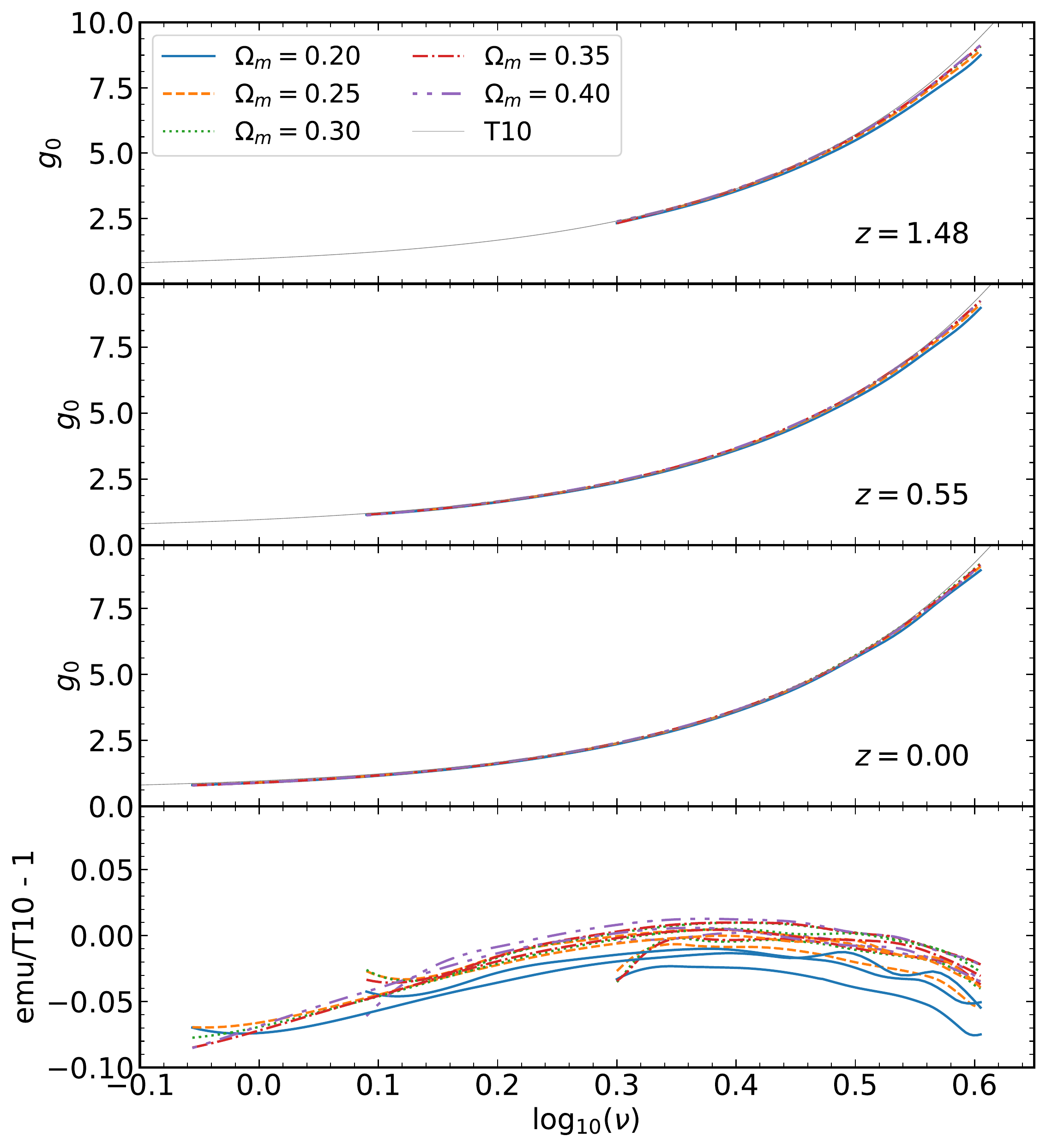}
\end{center}
\caption{Halo bias as a function of the peak height for different cosmologies at various redshifts as shown in the legend. We
vary $\Omega_{\mathrm{m}}$, but kept $\sigma_8$ and other cosmological parameters fixed to their fiducial values.
We also show by the thin solid line the fitting formula by \citet{Tinker10}.\label{fig:bias_nu}
}
\end{figure}

Once the spline interpolator is ready, we can compute the halo bias as a function of the halo mass or peak height, if one prefers, by first using the halo mass function module to convert the number density to the minimum halo mass and then taking a finite difference derivative to have the bias at a specific, desired halo mass scale. For this derivative, we employ $\pm 1\%$ changes in the mass as the default step size. The dependence of the results on the step size is much weaker than the typical accuracy of the emulator as shown in Fig.~\ref{fig:err_stepsize}.
The result is shown in Fig.~\ref{fig:bias_nu} as a function of the peak height $\nu\equiv\delta_\mathrm{c}/\sigma_M$, with $\delta_\mathrm{c}=1.686$. Now, we show the results for different cosmologies at three different redshifts. We vary $\Omega_\mathrm{m}$ keeping the flatness and fixing the present-day
amplitude of the linear matter power spectrum, $\sigma_8$, to its fiducial value. We compare the results with the fitting formula by \citet{Tinker10} as denoted by
the thin solid line,
which is independent of redshift or cosmology. The fractional difference from \citet{Tinker10} at the three redshifts are plotted together in the bottom panel. Our emulator prediction is overall consistent with the fitting formula with the accuracy no worse than $10\%$
over all the ranges examined here.
Such an inaccuracy of \citet{Tinker10} is also pointed out in the previous work \citep[e.g.,][]{2016PhRvD..93f3507L}.
We confirm that the bias function is rather universal with little dependence on cosmology or redshift. 
\begin{figure}[h]
\begin{center}
\includegraphics[width=8cm,angle=0]{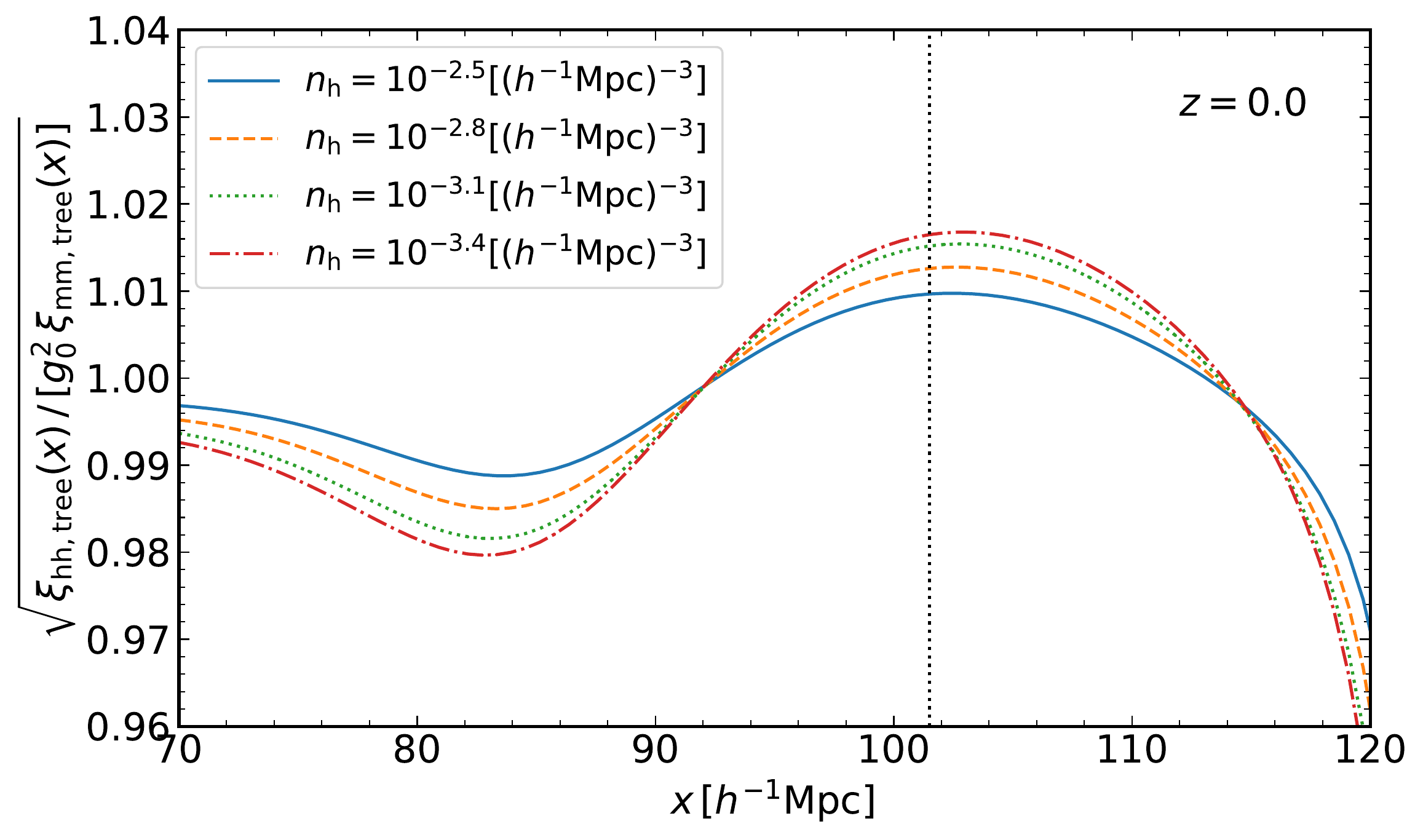}
\end{center}
\caption{Scale dependence of halo bias\label{fig:BAO_scale_dep} around the BAO scale (the vertical dotted line). We
show the square root of the ratio of the halo and matter correlation functions, with the latter scaled by the square of the linear bias factor $g_0$. We show with different lines four mass threshold halo samples with the number density listed in the figure legend, corresponding respectively to the threshold mass of $1.58\times10^{12}, 3.34\times10^{12}, 6.86\times10^{12}$ and $1.36\times10^{13}\,h^{-1}M_\odot$ at $z=0$.}
\end{figure}

Another interesting feature of bias is its scale dependence around the BAO scale. We implement the same spline interpolation for the fitting parameters $g_2$ and $g_4$ in the propagator. This allows us to estimate the propagator at any halo mass. The tree-level calculation, Eq.~(\ref{eq:xi_tree}), gives us a prediction of the correlation functions and we already show that BAO scale is well described by this simple model as illustrated in Figs.~\ref{fig:BAO_ximm}--\ref{fig:BAO_xihh}. We show in Fig.~\ref{fig:BAO_scale_dep} the square root
of the ratio of the halo and matter correlation functions at $z=0$. We normalize it by the linear bias factor $g_0$ such that the ratio becomes unity when the bias is independent of scale. We consider four halo samples with different number densities as written in the figure legend. The result indicates that the BAO peak structure can be boosted for low number density samples (i.e., when only massive halos are included in the sample). This is fully consistent with the expectation by the peak model \citep[compare our results with Figs.~7 and 8 of][]{desjacques10}. Also, this feature was previously found in numerical simulations \citep[e.g.,][]{angulo:2014lr,Crocce15}. This kind of prediction is possible because our model has a freedom to control the damping of BAO feature in terms of the two free parameters, $g_2$ and $g_4$, in the propagator, Eq.~(\ref{eq:prop_model}), in addition to the damping due to the typical random displacements of matter, $\sigma_\mathrm{d,lin}$.

\begin{figure*}
\begin{center}
\includegraphics[width=18cm,angle=0]{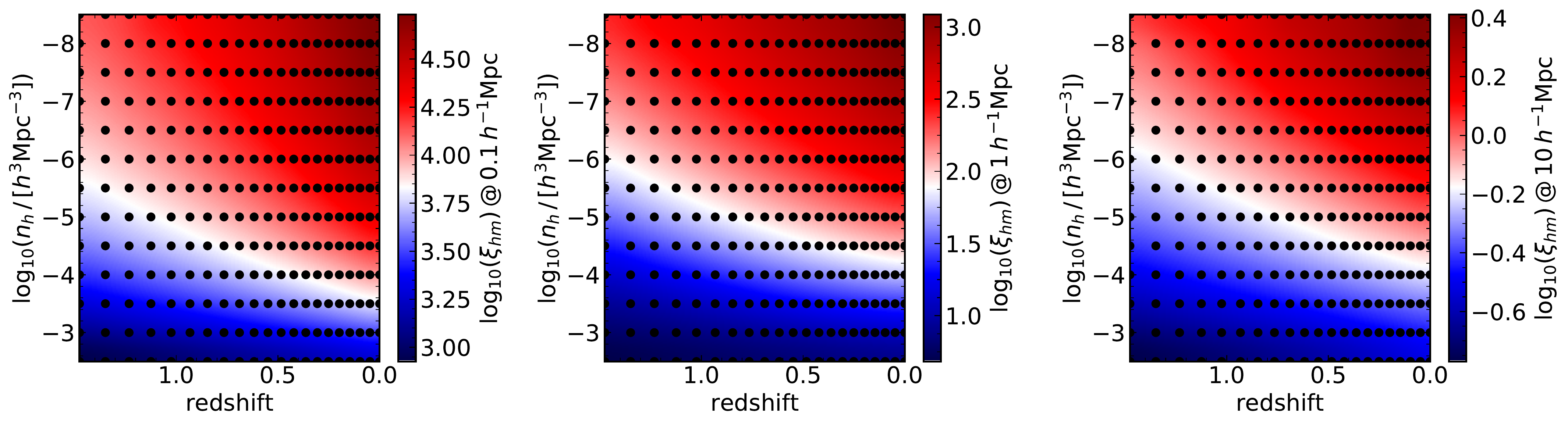}
\end{center}
\caption{Interpolation of the tabulated halo matter cross correlation function over the halo number density and redshift. Each panel shows the interpolated result (color scale) at a fixed separation as shown in the color bar label, and the dots show the location where the data table is available.\label{fig:xicross_interp}
}
\end{figure*}

\begin{figure*}
\begin{center}
\includegraphics[width=18cm,angle=0]{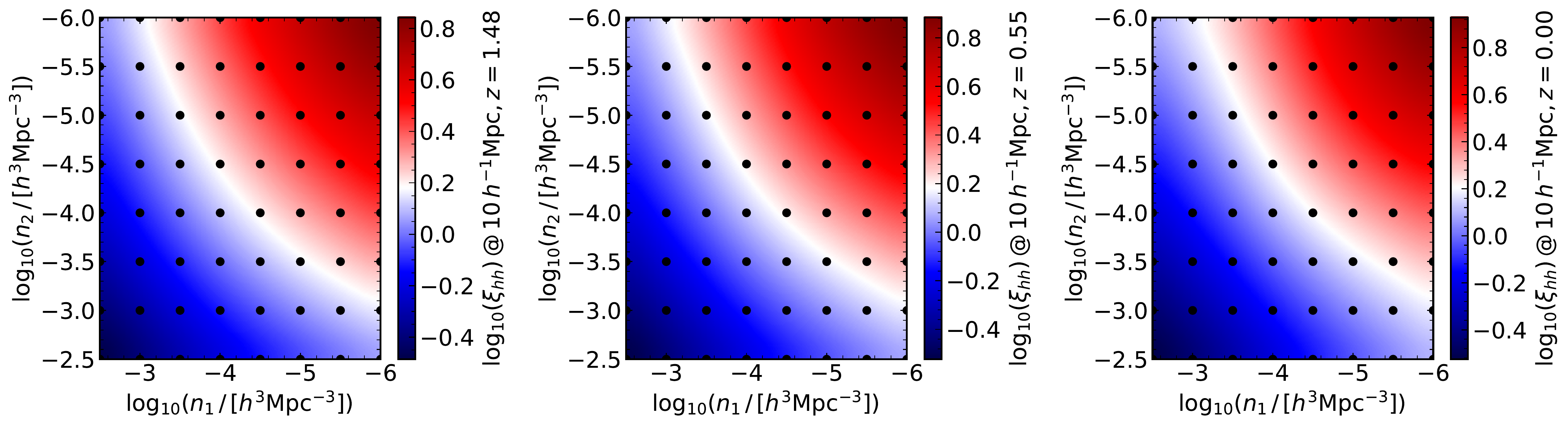}
\end{center}
\caption{Similar to Fig~\ref{fig:xicross_interp}, but for the halo auto correlation function. We now fix both separation and redshift in each panel and show the interpolation over the two number densities, $n_1$ and $n_2$.\label{fig:xiauto_interp}
}
\end{figure*}

We implement the same cubic spline interpolation for other quantities, such as $\xi_\mathrm{hm}(r;n_h)$ or $\xi_\mathrm{hh}(r;n_1,n_2)$, with the latter using the bivariate cubic spline for two number densities, $n_1$ and $n_2$.
We do not find any sizable error originating from this interpolation as all the quantities
vary rather smoothly with (the logarithm of) the halo number density. Analogously, the redshift dependence is interpolated with the cubic spline function.
As is clear from Fig.~\ref{fig:bias_nu}, the dependence of bias
on redshift is
weak and thus the same interpolation scheme works fine. The situation is similar for the other interpolated quantities. We show in Fig.~\ref{fig:xicross_interp} or \ref{fig:xiauto_interp} our interpolation of the halo-matter cross or halo auto correlation function over the number density and redshift. We first construct a data matrix at the locations as depicted by the dots based on the Gaussian Process and PCA methods, and then perform a cubic spline interpolation for each dimension.

While the halo matter cross correlation function (and also the propagator) are very well determined down to a quite low halo number density, $10^{-8.5}\,(h^{-1}\mathrm{Mpc})^{-3}$,
corresponding to very massive halos,
thanks to the fact that they are given by the cross correlation with the matter field, the halo auto correlation function instead suffers from a severe Poisson noise especially at such a high mass end. We thus switch to a simple scaling, $\xi_{\rm hh}(r; n_1, n_2) = [g_0(n_1) / g_0(n_\mathrm{min})]\,\xi_{\rm hh}(r; n_\mathrm{min}, n_2)$, when the number density $n_1$ is below the minimum number density $n_\mathrm{min} = 10^{-5.75}\,(h^{-1}\mathrm{Mpc})^{-3}$. In the above, the bias factor, $g_0(n_i)$, is computed again in the module that computes the propagator (i.e., the function plotted in Fig.~\ref{fig:bias_ndens}). We do the same when $n_2$ is below the threshold; we simply multiply the ratio of the large-scale bias one more time. While this might be a reasonable approximation on large scales, it can not properly reproduce the correlation functions around scales
where the halo exclusion effect is not negligible.
Nevertheless, our current implementation does not lead to a severe error for a sample of galaxies such as the CMASS sample, because the small-scale correlation function is mainly described
by the one halo term.

\begin{figure}[h]
\begin{center}
\includegraphics[width=8.8cm,angle=0]{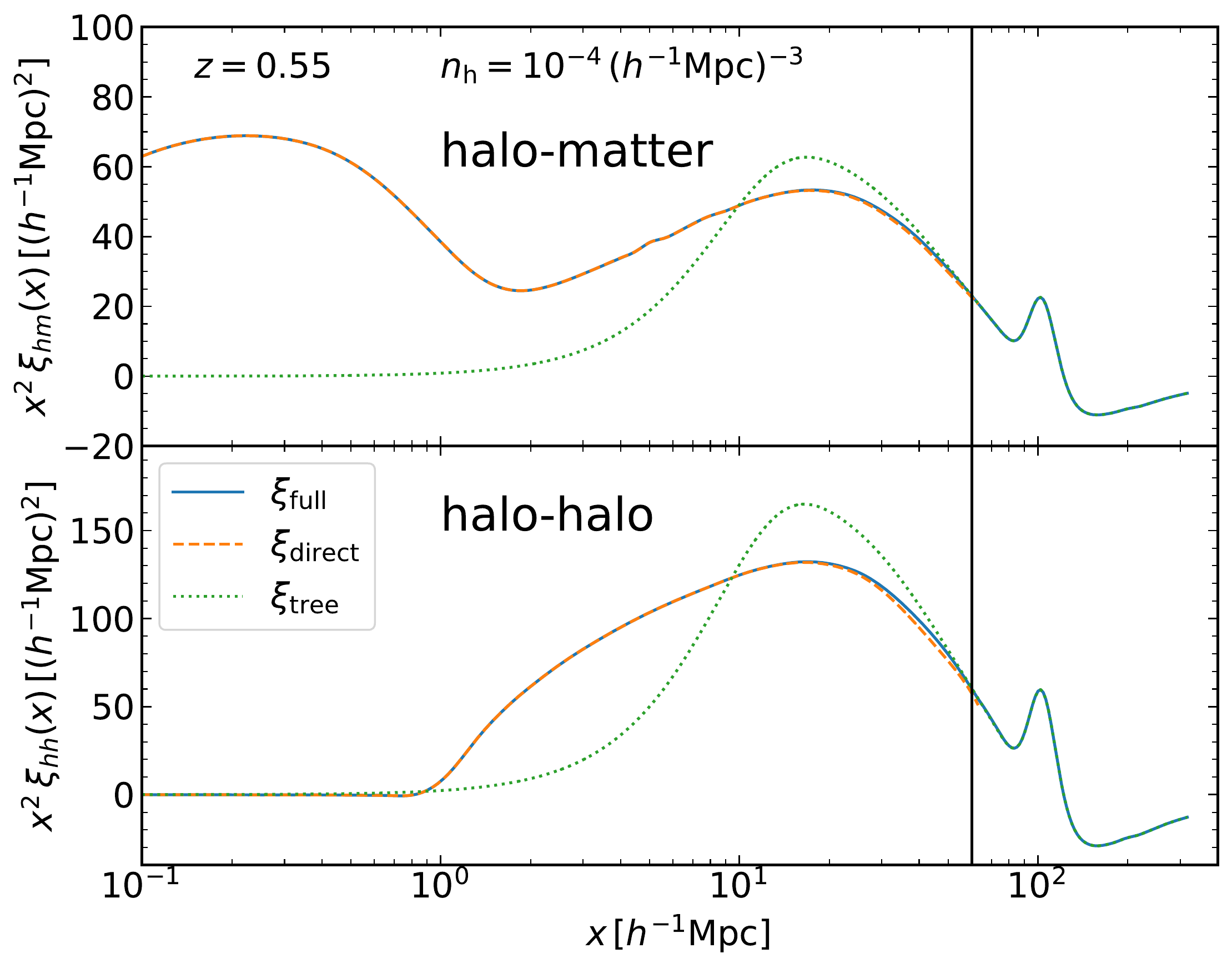}
\end{center}
\caption{Stitching of the large and small scale predictions. The direct output of the PCA-Gaussian Process modeling is shown by the dashed lines ($\xi_\mathrm{direct}$), the large-scale model based on the emulated propagator is by the dotted line ($\xi_\mathrm{tree}$) and the final prediction is by the solid line ($\xi_\mathrm{full}$). We show in the upper panel the halo matter cross correlation function and in the lower panel the halo auto correlation function, both for the halo number density $10^{-4}\,h^3\mathrm{Mpc}^{-3}$ at $z=0.55$. The vertical solid line marks the stitching scale $x_\mathrm{switch}=60\,h^{-1}\mathrm{Mpc}$. \label{fig:matching}
}
\end{figure}

Now we have predictions of $\xi_{\rm hm}$, $\xi_{\rm hh}$ and the propagator given as a function of halo number
density and redshift within the ranges relevant for the resolution and output redshifts of our simulations.
We combine all these predictions to obtain a well-behaved prediction over a wide range of separations.
We do this by smoothly stitching the two predictions as
\begin{eqnarray}
\xi_{{\rm ab},\mathrm{full}}(x) = D(x)\,\xi_{{\rm ab},\mathrm{direct}}(x) + \left[1-D(x)\right]\xi_{{\rm ab},\mathrm{tree}}(x),
\end{eqnarray}
where ``ab'' is either ``hm'' or ``hh'', and we use the damping function defined as
\begin{eqnarray}
D(x) = \exp\left[-\left(\frac{x}{x_\mathrm{switch}}\right)^4\right].
\end{eqnarray}
We find that $x_\mathrm{switch}=60\,h^{-1}\mathrm{Mpc}$ provides a reasonably good model for both the auto and cross correlation functions over the range of scales as demonstrated in
Fig.~\ref{fig:matching}.

In summary, the current implementation of our emulator works in a parameter sampler as follows. When a new cosmological model is proposed, the code first calls a GP interpolator to evaluate the coefficients for the PCs for all the statistics we consider here. Then, combining these coefficients with the eigenvectors, it computes the statistics at all the redshifts, mass and separation bins to form a data table. One function call of our high-level interface to set the cosmological parameters does all the tasks up to here internally. Now, an user can further call other high-level functions prepared for each statistics. These functions accept a redshift, a halo mass (either a threshold mass or a target mass scale) and a set of separations at which the correlation function should be evaluated. In this final step, the code finds the values by calling a spline interpolator over the table created in the previous step. For users who wish to compute galaxy statistics, a separate module can be used with additional parameters describing the HOD model. This module internally calls the functions for the halo statistics and integrate them over the halo mass with the product of the mean HOD and the halo mass function as a weight. We also prepare functions computing projected statistics, which works similarly.

\subsubsection{Demonstrations}
\label{subsubsec:demo}

Now we can compute the three main halo statistics, the halo mass function, the halo matter cross-correlation function and the halo auto-correlation function for an arbitrary cosmological model that is covered by our sampled cosmological models within the flat $w$CDM cosmologies.
Using the results, we can obtain how these halo statistical quantities vary with cosmological parameters as
demonstrated in Fig.~\ref{fig:demo1}, which gives their dependences on $\Omega_\mathrm{m}$.

We can further predict in detail, for instance, the density profile of dark matter halos from our emulator.
Properties and cosmological dependences of the mass density profiles around halos have been extensively studied
\cite[e.g.,][]{NFW}. The emulator output of the halo-matter cross correlation on small separations can be used
to study the mass density profiles for halos whose masses are in the range supported by the emulator\footnote{Notice that the average spherical density profile of halos is equivalent to the positional cross correlation function between halos and matter by definition.}.

\begin{figure}[h]
\begin{center}
\includegraphics[width=8.8cm,angle=0]{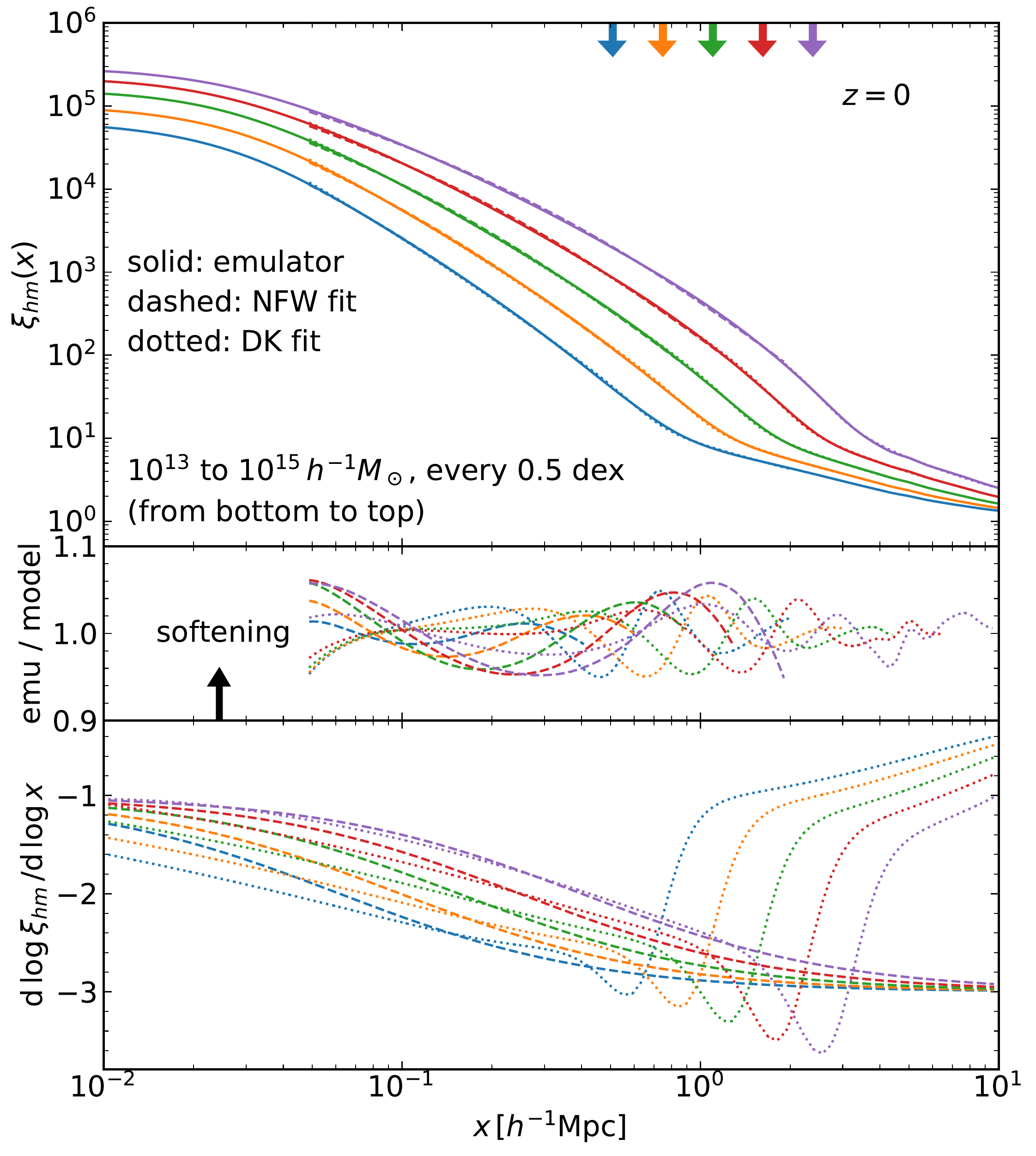}
\end{center}
\caption{Top: Model fit (dashed: NFW, dotted: DK) to the density profile around halos predicted by the emulator (solid). We stick to the fiducial {\it Planck} cosmology at $z=0$, and consider various halo masses as shown in the legend. Middle: we show the ratio of the emulator to the model fit in the top panel. Bottom: logarithmic derivative of the profile for the analytical fit. The radius $R_{200\mathrm{m}}$ for each sample is indicated by the the downward arrows in the top panel. The softening length is also shown by the upward arrow in the middle panel. \label{fig:radial_profile}
}
\end{figure}

In Fig.~\ref{fig:radial_profile} we compare the profiles from the emulator (solid) with the best-fit NFW profiles (dashed) for halos of different masses. For the fitting, we included the data over the range of radii from
twice the softening scale to $80\%$ of $R_\mathrm{200\mathrm{m}}$.
The middle panel shows that the NFW profile gives a good fit to a fractional accuracy better than about 5\% up to
the virial radius ($R_{200\mathrm{m}}$), beyond which the NFW profile no longer reproduces the simulation results.

\begin{figure}[h]
\begin{center}
\includegraphics[width=8.8cm,angle=0]{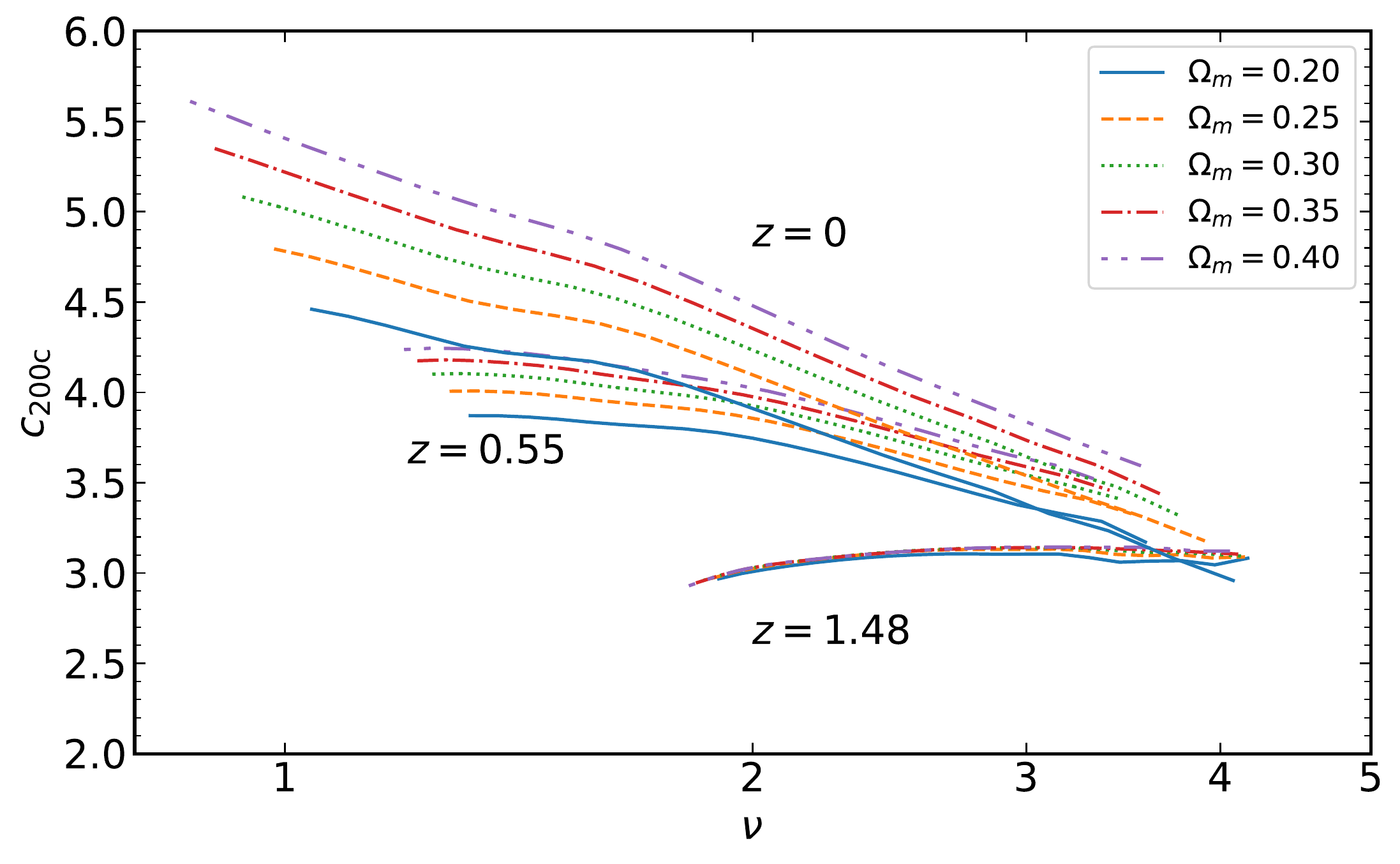}
\end{center}
\caption{Concentration-peak height relation at various cosmological models and at different redshifts.\label{fig:cM}
}
\end{figure}
Using the NFW fitting results, we can also study how the concentration parameter varies with halo mass as well as
cosmological models over different redshifts, where we define the concentration parameter, $c_\mathrm{200c}$, by
the ratio of the radius within which the density is $200$ times the critical density to the scale radius determined
by the NFW fit.
Fig.~\ref{fig:cM} shows $c_{\rm 200c}$ as a function of the peak height, $\nu=\delta_c/\sigma_M$, for cosmologies
with different $\Omega_\mathrm{m}$ at three redshifts. Similarly to the previous plots, we keep spatial flatness
and vary the normalization $A_\mathrm{s}$ such that $\sigma_8$ is kept unchanged for models with different
$\Omega_\mathrm{m}$. While the relation seems to be universal at high redshift with little dependence on
$\Omega_\mathrm{m}$, we can see clear dependence at lower redshifts. The increasing trend of $c_{\rm 200c}$ as
decreasing redshift, as well as its positive $\Omega_\mathrm{m}$ dependence can be found in
\citet{2015ApJ...799..108D}.
Further study on the dependence of the concentration-mass relation on the cosmological parameters can be found in
\citet{Kwan13}\footnote{We can not make a direct comparison with their emulator because the Hubble parameter is
automatically determined to match to the CMB constraint given the other parameters in their code. On the other
hand, we here vary $\Omega_\mathrm{de}$ keeping the spatial flatness and $h$ is simultaneously changed to keep
$\omega_\mathrm{b}$ and $\omega_\mathrm{c}$ fixed.}.
In this way our emulator approach automatically incorporates a possible non-universality of the concentration-mass
relation. This is quite different in the standard analytical halo-model approach, where one usually employs a
simulation-calibrated scaling relation for the concentration.
We would also like to notice that such a calibration of the concentration
is often done for a specific cosmological model.

Now we focus on the halo mass density profiles at radii larger than $R_{200\mathrm{m}}$ in
Fig.~\ref{fig:radial_profile}, where NFW no longer gives a good fit.
The figure shows a clear feature of the transition from the one halo to the two halo regime.
This feature recently draws attention as a possible ``physical'' outer boundary of a halo associated with the first
orbital apocenter of accreted matter after its infalling, dubbed as the ``splashback'' feature
\citep[\citealt{Diemer14}, also see][for the first detection from observational data]{2016ApJ...825...39M}.
This feature has already been studied from our \textsc{Dark Quest} simulation suite in \citet{Okumura17,Okumura18}
with particular attention to the feature in the velocity statistics around halos.

The feature can be found from the bottom panel of Fig.~\ref{fig:radial_profile} where we show the logarithmic slope
of mass density profile. We obtain this by first fitting the emulator results by the functional form proposed in
\citet{Diemer14} (DK fit, hereafter), and then take the derivative. We do this for the separation range again
from twice the softening scale but to four times the radius $R_\mathrm{200\mathrm{m}}$, to cover both the one-
and two-halo regimes. The best-fit model is shown in the top panel by the dotted lines (but they are difficult to
distinguish from the solid lines; they are almost on top of each other), and the ratio to the emulator results are
shown in the middle panel. The accuracy of the fit is similar or better than the NFW form, and it remains to be a
good fit to much larger scales. Now, we can see in the bottom panel, that the derivative based on the DK fit shows
a sharp dip with a slope steeper than the outer NFW slope (i.e., $-3$), marking the location of the splashback
radius.

\begin{figure}[h]
\begin{center}
\includegraphics[width=8.8cm,angle=0]{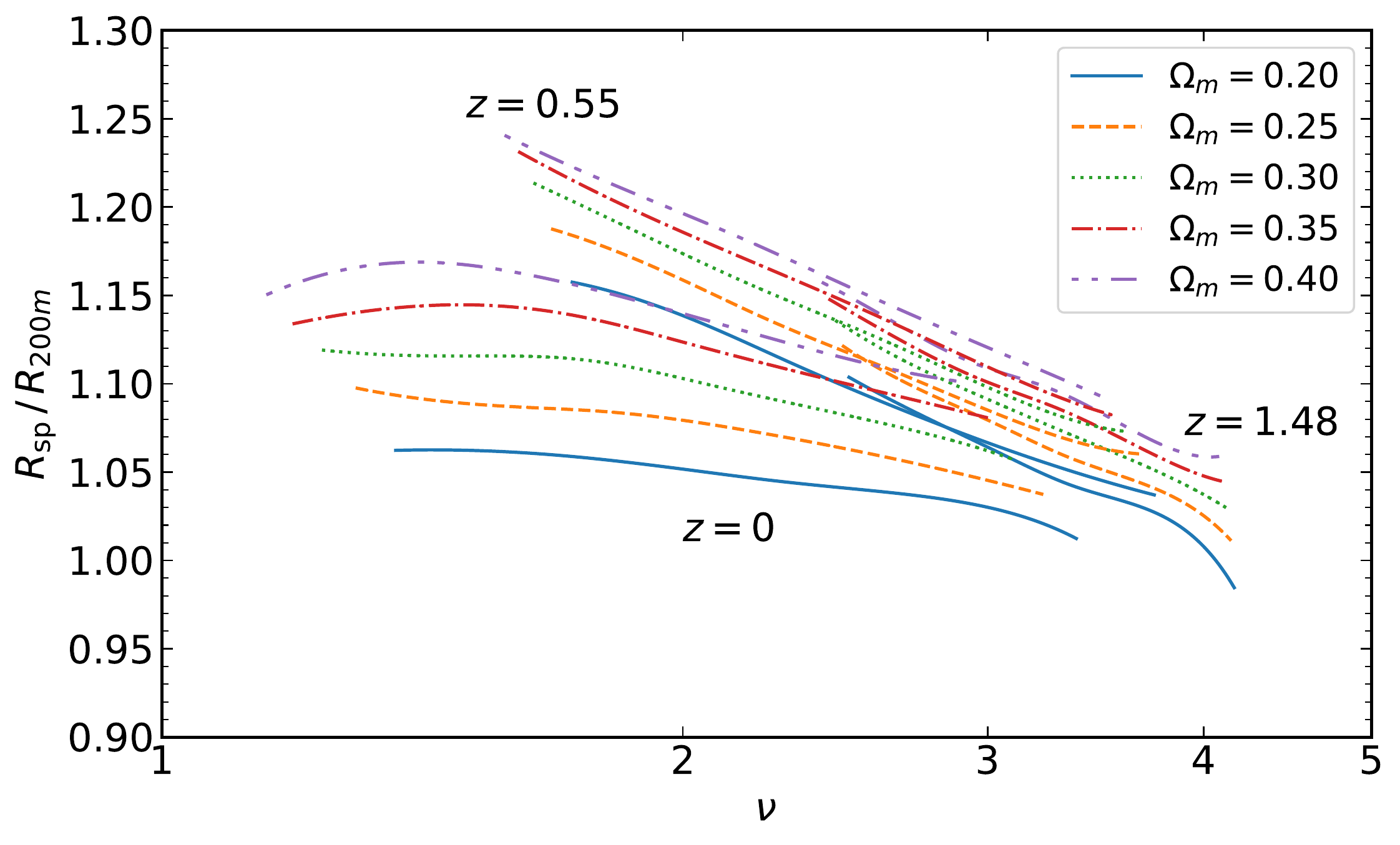}
\end{center}
\caption{Splashback radius divided by $R_{200\mathrm{m}}$.\label{fig:splashback}
}
\end{figure}
\begin{figure*}[htb]
\begin{center}
\includegraphics[width=8.5cm,angle=0]{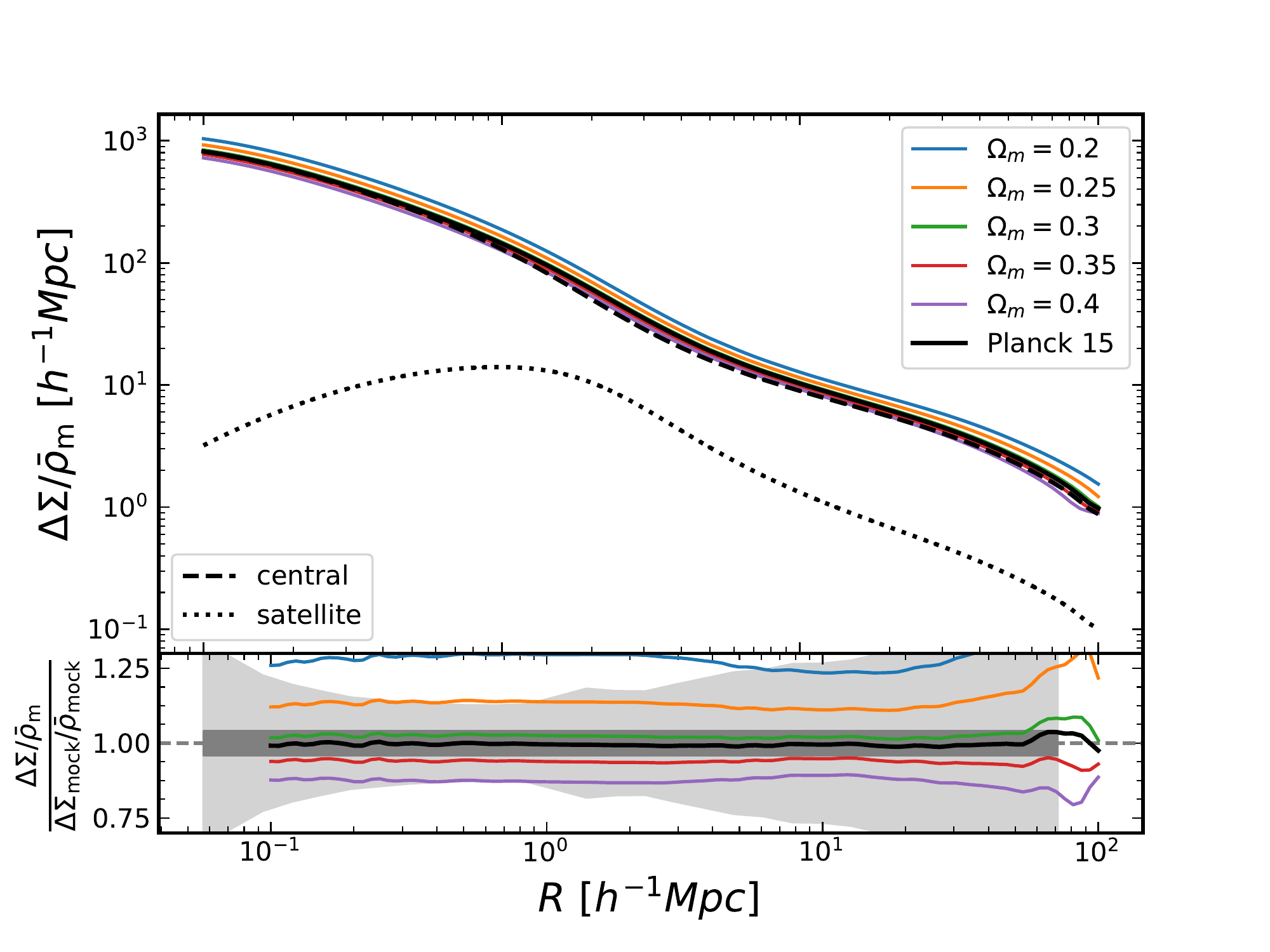}
\includegraphics[width=8.5cm,angle=0]{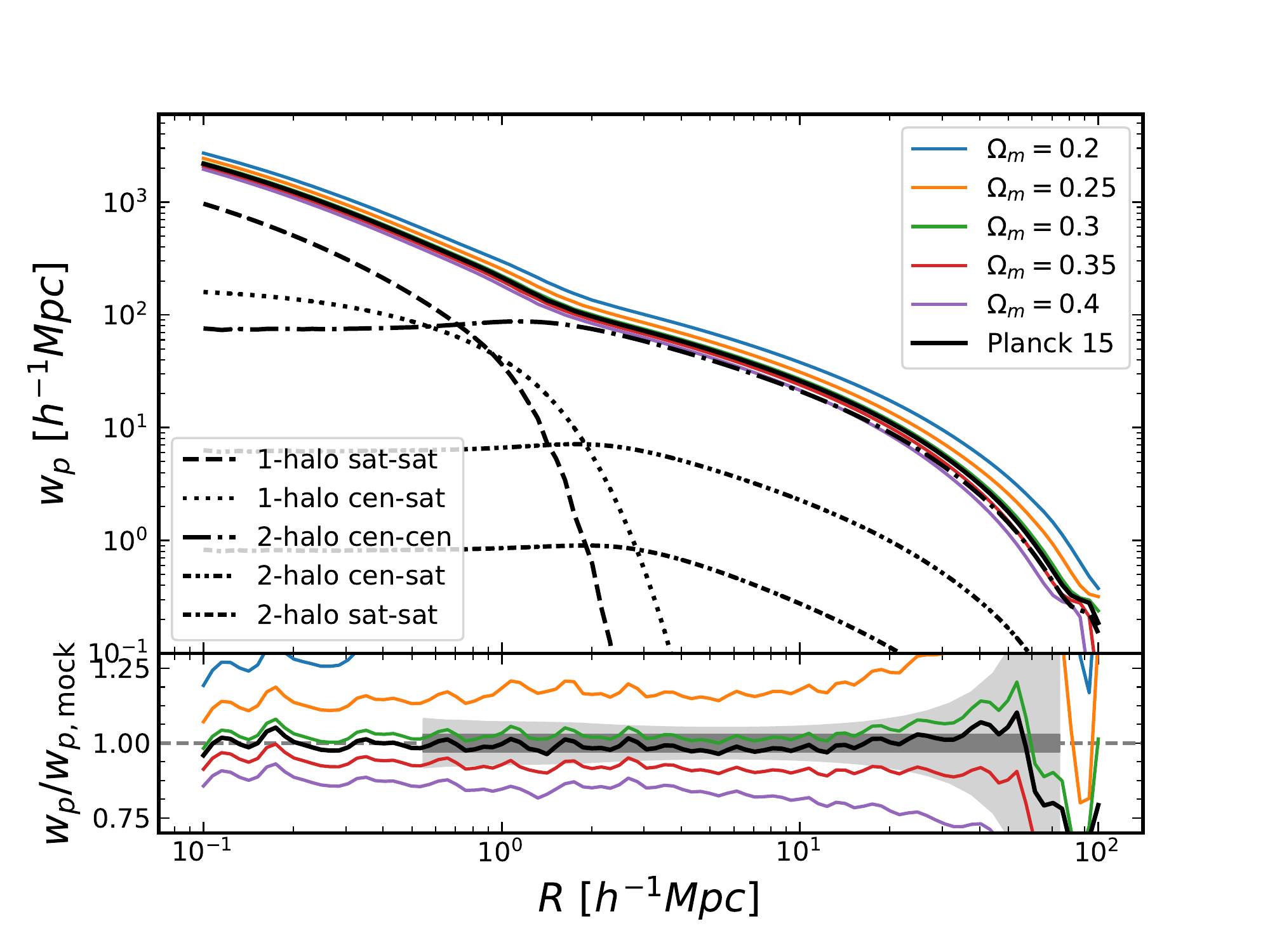}
\end{center}
\vspace*{-2em}
\caption{Example usage of \textsc{Dark Emulator} in combination with an HOD model for making model predictions of
galaxy clustering observables. {\it Left panel}: the excess surface mass density profile from galaxy-galaxy weak
lensing ($\Delta\Sigma$).
{\it Right panel}: the projected correlation function of galaxies ($w_{{\rm p}}$). In the upper panels, the solid
lines show how the prediction varies with $\Omega_\mathrm{m}$, but keeping other parameters fixed to the fiducial
{\it Planck} values. The other lines show the different contributions to the total power as indicated by the figure
legend (see text for details).
The lower panels compare the emulator-based predictions with the ``mock'' signals measured from 72 mock
realizations of projected maps of CMASS-type galaxies that are generated from the halo catalogs for the fiducial
{\it Planck} cosmology (see text for details).
The two results obtained from totally different methods are in remarkably nice agreement with each other.
In the left panel, the gray shaded region shows the measurement errors expected when combining the Subaru HSC
galaxies and the SDSS CMASS galaxies for background and foreground galaxies, respectively, where the overlapping
region is about 140~sq. degrees. In the right panel, we assume the measurement expected for the SDSS DR11 CMASS
galaxies around $z=0.484$ covering about 8,500~sq. degrees.
The dark gray regions around unity give a requirement on the overall uncertainty in the model prediction, which is
estimated from the inverse of the total signal-to-noise ratio over $0.057\le R\le 71~h^{{-1}}{\rm Mpc}$:
the requirements are about 0.04 and 0.029 corresponding to $S/N=25$ and 35 for $\Delta\Sigma$ and
$w_{{\rm p}}$, respectively. The black line shows that the emulator predictions safely meet the requirements over
the range of separation bins.
\label{fig:HOD}
}
\end{figure*}
Fig.~\ref{fig:splashback} shows
the splashback radius, $R_\mathrm{sp}$ for various cosmological models and redshifts, where we define
$R_\mathrm{sp}$ by the location of the minimum logarithmic slope of the DK fit.
For clarity we here plot the ratio, $R_\mathrm{sp}/R_{200\mathrm{m}}$, as a function of the peak height. Overall,
$R_\mathrm{sp}$ is similar to $R_{200\mathrm{m}}$, with a slight decreasing trend as a function of the peak height.
In addition, the ratio is higher for cosmological models with larger $\Omega_\mathrm{m}$. These trends are in
qualitative agreement with the fitting formulae in \citet{More15}, in which the dependence is encoded in the
redshift-dependent density parameter $\Omega_\mathrm{m}(z)$ in addition to the peak height $\nu$
\citep[see also][]{Adhikari14}\footnote{Note, however, it was argued that the majority of the dependence of
$R_\mathrm{sp}/R_{200\mathrm{m}}$ comes from the accretion rate, and its distribution at different redshifts and
for cosmologies should depend on how to define distinct halos and their mass accretion histories from $N$-body
simulation outputs in quite detail. This is beyond the scope of this paper.}

\subsection{Projected galaxy clustering statistics}
\label{subsec:projected}

We have introduced the emulation of halo clustering statistics in the previous sections.
Since our emulator's accuracy depends on the mass of halos, it would be useful to examine the accuracy for a \textit{galaxy} sample, whose clustering statistics is approximately given as a weighted sum of those of halos. Here we consider the following HOD parameters to make a representative galaxy mock catalog similar to the BOSS CMASS sample (based on a conservative volume-limited selection): $M_\mathrm{min} = 10^{13.94} \, h^{-1} M_\odot, \sigma_{\mathrm{log}M} = 0.63, M_1 = 10^{14.49}\,h^{-1} M_\odot, \alpha = 1.19$ and $\kappa = 0.60$ for the HOD parameters (see Appendix~\ref{sec:HOD} for definitions).

Our \textsc{Utility Modules} combines the outputs of \textsc{Halo Emulators} to first make the galaxy clustering signals in three dimensions, and then project them along the line of sight to obtain the relevant signals based on the FFTLog algorithm \citep{Hamilton00}.
To test the accuracy of \textsc{Dark Emulator} to predict galaxy clustering, we also generated the {\it mock} catalogs of galaxies;
we populate central and satellite galaxies into halos taken from the halo catalog in each of 24 HR realizations of the fiducial
{\it Planck} cosmology, and then measure the galaxy-galaxy weak lensing and the projected correlation function of galaxies from the mock catalogs.
To be more precise,
assuming the plane parallel approximation, we project the matter and galaxy distributions along one of the three axes in each realization
and then measure the galaxy-matter cross and galaxy auto-correlation functions using the two-dimensional FFT,
respectively. We use, as the prediction of the mock catalogs,
the average of the $72$ measurements ($24$ realizations times $3$ projection directions).
Note that the fiducial {\it Planck} cosmology is not used in the GPR and thus it should serve as a cross validation test after the additional ingredients in \textsc{Utility Modules}.

The measurements from the mock catalogs
are compared with the emulator predictions in Fig.~\ref{fig:HOD} for the galaxy-galaxy lensing (left panel) and the projected galaxy correlation function (right).
In the upper panels, we show the emulator predictions by the solid lines for models with different $\Omega_\mathrm{m}$ as indicated in the figure legend. For the fiducial {\it Planck} cosmology, the dashed, dotted and dot-dashed lines show the different contributions in the model calculations; the contributions from central and satellite galaxies are shown for the galaxy-galaxy lensing,
while the one- and two-halo term contributions are for the projected correlation function of galaxies.
For the galaxy-galaxy lensing profile, we plot $\Delta\Sigma/\bar{\rho}_{{\rm m}}$ for each of different $\Omega_{\mathrm{m}}$ models because
it becomes the same dimension as that of $w_{{\rm p}}$ in the right panel. With this definition, both the $\Delta\Sigma$ and $w_{{\rm p}}$ display
a similar dependence on $\Omega_{{\rm m}}$; increasing $\Omega_{{\rm m}}$ leads to a smaller amplitude
(we vary $\Omega_\mathrm{de}$ and $A_\mathrm{s}$, while the other four input cosmological parameters are fixed, to keep the spatial flatness as well as the value of $\sigma_8$).

The lower panels explicitly compares the emulator-based predictions with the mock measurements, showing the ratio for each galaxy observable,
for the fiducial {\it Planck} cosmology.
The gray shaded region around unity shows the statistical errors expected for the measurements. In the left panel,
we assume, for
the galaxy-galaxy weak
lensing measurement, the Subaru Hyper Suprime-Cam 1st year shape catalog \citep{2018PASJ...70S..25M}
and the SDSS DR11 CMASS galaxies at redshifts around $z\simeq 0.484$ \citep{2015ApJS..219...12A}
for background galaxy shapes and foreground lensing galaxies, respectively, where the overlapping region of the two data sets is about
140~sq. degrees. In the right panel, we assume the projected correlation function of the CMASS galaxies for about 8,500~sq. degrees \citep{2015ApJ...806....2M}.
Each panel shows that the ratio is very close to unity, meaning a remarkable agreement between
the emulator-based prediction and the mock measurement, for each observable.
Most importantly, the emulator-based predictions take just a CPU time of a few seconds.
The wiggly features in the ratio, especially for the projected correlation function, is due to an imperfect accuracy in the numerical calculation
such as the numerical integration of the emulator outputs over halo masses.
The gray shaded region gives statistical errors at each radial bin, estimated from the mock catalogs, where we assumed 30 bins over the range of
$0.057\le R\le 71~h^{{-1}}{\rm Mpc}$ corresponding to $\Delta\log_{{10}}R\simeq 0.1$. The dark shaded region gives an {\it overall} requirement on
the uncertainty in the model prediction of each observable. The requirement is estimated from the inverse of the total signal-to-noise ratio integrated over
all the radial bins. Sine we find $S/N\simeq 25$ and $35$ for the weak lensing and the projected correlation function, respectively, the requirement
on the overall factor in the model prediction, i.e. $m$ for
$\Delta \Sigma =(1+m)\Delta\Sigma_{{\rm emulator}}$ or
$w_{\rm p}=(1+m)w_{{\rm p}}{}_{{\rm emulator}}$, is $m\lesssim 0.04$ or $m\lesssim 0.029$, respectively,
 such that an uncertainty in the model prediction does not exceed the overall statistical error by more than $1\sigma$. The figure shows that the accuracy of \textsc{Dark Emulator} safely meets the requirements
for the Subaru HSC and SDSS measurements.
We note that, since variations in cosmological parameters cause a scale-dependent change in these observables,
the requirements for such changes are less stringent.

\subsection{Cross-correlation coefficient}
\label{subsec:r_hm}
\begin{figure}[htb]
\begin{center}
\includegraphics[width=8cm,angle=0]{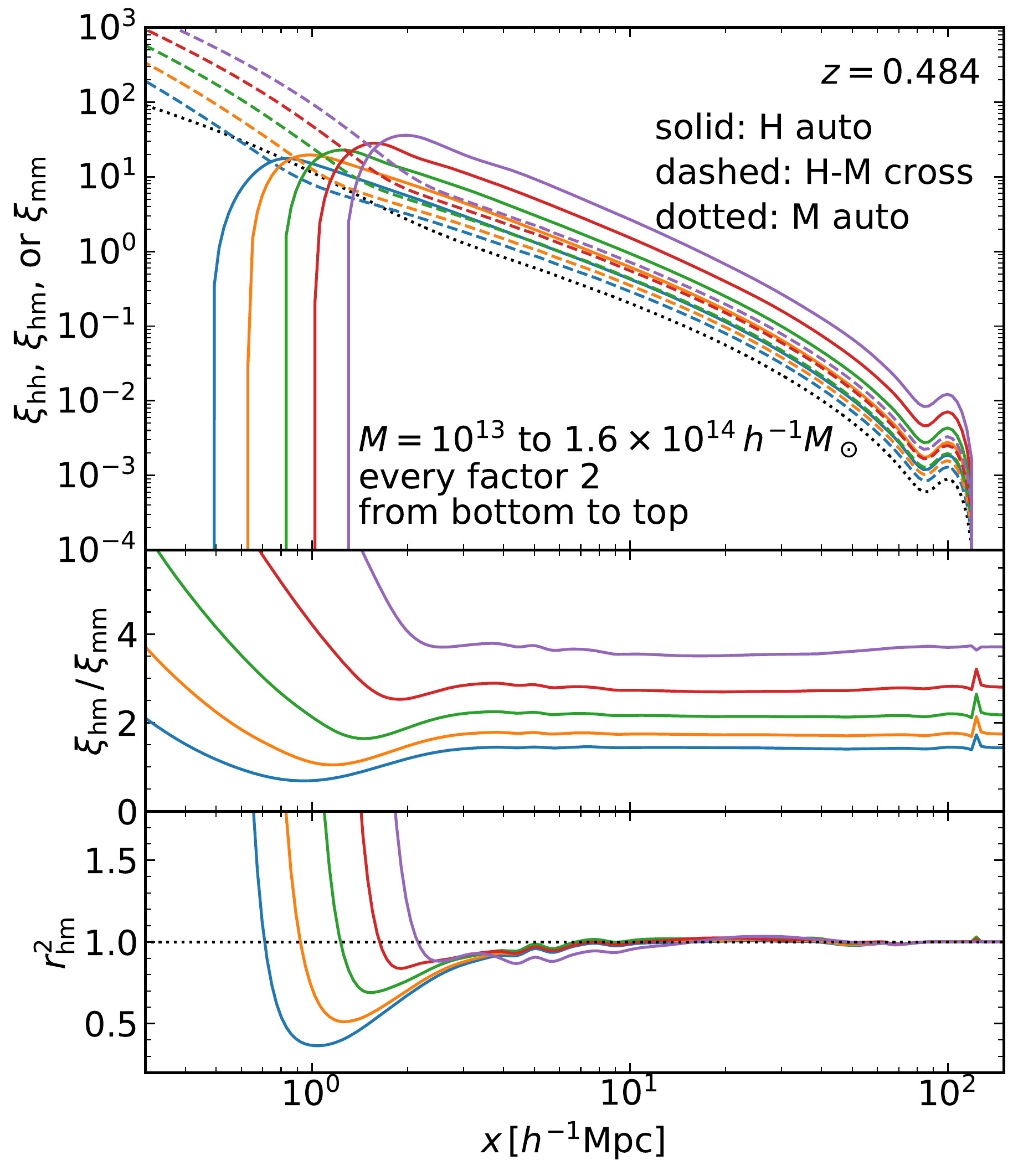}
\end{center}
\caption{Various halo clustering statistics at $z=0.484$. In the top panel, we show by the solid (dashed) lines the halo auto (halo-matter cross) correlation function for the masses indicated in the figure legend. We also show by the dotted line the matter correlation function. The middle panel depicts the halo bias defined by the ratio $\xi_\mathrm{hm}\,/\,\xi_\mathrm{mm}$. The bottom panel shows the square of the cross-correlation coefficient for the halo samples.
\label{fig:rhm}
}
\end{figure}
\begin{figure}[htb]
\begin{center}
\includegraphics[width=8cm,angle=0]{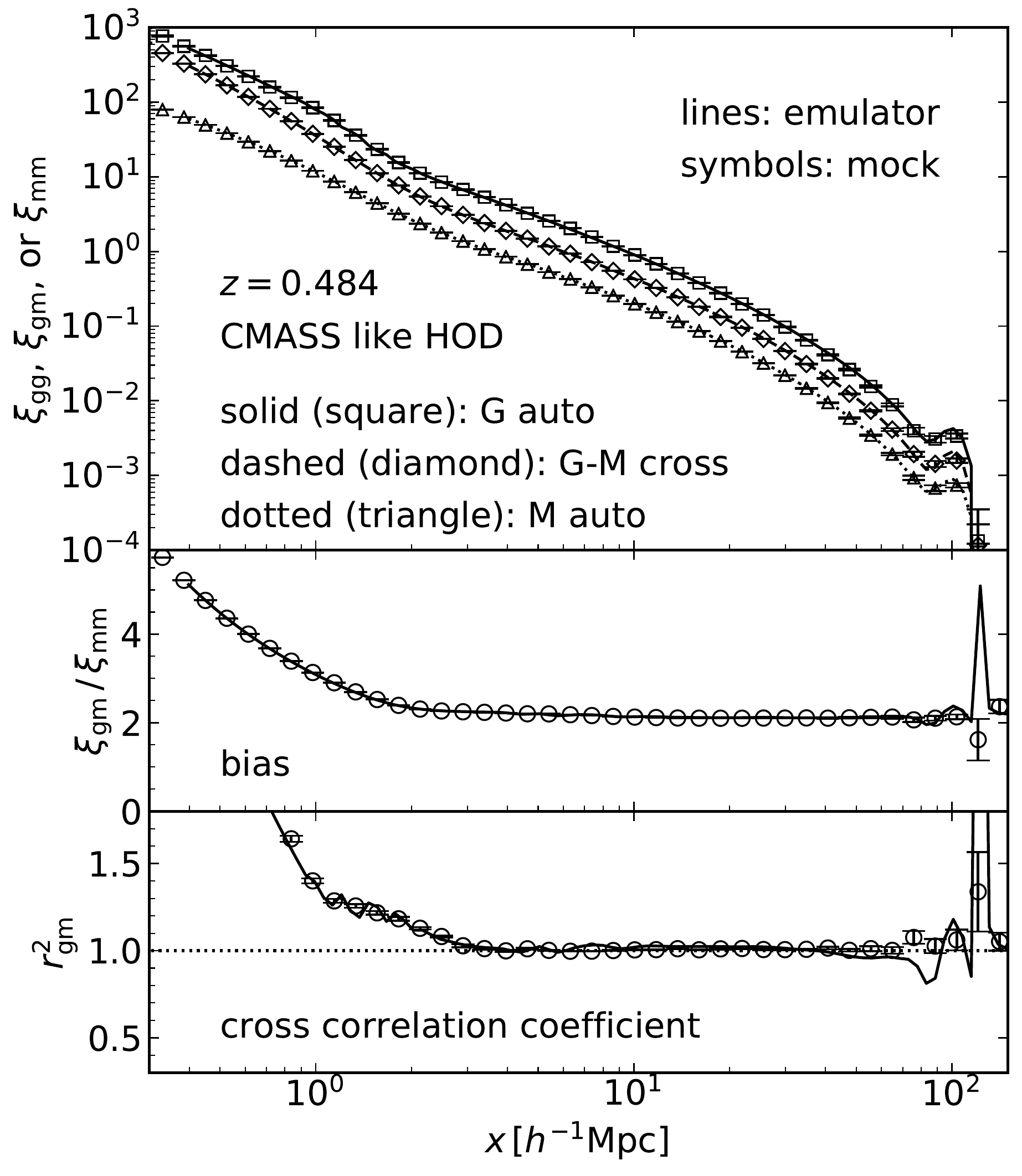}
\end{center}
\caption{Similar to Fig.~\ref{fig:rhm} panel, but for galaxy clustering with the HOD prescription that is the same as in Fig.~\ref{fig:HOD}. We also show here the measurements of the same quantities from the mock galaxies distributed following the same HOD model (circles with error bars).
\label{fig:rgm}
}
\end{figure}

There remains another interesting and important check. It would be of great practical use if we can infer the underlying matter clustering properties from biased fields alone. One can compute the cross-correlation coefficient between matter and halo, or between two different halo samples, from our emulators\footnote{To this end, we need an emulator to compute the matter auto correlation function, which is not supported in the current version of \textsc{Dark Emulator} specifically designed for biased tracers. While the optimization for implementation detail or the final accuracies are not tested as stringently as the other modules, we have a development version of a module to do this. The figures in this section are based on this version, but it would be sufficient for demonstration purposes.}.

First, we show in Fig.~\ref{fig:rhm} the cross-correlation coefficient for halo samples selected by mass at the fiducial {\it Planck} cosmology at $z=0.484$. The curves are computed by \textsc{Dark Emulator}. We consider five different halo masses from $10^{13}$ to $1.6\times10^{14}\,h^{-1}M_\odot$. We show in the top panel three quantities, $\xi_\mathrm{hh}(x)$, $\xi_\mathrm{hm}(x)$ and $\xi_\mathrm{mm}$. We then take the ratio $\xi_\mathrm{hm}(x)/\xi_\mathrm{mm}(x)$
to examine the scale dependence of bias in the middle panel. Finally, the bottom panel depicts the cross-correlation coefficient. A similar plot can be found in Fig.~\ref{fig:rgm} for galaxies based on the HOD model described in the previous section. Here, we also show by the symbols with error bars the measurements from the mock galaxies distributed in the simulated halos based on the same HOD prescription. The emulator predictions are in agreement with the measurements from the mock galaxies.

The middle panels indicate that the scale dependence of bias is rather weak on scales larger than several $h^{-1}\mathrm{Mpc}$ for all the cases investigated here. On these scales, the cross-correlation coefficient is close to unity within $\sim 10\%$. Based on the results, we may extract the underlying matter clustering statistics by combining the auto and cross correlation functions of  biased tracers. This statement would be true as long as we consider a simple model for the galaxy-halo connection as the HOD model considered here. Further studies are warranted to fully explore the potential to reconstruct the underlying matter clustering from real data for a wider class of galaxy populations.

\subsection{Summary of the current code}
\label{subsec:summary}
Finally, we summarize the functionalities of \textsc{Dark Emulator} in this section.

\begin{deluxetable*}{l|llll}
\tablecolumns{5}
\tablewidth{0pt}
\tabletypesize{\footnotesize}
\tablecaption{%
Summary of model parameters
}
\tablehead{\colhead{Class} &
\colhead{Parameter}
&\colhead{Prior range}
&\colhead{Definition}
}
\startdata
Cosmology & $\Omega_{\rm b}h^2$ & $[0.0211375, 0.0233625]$ & physical baryon density parameter & \\
& $\Omega_{\rm cdm}h^2$& $[0.10782, 0.13178]$ & physical CDM density parameter & \\
& $\Omega_{\rm de}$ & $[0.54752, 0.82128]$ & DE density parameter  & \\
& $\ln (10^{10} A_\mathrm{s}) $ & $[2.4752, 3.7128]$ & the amplitude of primordial power spectrum
& \\
& $n_\mathrm{s}$ & $[0.916275, 1.012725]$ & the spectral tilt of primordial power spectrum
& \\
& $w$ & $[-1.2, -0.8]$ & equation of state parameter of DE & \\ \hline
Common & $z$ & $[0, 1.47619]$ & redshift & \\ \hline
Halo & $M$ or $M_{\rm th}$\tablenotemark{a} & $[10^{12}, 10^{16}]$
& halo mass\tablenotemark{b} in $[h^{-1}M_\odot]$ & \\
& $M'$ or $M_{\rm th}^\prime$ & $[10^{12}, 10^{16}]$
& halo mass of the second halo for the halo-halo power spectrum  & \\
& $n$ & $[10^{-8.5}, 10^{-2.5}]$\tablenotemark{c} & halo number density\tablenotemark{c} in $[(h^{-1}\mathrm{Mpc})^{-3}]$ & \\
& $n'$ & $[10^{-8.5}, 10^{-2.5}]$ & number density of the second halo for the halo-halo power spectrum  & \\
\hline
HOD\tablenotemark{d} & \multicolumn{3}{l}{7 parameter model (default)
$\{M_{\rm min},\sigma_{\log M},\alpha_{\rm inc},M_{\rm inc},\kappa,M_1,\alpha\}$}
 \\ \hline
 Profile\tablenotemark{e} & \multicolumn{3}{l}{NFW model (default) $\{c(M,z),f_{\rm off},R_{\rm off}\}$}
\enddata
\tablenotetext{a}{The halo-matter and halo-halo power spectrum can be output for
a sample of halos with a given number density $n$,  for halos
at a given mass $M$ or for halos with masses greater than a given mass threshold $M_{\rm th}$ (see text).}
\tablenotetext{b}{The emulator employs $M_{200\mathrm{m}}$ for halo mass definition.}
\tablenotetext{c}{A warning message can be output if the input number density is too high for an input set of
cosmology parameters and redshift, i.e. if the input number density is outside the support of emulator.}
\tablenotetext{d}{We employ the halo occupation distribution (HOD) given by 7 parameters
as a default prescription for halo and galaxy connection. A user can replace this module with another
prescription if needed.}
\tablenotetext{e}{We assume that the distribution of satellite galaxies in their host halo follows
a normalized Navarro-Frenk-White (NFW) model, where the halo mass and concentration follows the fitting formula
in \citet{2015ApJ...799..108D}, as our default model. Another option is to distribute satellite galaxies following the matter distribution around a halo as predicted by the halo matter cross correlation function. We also include a possibility that a fraction $f_{\rm off}$
of central galaxies is offset from the true halo center and assume that the normalized distribution, with respect to the true center,
is a Gaussian with width radius $R_{\rm off}$.
A user can replace this module if needed.}
\label{tab:galaxy_parameters}
\end{deluxetable*}

First, the input parameters for the code are listed in Table~\ref{tab:galaxy_parameters}.
The items in ``Cosmology'' class are the six cosmological parameters of $w$CDM cosmologies
considered in this paper.
In ``Common'' class, we have redshift $z$ as a common parameter for all the modules. Quantities calculated from linear theory are evaluated at $z=0$, and then properly scaled by the linear growth factor. The third class is ``Halo'' relevant for \textsc{Halo Modules}. The parameters in this class
specify a halo sample, either in terms of a mass range or a specific mass scale
at which the desired halo clustering quantities
are evaluated. The number density and the mass threshold can be converted
to each other using the module that computes the halo mass function. Next, we have ``HOD'' class. The parameters here determine how many galaxies (centrals and satellites) are populated in the halos. Finally, we have a set of parameters that
model variations in the locations of the galaxies inside a halo. For instance we
allow central galaxies to be off from the true halo center using
the two off-centering parameters. The satellite galaxies are assumed to follow either the NFW profile or
the average halo mass profile, where the latter is equivalent to the halo-mass cross-correlation for halos of each
mass range that is an output of our emulator.
In case of the NFW profile we assume the mass-concentration relation calibrated by \citet{2015ApJ...799..108D}.

The HOD model as well as the profile of galaxies can be modified easily when needed. This way we allow the code to have flexibility to support various galaxy populations possibly beyond the model currently implemented. It has been suggested that the clustering
statistics could depend
on a secondary parameter beyond the halo masses. The so-called halo assembly bias is not implemented in the current code. We study the impact of such effects to cosmological analyses in a separate paper. Our code can also account for the effect of the residual redshift space distortions to the projected statistics with a finite projection width assuming linear theory as well as a modification of the mass profile around halos due to baryonic effects in a parametric manner. Again, these effects are studied in full detail in a separate paper.

\begin{deluxetable*}{l|lll}
\tablecolumns{4}
\tablewidth{0pt}
\tablecaption{%
Summary of the emulator output. \label{tab:emulator_out}
}
\tablehead{\colhead{Class} &
\colhead{Output}
&\colhead{Definition}
}
\startdata
Linear & $\sigma^2(M,z)$ & linear mass variance & \\
 & $\sigma_\mathrm{d}(z)$ & rms linear displacement in one dimension & \\
 & $P_\mathrm{lin}(k,z)$ & linear matter power spectrum & \\ \hline
Primary & $n(M_\mathrm{min},M_\mathrm{max}; z)$ & number density of halos in the mass range $[M_\mathrm{min}, M_\mathrm{max})$& \\
& $\xi_{\rm hm}(x; M,z)$& the 3D halo-matter cross-correlation for halos  & \\
& $\xi_{\rm hh}(x; M, M', z)$& the 3D halo auto-correlation for halos of masses $M$ and $M'$ & \\
& $G_\mathrm{m}(k;z)$ & the propagator for the matter density field & \\
& $G_\mathrm{h}(k;M,z)$ & the propagator for the halo density field specified by mass $M$& \\
\hline
Derived & $\Sigma_{\rm hm}(R; M, z)$ & the surface mass density profile of halos of $M$ & \\
& $\Delta \Sigma_{\rm hm}(R; M, z)$ & the excess surface mass density profile around halos of $M$ & \\
& $\Delta\Sigma_{\rm gg}(R;z)$ & the excess surface mass density profile around galaxies & \\
 & $w_{\rm gg}(R;z)$ & the projected correlation function of galaxies &
\enddata
\end{deluxetable*}

Finally, the outputs of the emulator are summarized in Table~\ref{tab:emulator_out}.
The first set of outputs are the three linear quantities based on \textsc{Linear Modules}. The primary outputs of
the emulator are the abundance and the clustering of halos and matter. These include the abundance of halos in a
given mass range, the halo matter cross-correlation function, halo auto-correlation function and the propagators of
matter and halo. These quantities are then combined and projected based on analytical calculations
to eventually have the items in the ``Derived'' class. The connection between halos and galaxies as specified in
the \textsc{Utility Modules} (i.e., HOD and profile) is reflected to the final galaxy statistics.

\section{Summary}
\label{sec:summary}
In this paper, we have performed an $N$-body simulation ensemble, dubbed as \textsc{Dark Quest}, and
then developed an emulator enabling a fast computation of halo clustering quantities from the simulation outputs,
named \textsc{Dark Emulator}.
The main features of our products are
\begin{itemize}
\item $2048^3$ particles were employed in either $1$ or $2~$Gpc/$h$ comoving boxes, covering $100$ six-parameter $w$CDM cosmological models sampled via the Sliced Latin Hypercube Design around a fiducial $\Lambda$CDM cosmology.
The mass density fields and the catalogs of halos with $M_{200}\gtrsim 10^{12}\,h^{-1}M_\odot$ (slightly depending on cosmological models)
were extracted at $21$ redshifts in the range of $z=[0,1.48]$.
The parameter space covers a sufficiently broad range of parameters that are consistent with the existing cosmology datasets.
\item We used the \textsc{Dark Quest} datasets to build \textsc{Dark Emulator}. It models the halo mass function, halo-matter cross-correlation and halo auto-correlation based on the Gaussian Process regression after significant dimension reduction via the principal component analysis. The predicted halo clustering properties are easily combined assuming a model for the halo-galaxy connection, such as an HOD prescription, to compute the galaxy statistics.
\item We carefully validated the accuracy of the Dark Emulator predictions (outputs) using validation samples of
$N$-body simulations for cosmological models that are not used in the emulator development.
The validation samples are also located following the Latin Hypercube Design, and are maximin design by themselves,
in combination with the training samples. Thus they allow us to test the accuracy at distant points from the
nearest training data, and at the same time, covering uniformly the whole domain of the parameter space.
\item We achieved $1 \text{--} 2\%$ accuracy for the halo mass function (the rms error over the $20$ models) except for the massive end ($M\gtrsim 10^{14}h^{-1}M_\odot$) where the Poisson error is significant both in the training and the validation sets. The accuracy for the halo-matter cross correlation function for a halo sample with number density $10^{-4} (h^{-1}\mathrm{Mpc})^{-3}$,
which resembles typical host halos of LRGs or CMASS-like galaxies,
was shown to be $\sim 2\%$ over the comoving separation $0.1\,h^{-1}\mathrm{Mpc} < x < 30\,h^{-1}\mathrm{Mpc}$ (again in terms of the rms error). The halo-halo auto correlation function for the same halo sample has a slightly larger error, $\sim 3 \text{--} 4\%$, reflecting the shot noise error. The accuracy gets worse at $x\lesssim 1\,h^{-1}\mathrm{Mpc}$, where the halo exclusion effect is significant and thus do not contribute much to
galaxy clustering signals.
In all the cases, the biggest discrepancy between the prediction and the validation set is not
worse than $5\%$ over the ranges of halo masses and separations.
\item The accuracy of the emulator depends on the halo mass and slightly on the redshift. This can be checked in Appendix~\ref{sec:extra_dependence}. We find overall that the validation accuracy scales consistently with the sample variance error estimated from the multiple random realizations prepared for the fiducial cosmology. Thus a
further significant improvement of the accuracy would be possible only by
using more simulations (with a larger box size in addition), and further refinement of the implementation detail would not at this moment.
\item We introduced a special treatment based on the propagator to large-scale clustering signals where the large sample variance prevents us from an accurate modeling or an accurate validation test. The propagator encodes the large-scale bias as well as the damping of the BAO feature and it can be measured accurately as the sample variance mostly cancels in its estimator. A module that emulates the propagator is also trained and validated to ensure the accuracy of the predictions of the correlation functions on large separations.
\item We demonstrated that the \textsc{Dark Emulator} outputs can be used to study detailed properties of
the mass density profiles around halos such as the concentration-mass relation and the splashback feature.
The emulator can predict their dependence on redshift, halo mass and cosmological models.
\item We also demonstrated that the emulator outputs can be used to predict, as an example, the projected galaxy
correlation function and the galaxy-galaxy weak lensing profile when combined with a prescription of the
halo-galaxy connection such as an HOD model.
In doing this we can easily incorporate variants of the small-scale effects such as the off-centering of
galaxies with respect to the halo center, the incompleteness selection of galaxies,
and the distribution of satellite galaxies in their host halo based on a Fourier-space implementation.
The \textsc{Dark Emulator} modules extensively uses the FFTLog algorithm that enables a fast computation of
converting the three-dimensional correlation functions to the projected correlation functions.
\item The evaluation time for the halo and galaxy statistics is typically of order $\sim 100$ milliseconds and a few seconds, respectively, on a standard laptop computer available today. The latter is slower because it usually involves integrals over the halo masses.
\item The cross-correlation coefficient between halos and matter are shown to be quite close to unity on large scales. This remains the same for galaxies populated into halos based on the HOD description.
\end{itemize}

The current implementation and accuracy are likely sufficient for the ongoing wide-area galaxy surveys such as the
Subaru HSC survey (see Fig.~\ref{fig:HOD} for its validation). It is still not clear how the cosmological
information can be extracted from the cosmological dependences of halo clustering quantities that are measured from
such a galaxy survey, even after marginalizing over nuisance parameters that model the small-scale clustering in
the 1-halo term. To address this, one has to use realistic mock catalogs that resemble the actual galaxy survey,
measure clustering observables of interest from the mock catalogs including all realistic small-scale
effects, and then make a hypothetical parameter inference from the comparison of the emulator predictions with the
mock measurements, including marginalization of the nuisance parameters
\citep[][for a similar discussion]{2017JCAP...10..009H}.
This kind of study can assess the power and usefulness of the \textsc{Dark Emulator} for precision cosmology,
and gives a validation of the parameter inference method -- cosmology challenges.
This is our ongoing project and will be presented in the future (Miyatake et al. in preparation).
Our implementation to incorporate the residual redshift-space distortion as well as the baryonic effects to the
mass profile around halos will also be presented and tested in that paper.

However, the current emulator would not meet an accuracy required for future surveys such as LSST, Euclid and
WFIRST. As already mentioned above, a naive and straightforward way is to accumulate more simulation data and reduce the statistical uncertainties on the training data. Since the current implementation of \textsc{Dark Emulator} includes several approximate treatments, the systematic error from them can be a problem with the improved statistical error. These include
\begin{itemize}
\item The sample variance error on the measured statistical signals is assumed to be diagonal (i.e., no off-diagonal covariance) and independent of cosmological models.
\item The PCA coefficients are modeled by GPR one by one ignoring the correlation between them.
\item The metric in the cosmological parameter space is assumed to be stationary (i.e., independent of the location in the space).
\item The functional forms assumed in the HMF and the propagator might be insufficient for ultimate precision.
\item Although the damping of BAO peak is already included, a possible ``shift'' of the BAO scale due to nonlinearity is ignored.
\item Extra dependence of the halo clustering properties other than the mass dependence, i.e., halo assembly bias, is not considered at all.
\item{The current emulator supports halos with mass $\gtrsim 10^{12}h^{-1}M_\odot$. This should be improved to model e.g., emission line galaxies which form in less massive halos.}
\item{The suite of $N$-body simulations and halo catalogs can also be used to study intrinsic alignments (IA) of halo shapes and their dependences on cosmological models, halo mass and redshifts. IA is not only one of the major systematic errors in high-precision weak lensing measurements,  but also can be a new cosmological probe as it arises from large-scale structures. This is our future project, and will be presented elsewhere.}
\end{itemize}

Nevertheless, we are optimistic for such a challenge. As we stressed, we designed the \textsc{Dark Emulator} to
cover a sufficiently broad range of cosmological models within $w$CDM cosmologies, which are much broader than the
models favored by the {\it Planck} CMB measurements, because we want to keep a broader range of applications of
the \textsc{Dark Emulator} to problems which users might want to study.
If the range of cosmological parameters are narrowed down and if specific requirements for given clustering
observables for a future survey under consideration are given, some of the approximations, such as the
cosmology-independent modeling of the statistical error or the stationary metric, would be even more appropriate.
We can also design a new set of $N$-body simulations to run in the new narrower parameter space and then construct
an emulator that can meet the requirements.
We believe that the methods and techniques developed in this paper would be useful to explore such an $N$-body
simulation suite and then develop a sufficiently accurate emulator enabling to predict the clustering observables
that one wants to use for precision cosmology. The current version of \textsc{Dark Emulator} will be made public in the near future.

\acknowledgments
We appreciate useful comments by Salman Habib and Katrin Heitmann on the numerical simulations and Gaussian process.
We also thank Surhud More for useful discussion during the early stage of this work and providing us with the HOD model for the SDSS galaxies. We appreciate useful discussion with Shiro Ikeda and Naonori Ueda on Bayesian techniques and experimental design.
This research was supported by World Premier International Research Center Initiative (WPI), MEXT, Japan.
This work was in part supported by MEXT Grant-in-Aid for Scientific Research on Innovative Areas (No.~JP15H05887, JP15H05892, JP15H05893, JP15K21733),
by Japan Science and Technology Agency CREST JPMHCR1414, by MEXT Priority Issue 9 on Post-K Computer
(Elucidation of the Fundamental Laws and Evolution of the Universe), and by JICFuS.
This work was also supported by JSPS KAKENHI Grant Numbers JP17K14273 (TN), JP15H03654 (MT), JP17H01131 (RT), JP16J01512 (KO), JP18H04358 (MS), JP18H04350 (HM), JP18K03693 (MO) and JP17J00658 (RM).
Numerical computations were carried out on Cray XC30 and XC50 at Center for Computational Astrophysics, National Astronomical Observatory of Japan.

\bibliographystyle{yahapj}
\bibliography{lssref}

\appendix

\section{Linear modules}
\label{sec:linear}
While the linear-theory predictions of cosmological structure formation can be obtained accurately and quickly using public Boltzmann codes such as \textsc{CMBFAST}~\citep{cmbfast}, \textsc{CAMB}~\citep{camb}, and
\textsc{CLASS}~\citep{class1,class2}, the computation time is still
non-negligible, e.g. for parameter inference using Markov chain Monte Carlo
in a high-dimensional parameter space.
Within \textsc{Halo Modules},
we need to evaluate the linear power spectrum $P_\mathrm{lin}(k)$, the mass variance $\sigma_M$, and the rms displacement $\sigma_\mathrm{d}$ for the input set of cosmological parameters, halo mass, and redshift.
In particular, the latter two involve an integral over wavenumber.
To speed up the computations, we develop an emulator module that allows for a quick computation of these quantities based on PCA and GPR methods as we did in \textsc{Halo Modules}\footnote{See \citet{PICO1,PICO2} for a similar attempt to speed up the calculation of linear power spectra.}

We first sample $400$ sets of cosmological parameters
each of which is taken within the range of parameter given by
Eq.~(\ref{eq:params}), using the \textsc{SLHD} scheme.
In this case, we generate ten slices with $40$ samples each.
We use \textsc{CLASS} code to compute the relevant linear-theory quantities
for each model, although we have used \textsc{CAMB} for the initial
conditions of our $N$-body simulations.
While it is known that the results of \textsc{CAMB} and \textsc{CLASS} can differ slightly depending on their accuracy parameters, the difference is typically at a sub-percent level,
which is much below our target accuracy here. We use the $360$
cosmological models in nine slices as the training data and test the accuracy of the emulator
using the remaining $40$ models in the last slice as a validation sample.
We reduce the dimensionality of the data vector for
$P_\mathrm{lin}(k)$ and $\sigma_M$ by keeping only the most significant principal components.
We originally sample $P_\mathrm{lin}(k)$ ($\sigma_M$) by $200$ points over wavenumber ($401$ points over mass) and keep only $13$ ($4$) PCs. Since $\sigma_\mathrm{d}$ is a single number, we use it as it is. Finally, the coefficients of the PC eigenvectors (or $\sigma_\mathrm{d}$) are modeled by GP regression.

\begin{figure}[h]
\begin{center}
\includegraphics[width=10cm,angle=0]{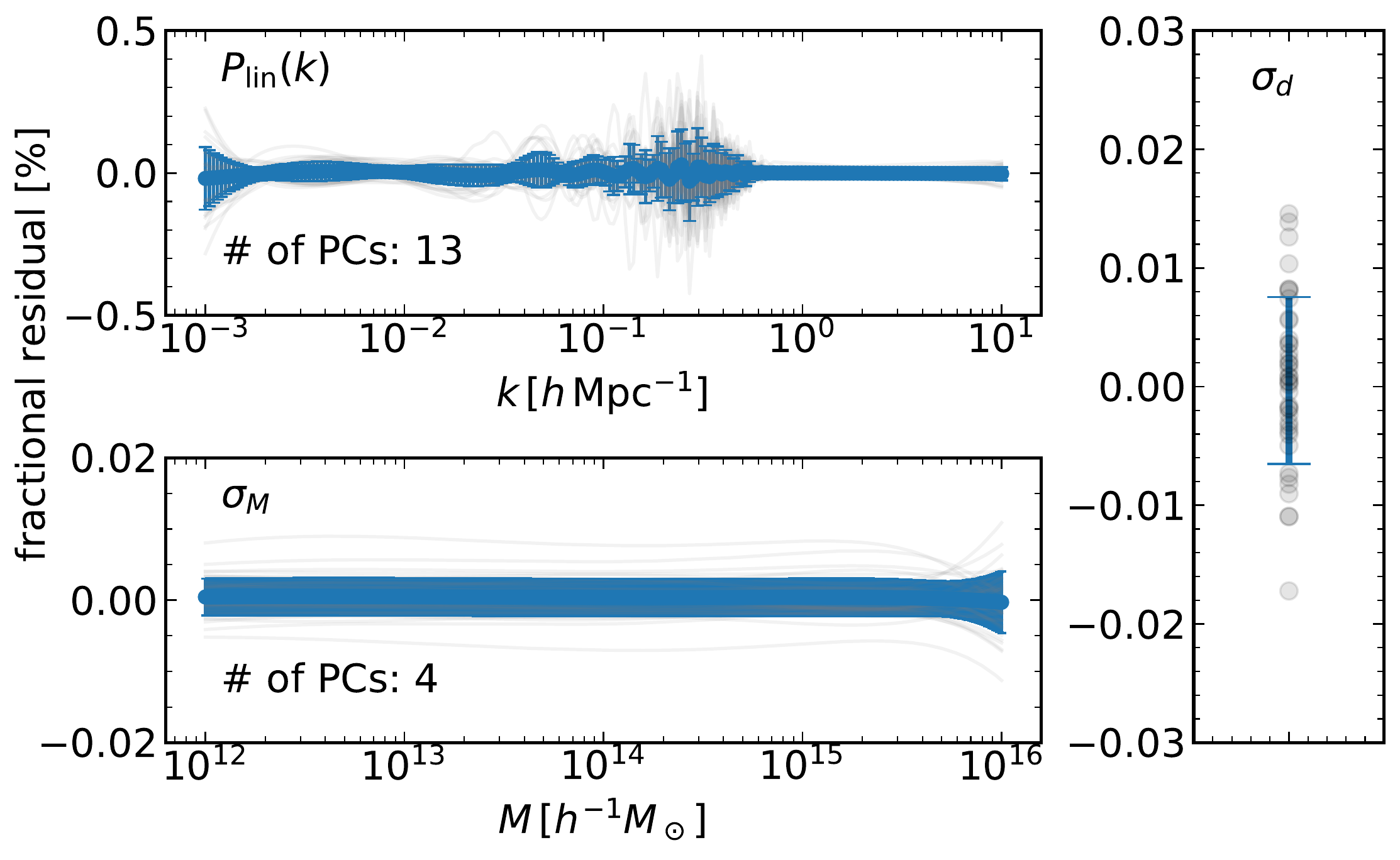}
\end{center}
\caption{Cross validation study of the \textsc{Linear Modules}. We show the fractional residual in percent
comparing the emulator prediction to the direct evaluation by \textsc{CLASS} for each of
the $40$ validation cosmological models in a \textsc{SLHD} slice by gray lines or circles in each panel (left upper: $P_\mathrm{lin}(k)$, left lower: $\sigma_M$ and right: $\sigma_\mathrm{d}$). In each panel, we also show by the error bars the scatters among the $40$ models ($1\sigma$ level).
\label{fig:linear}
}
\end{figure}

The accuracy of our model is assessed by cross validation and is shown in Fig.~\ref{fig:linear}. The accuracy is always better than $1\%$, with $P_\mathrm{lin}(k)$ generating the biggest error of $\sim 0.3 \text{--} 0.4\%$. The accuracy for the other two quantities is even better, and is typically $0.01\%$ level. The bigger error on $P_\mathrm{lin}(k)$ is attributed to the characteristic features of the baryon acoustic oscillations. This is contrasted to the rather smooth and monotonic dependence of $\sigma_M$ on $M$. The current implementation of the \textsc{Linear Modules} enables us to evaluate all these quantities in a few milliseconds for
an input cosmological model,
which is a negligible time in the whole calculation of \textsc{Dark Emulator}.

\section{Initial conditions of $N$-body simulations}
\label{sec:IC}
\subsection{Optimal initial redshift}
\label{subsec:zini}
In this appendix, we discuss how the choice of redshift used to set the initial conditions of $N$-body
simulation affects the results of late-time clustering.
While a higher initial redshift is preferable to reduce the transient effect arising from the fact that the initial conditions do not follow the growing solution precisely at higher orders \citep{crocce06b}, the regular lattice pre-initial configuration can excite spurious modes \citep[e.g.,][]{Marcos06,Joyce07,Garrison16} if the starting redshift is very high. The latter effect can be understood via ``particle linear theory'' \citep{Marcos06}: the growing solution of particles close to the lattice configuration is different from the fluid growing solution in a direction-dependent manner. The net effect after averaging over the direction is to slow down
the growing modes compared to what the \textit{fluid} linear theory predicts, and this
becomes more important toward larger wavenumbers. While a numerical method to correct for this effect was recently proposed by \citet{Garrison16}, we here adopt a simpler approach. Since both the effects suppress the structure growth, we choose an initial redshift so that the $N$-body simulation produces the highest power spectrum at late times.

We first focus on the evolution of the matter density contrast at early epochs. We generate
particle distributions from an identical random realization of the linear density field at different
initial redshifts ($1+z_\mathrm{in}=15$, $30$, $60$, $120$, and $240$)
using either of the Zel'dovich approximation or the 2LPT.
We implement this numerical experiment using $N$-body simulations in a cubic volume
with side length of $L=250\,h^{-1}\mathrm{Mpc}$ employing
two different resolutions, one with $512^3$ and the other with $256^3$ particles, corresponding to the resolution of \textsc{HR} and \textsc{LR} runs, respectively.
We run $N$-body simulations assuming the different initial conditions, and then measure
the matter power spectrum at later epochs.
In doing this, we store the snapshots of $N$-body simulations at different epochs
starting from $z=49.75$ corresponding to the linear growth factor $D_+=0.025$ down to $z=0$ at every interval of $\Delta D_+ =0.025$ ($40$ snapshots in total).

\begin{figure}[h]
\begin{center}
\includegraphics[height=10.4cm,angle=0]{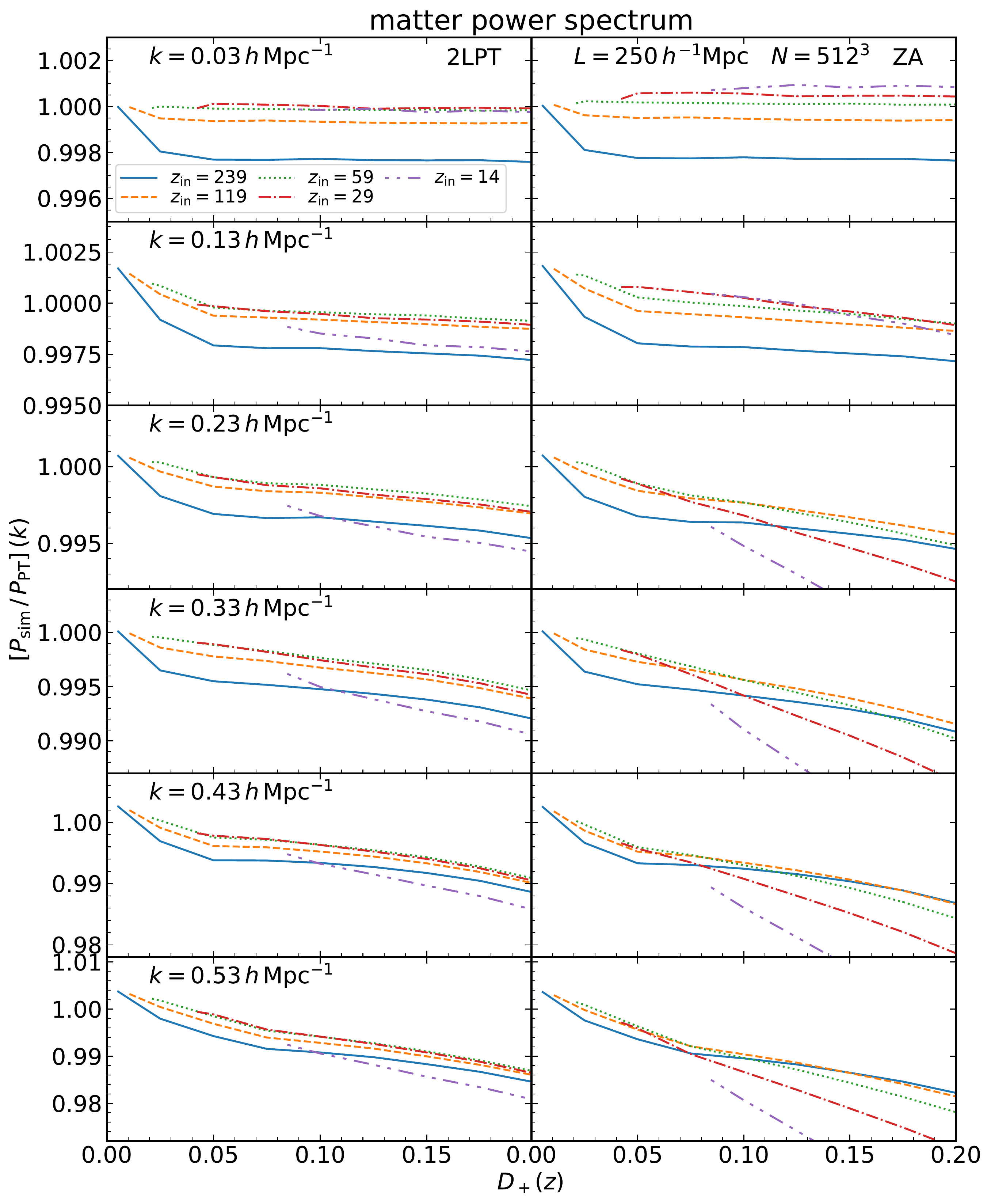}
\includegraphics[height=10.4cm,angle=0]{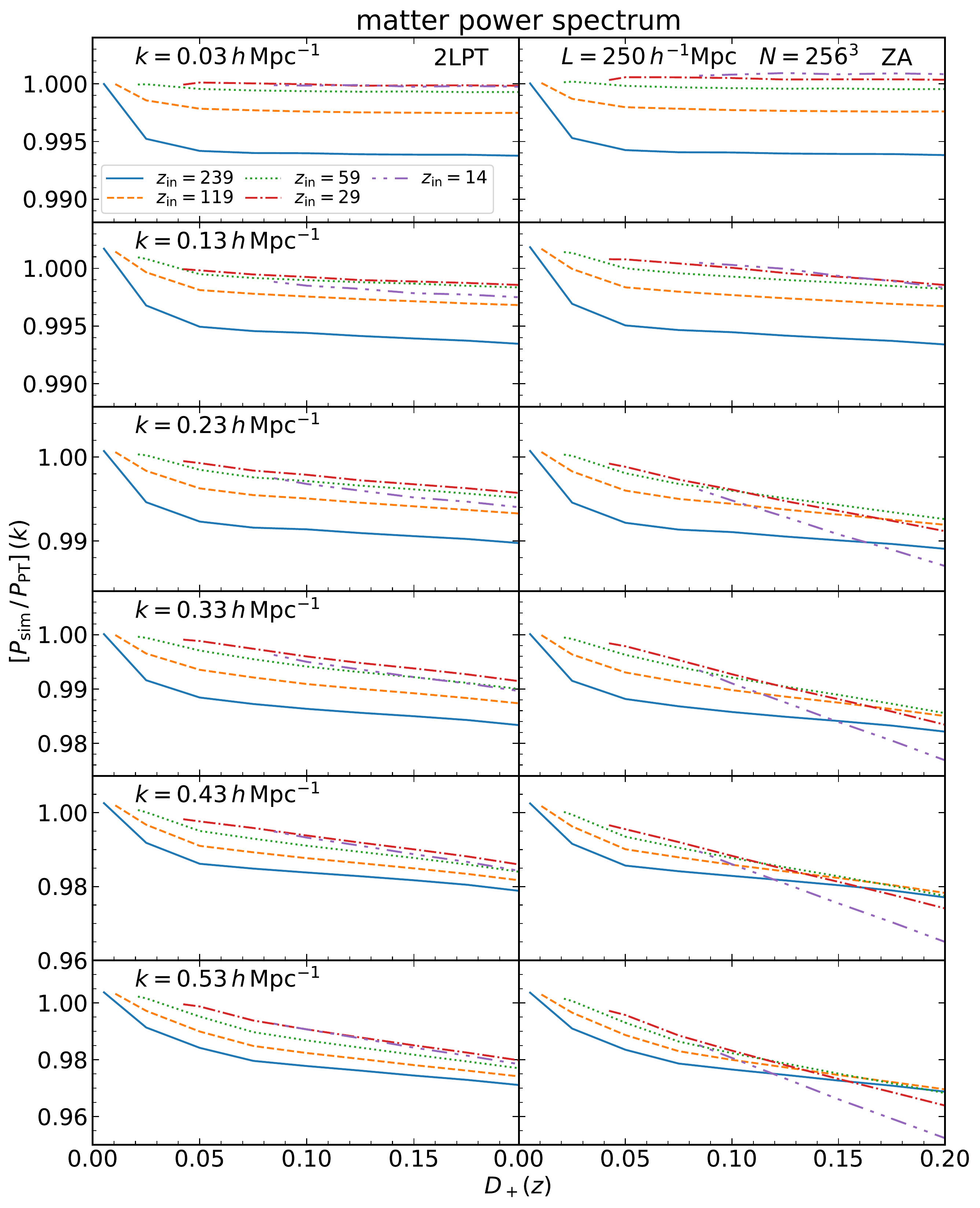}
\end{center}
\caption{Time evolution of the matter power spectra in $N$-body simulation relative to
the perturbation-theory predictions
(see text for detail), where we use the simulations with $512^3$ or $256^3$ in a cubic
volume with side length of $250\,h^{-1}\mathrm{Mpc}$ in the left or right panel, respectively.
Note that the simulations have
the same resolution as those of the \textsc{HR} (\textsc{LR}) runs.
We plot the ratio as a function of the linear growth rate that is normalized to unity at present.
Different type lines correspond to different initial redshifts of the simulations as denoted in the legend.
In each panel, we show the results of 2LPT and ZA initial conditions in the left and right column, respectively, which are used to set up the initial displacements of $N$-body particles.
\label{fig:pnl_evol}
}
\end{figure}

\begin{figure}[h!]
\begin{center}
\includegraphics[height=16cm,angle=0]{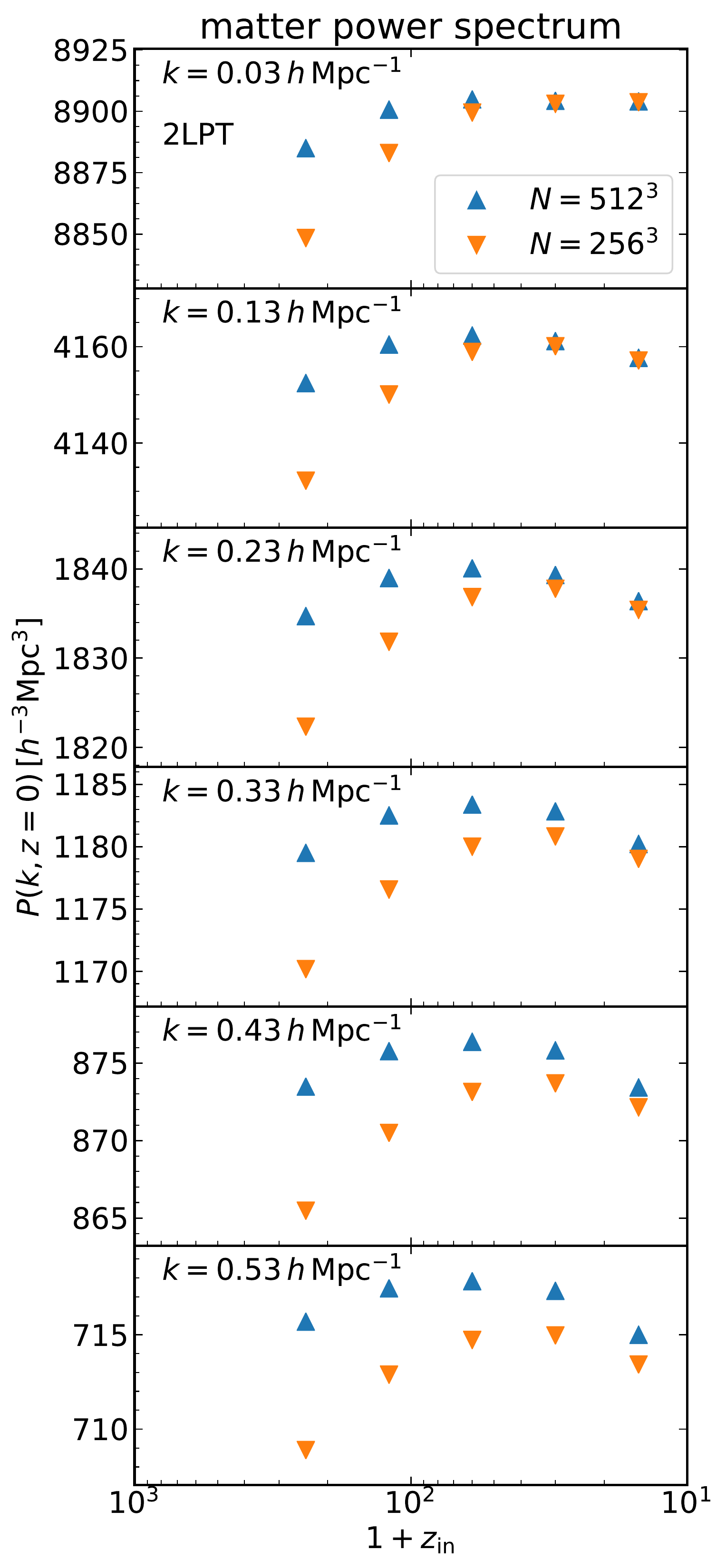}
\includegraphics[height=16cm,angle=0]{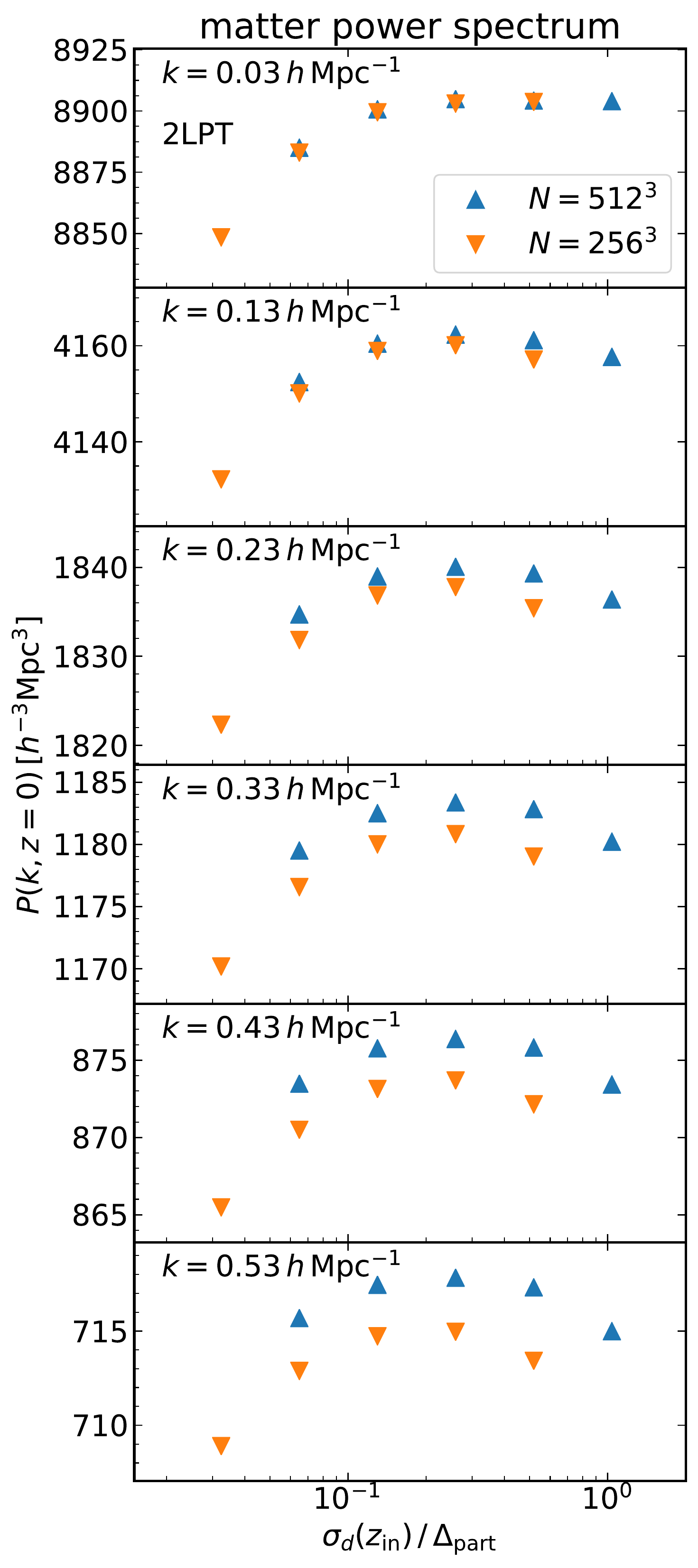}
\end{center}
\caption{Dependences of the matter power spectrum amplitudes at $z=0$ on the initial redshift of simulation for
different wavenumbers. The left plot shows the results as a function of the initial redshift of simulation,
whereas the right plot shows the results as a function of $\sigma_\mathrm{d}/\Delta_\mathrm{part}$, where
$\sigma_\mathrm{d}$ is the rms of the initial particle displacements at the initial redshift
and $\Delta_\mathrm{part}$ is
the mean inter-particle distance. Even for the same initial random seeds, the power spectrum amplitudes
vary for different initial redshifts, and the amplitude peaks at a particular initial redshift (see text for details).
\label{fig:pnl_zin_z0}
}
\end{figure}
Fig.~\ref{fig:pnl_evol} shows the ratio of the measured power spectra to the prediction by an Eulerian perturbation theory that is computed on the grid basis assuming the same random realization of the simulation
up to the 2-loop order using Fast Fourier Transform
\citep[\textsc{GridSPT} in][]{2018arXiv180704215T}.
Note that in the computation we have included odd-order contributions to the power spectrum, which should vanish
in the ensemble average sense but are
present in a finite volume (or a given realization) where we
have a limited number of Fourier modes.
We show the evolution as a function of the linear growth factor $D_+(z)$ normalized to unity at present.
Each line starts at the initial growth rate corresponding the initial redshift denoted in the legend.
The figure shows two overall trends.
First, for some results,
the growth in the power spectrum is sharply suppressed compared to the perturbation-theory prediction
soon after the initial redshift,
even if the perturbation theory should be accurate at very early epochs, especially at small wavenumbers.
The higher initial redshift we start the simulation, the greater suppression
the growth has.
The figure also shows that the effect is more important at higher wavenumbers
(note the different plotting ranges in different panels): it is only
a $\sim0.2\%$ level at $k=0.03\,h\,\mathrm{Mpc}^{-1}$ and it reaches to $\sim1\%$ at $k=0.53\,h\,\mathrm{Mpc}^{-1}$ for the simulation with $512^3$ particles (left panel).
Comparing the left and right plots shows that
the sudden drop in the power is about twice larger for the case with $256^3$ particles than that with $512^3$.
The same trend can be seen both for the 2LPT and ZA initial conditions, and thus this effect is not associated with
the accuracy of the Lagrangian perturbation theory that is used to set up
the initial displacement field. Furthermore, although we do not show here, the suppression in the power
persists even if we choose more stringent parameters to control the accuracy of $N$-body simulations (both in the force computation and time stepping).
All these features indicate that this effect is ascribed to the particle discreteness effect.

Secondly, Fig.~\ref{fig:pnl_evol} shows that the ratio gradually decreases with time, after the first sharp decrease,
meaning that the structure grows slowly compared to the perturbation theory
(i.e., the slope of the ratio is negative).
While the slope is almost zero for $k=0.03\,h\,\mathrm{Mpc}^{-1}$, it
becomes increasingly
negative towards larger wavenumbers. The apparent slow growth of the power spectrum in the simulations is at least partly due to a breakdown of the perturbation theory at later
epochs, at $k\gtrsim 0.1 \,h\,\mathrm{Mpc}^{-1}$; that is, the perturbation theory over-predicts the power spectrum amplitudes at such high $k$'s.
More important is the difference in the slope of the curves for simulations with different $z_\mathrm{in}$. It can be seen that the dependence is subtle for 2LPT and quite significant for ZA. In general the slope is steeper for a smaller $z_\mathrm{in}$. This is the transient effect due to the fact that we truncate the perturbative calculation for the displacement field at a finite order.

Now, the question is how these systematic effects due to the initial conditions affect the outputs of $N$-body
simulations at late epochs, $z\lesssim 1.5$, in which we are most interested.
To explicitly study this,
Fig.~\ref{fig:pnl_zin_z0} shows the power spectrum at $z=0$ measured from the simulations at various wavenumbers
as a function of $1+z_\mathrm{in}$ (left panel).
For simplicity, we show only the results using the 2LPT initial conditions here.
Importantly, the figure shows that the systematic effects that we find at the early-time evolution persist even at late times.
For the simulations with $512^3$ particles (upward triangles),
the power spectrum has the greatest amplitudes at all the wavenumbers when the initial redshift $z_\mathrm{in}=59$ is employed.
The peak redshift is shifted toward lower redshift for the simulations with $256^3$ particles (downward triangles).
This peak structure is as a result of competition
of the two systematic effects that we have discussed above.
The suppressed power for the higher $z_\mathrm{in}$ than the peak redshift
is due to the particle discreteness effect,
while the inaccuracy for the lower $z_\mathrm{in}$ is due to the insufficient nonlinear evolution in simulations.
Since the former effect should scale as the typical initial displacement of particles from the regular lattice in units of the lattice interval, we plot in the right panel the same power spectrum as a function of $\sigma_\mathrm{d}(z_\mathrm{in})/\Delta_\mathrm{part}$, where $\sigma_\mathrm{d}$ is the rms of the initial particle displacements and $\Delta_\mathrm{part}$ is the mean inter-particle distance. The peak location in the power spectrum amplitude for the two resolutions are almost identical when plotted as a function of this combination.
To be more quantitative, the peak location appears when the rms displacement is about $20$ to $30\%$ of the inter-particle separation. We thus simply adopt $z_\mathrm{in}$ that gives $\sigma_\mathrm{d}(z_\mathrm{in})/\Delta_\mathrm{part}=0.25$ for the main simulations, \textsc{HR} and \textsc{LR}, presented in this paper; these correspond to $z_\mathrm{in}=59$ and $29$, respectively.
Note that these conclusions hold for neighboring cosmological models around the fiducial {\it Planck} model, but
$z_\mathrm{in}$ could vary significantly depending on cosmological modes.

\subsection{Impact on halo statistics}
\label{subsec:zini_halo}

\begin{figure}[h]
\begin{center}
\includegraphics[width=16cm,angle=0]{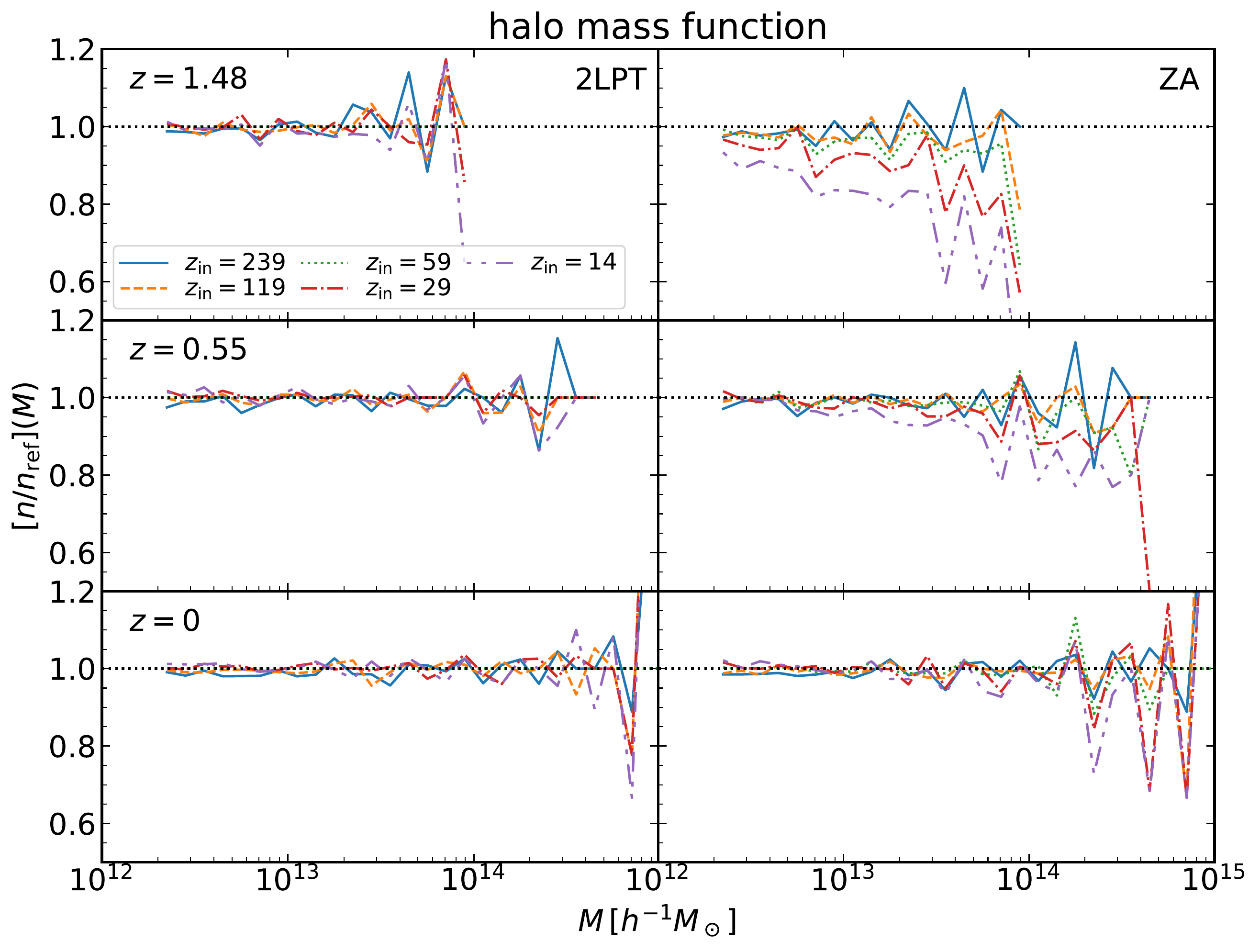}
\end{center}
\caption{Dependence of the halo mass function on the initial redshift.
For our fiducial setting we adopt $z_\mathrm{in}=59$ for the HR simulation (equivalent to $512^3$ particles
for a box of 250~$h^{-1}{\rm Mpc}$), and use 2LPT to set up the initial displacements.
The panels in the left column show the results for 2LPT and different initial redshifts
relative to the fiducial result, while those in the right column are
the results for the ZA initial conditions and different initial redshifts (but with the same resolution). The line styles indicate the initial redshift as shown in the legend.
\label{fig:HMF_zin}
}
\end{figure}
\begin{figure}[h]
\begin{center}
\includegraphics[width=16cm,angle=0]{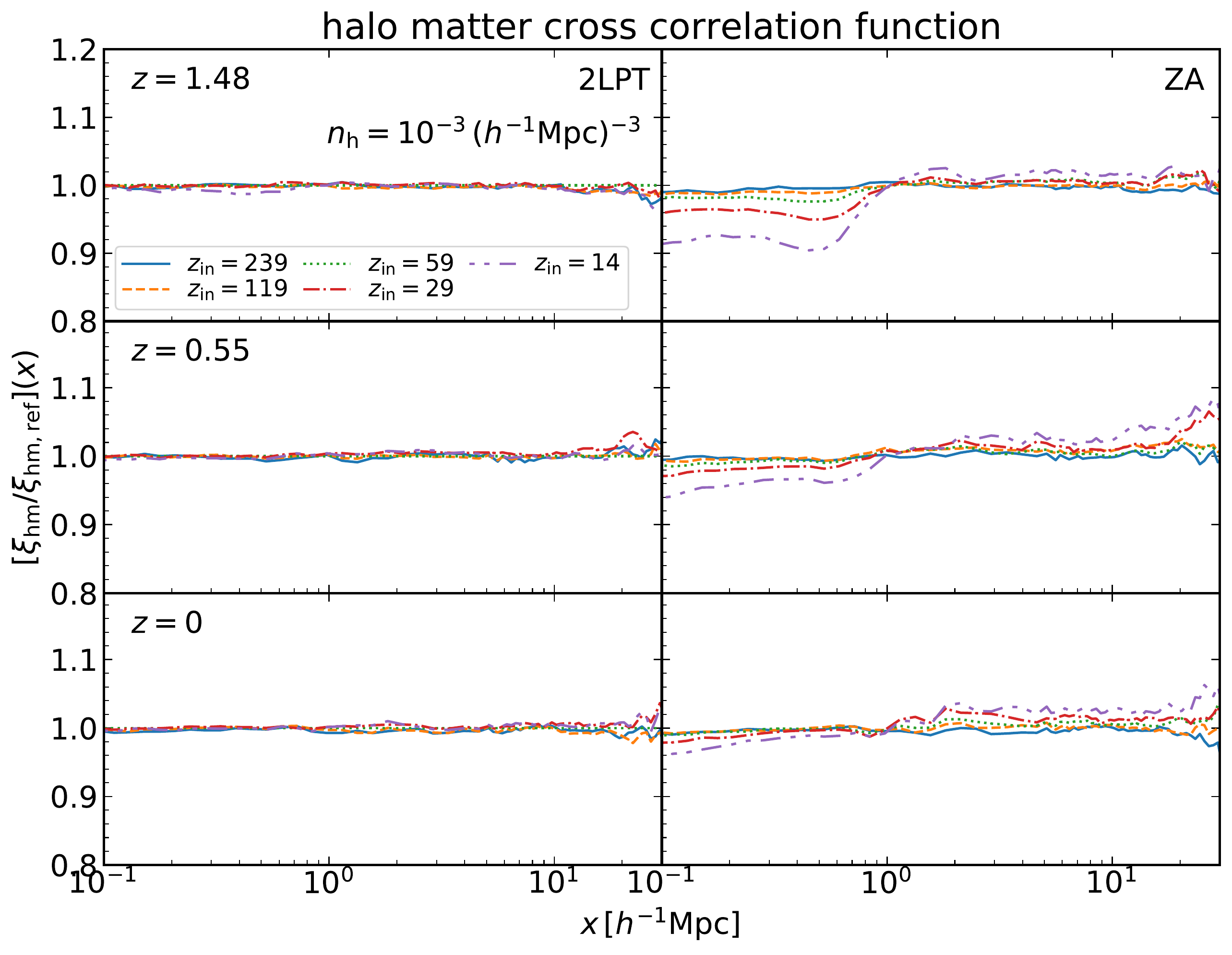}
\end{center}
\caption{Similar to Fig.~\ref{fig:HMF_zin}, but for the halo-matter cross correlation functions.
Here we consider a sample of halos with number density of $10^{-3} \, (h^{-1}{\rm Mpc})^{-3}$.
\label{fig:cross_zin}
}
\end{figure}
\begin{figure}[h]
\begin{center}
\includegraphics[width=16cm,angle=0]{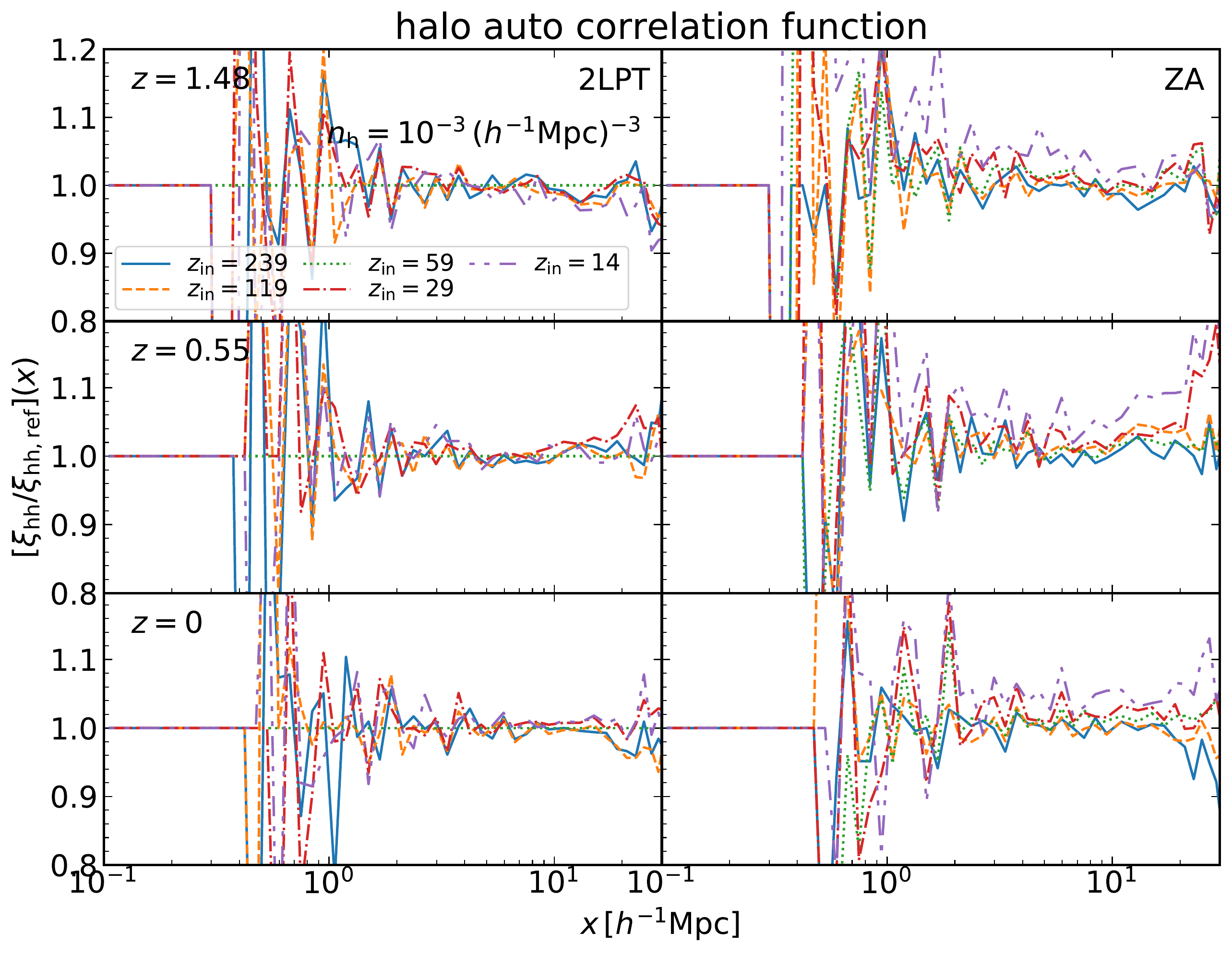}
\end{center}
\caption{Similar to Fig.~\ref{fig:HMF_zin}, but for the halo auto correlation functions.
Here we consider a sample of halos with $10^{-3} \, (h^{-1}{\rm Mpc})^{-3}$ in computation of the
auto correlation functions.
\label{fig:auto_zin}
}
\end{figure}
Fig.~\ref{fig:HMF_zin} shows how the halo mass function measured from simulations at late time vary with initial redshifts.
The halo mass function does not largely vary with different initial redshifts as long as the 2LPT instead of ZA
is used to set up the initial conditions.
Similarly Figs.~\ref{fig:cross_zin} and \ref{fig:auto_zin} show how the halo-matter cross and halo-halo auto
correlation functions vary when using simulations with different initial redshifts. Here we consider halo samples with number density of $10^{-3}\,(h^{-1}\mathrm{Mpc})^{-3}$ instead of the fiducial value of $10^{-4}\,(h^{-1}\mathrm{Mpc})^{-3}$ (i.e., including less massive halos compared to the fiducial analysis) to investigate the case where the systematic effects would be more important. The figures show that the results are well converged for a range of redshifts if the 2LPT initial conditions are used.

\section{Random vs fixed phase simulations}
\label{sec:seed}
For the Gaussian random initial conditions, the amplitude $|\delta_{\mathrm{lin},\mathbf{k}}|$ and the phase
$\theta_\mathbf{k}$ of the initial condition, when expressed as
$\delta_{\mathrm{lin},\mathbf{k}}=|\delta_{\mathrm{lin},\mathbf{k}}|e^{i\theta_\mathbf{k}}$,
follow the Rayleigh or uniform probability distributions, respectively, for each $\mathbf{k}$ mode, where
the width of Rayleigh distribution is set by the initial power spectrum.
A random number seed is often used to generate $|\delta_{\mathrm{lin},\mathbf{k}}|$ and $\theta_\mathbf{k}$
for each $\mathbf{k}$ mode, i.e. generate the initial density field in an $N$-body simulation for a given cosmological model.

When studying dependences of nonlinear structure formation on cosmological models with $N$-body simulations, there
is a choice of whether or not one keeps the same random seeds to set up the initial conditions for different cosmological models.
A possible advantage of using the common random seeds
is to reduce the sample variance contamination when comparing the clustering quantities between different
cosmological models. We expect the advantage for neighboring cosmological models
around a specific target model.
However, for two models that are sufficiently far from each other,
nonlinear evolution, via complex mode coupling, could produce significantly different results so that
the naively expected variance ``cancellation'' is ruined.
Thus there is no guarantee that using the same random seeds leads to converged
estimations of statistical quantities from a limited number of simulation realizations.

\begin{figure}[h!]
\begin{center}
\includegraphics[height=8cm,angle=0]{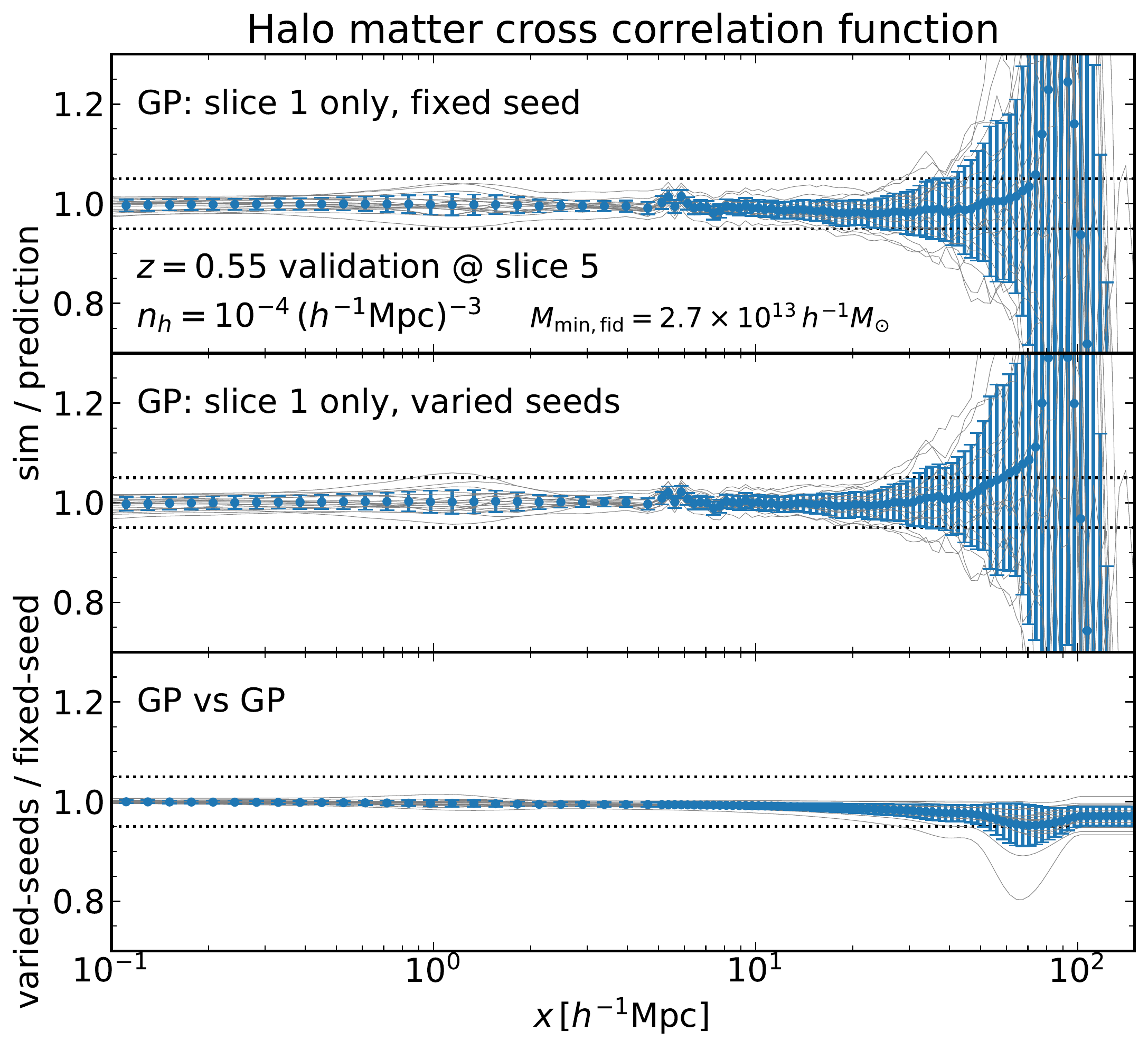}
\end{center}
\caption{Comparison of accuracy of the emulators built by using the 20 different cosmology
simulations with fixed random seeds or varied random seeds in Slice~1. Here we show the fractional
difference of the emulator prediction for the halo-matter cross correlation relative
to the direct measurement from each of 20 cosmological models
in Slice~5, which is a validation sample that is not used in building the emulator considered here.
\label{fig:seed}
}
\end{figure}
To test whether a development of the emulator benefits from the fixed-seed simulations,
we compare the Gaussian process regression results obtained from
two sets of $20$ simulations performed on Slice~$1$; one set contains 20 simulations using the fixed same seed,
while the other set is from 20 simulations with varied random seeds.
To do this we use the \textsc{HR} runs and compare the results for the halo-matter cross correlation function.
For simplicity, we do not repeat the hyperparameter optimization for this purpose and reuse the one optimized for our whole sample. In Fig.~\ref{fig:seed} we show a cross validation test for the GP models with the two sets
(the fixed and varied seeds, in the upper and the middle panel, respectively) at other $20$ models in Slice~$5$. The ratio of the GP models to the simulations in Slice~$5$
is generally very close to unity. The two panels look quite similar. We also show in the lower panel the ratio of the two GP models. The difference is mostly below $1\%$ level except at the very large scales ($\sim 70\,h^{-1}\mathrm{Mpc}$). Note that we use a different prescription to calibrate the clustering statistics on large scales (see Sec.~\ref{subsubsec:largescale}).

From this exercise, we conclude that the difference in the choice of the random number seeds does not
largely affect the accuracy of our emulator. For simplicity and to be more conservative, we employ the method using different random number seeds for each of cosmological models for our main results (development of the emulator).

\section{Effect of massive neutrinos}
\label{sec:neutrino}
Massive neutrinos can impact the growth of cosmological fluctuations, and thus the large-scale structure observables may provide us with a unique opportunity to constrain the sum of
the three mass eigenstates~\citep{Bond80}. While a proper treatment of
massive neutrinos including their impact on nonlinear structure formation
would be important for such cosmological tests~\citep[e.g., ][]{Saito08}, we here restrict ourselves to a cosmological model with
neutrinos of small mass scales as
implied from oscillation experiments, $\sum m_\nu = 0.06\,\mathrm{eV}$, and treat them only at the level of the linear transfer function. More precisely, we compute the transfer function of the total matter fluctuations including massive neutrinos at $z=0$ using \textsc{CAMB}~\citep{camb}, and multiply it with the linear growth factor to scale back to the initial redshift.
When we compute the linear growth factor, we ignore the scale-dependent growth due to the massive neutrinos
and assume the $w$CDM model with the neutrino density included in the matter content
throughout this paper.
After generating the initial particle distribution based on this scaled transfer function,
we consistently ignore massive neutrinos (and radiations/massless neutrinos)
and solve the time evolution of the particle distribution eventually down to $z=0$ in an $N$-body simulation.

\begin{figure}[h]
\begin{center}
\includegraphics[height=8cm,angle=0]{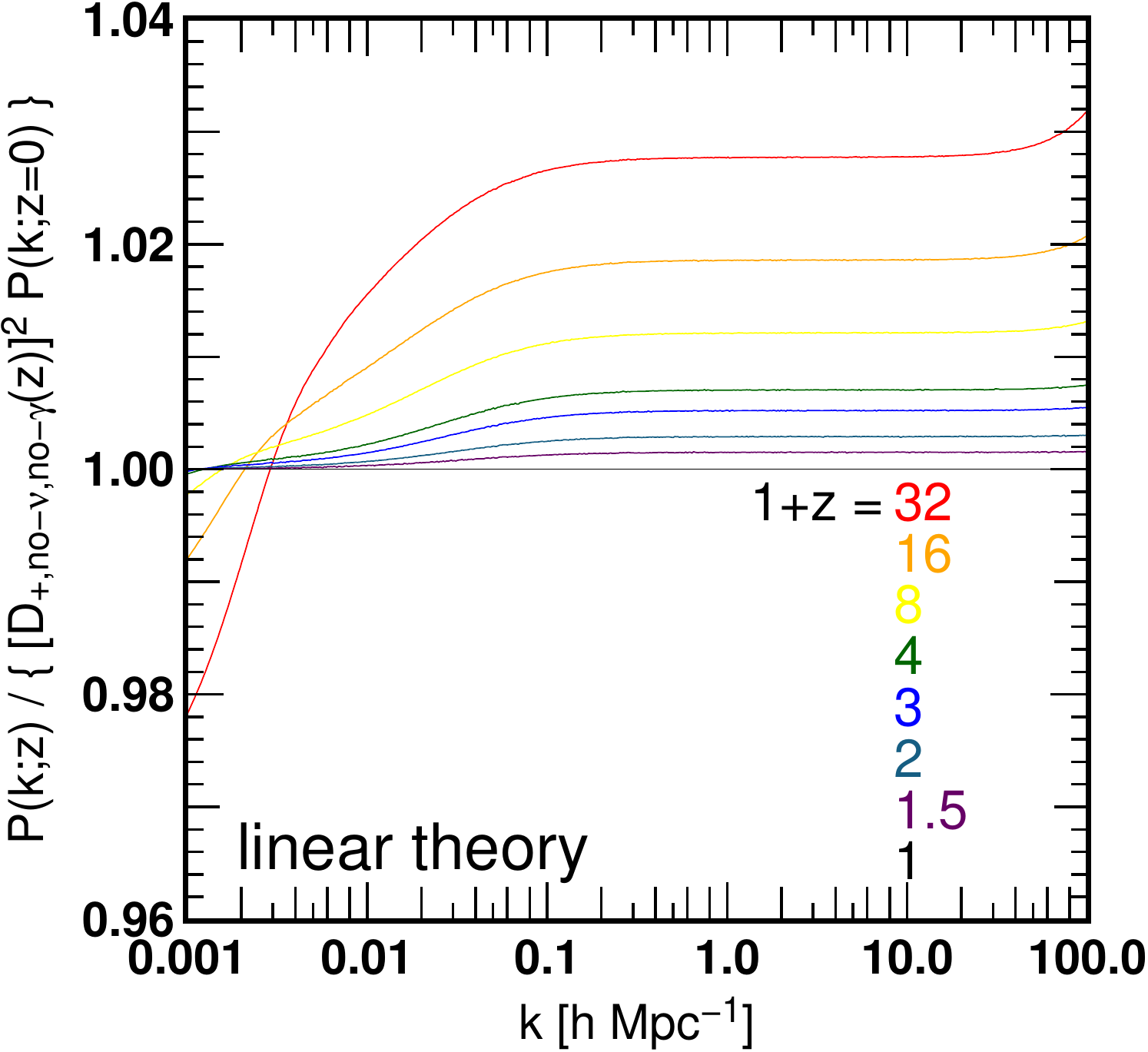}
\end{center}
\caption{Fractional ratio of the linear matter power spectra, where the effect of massive neutrinos
with $0.06\,\mathrm{eV}$ is included by an approximated method (see text for details), relative to the
spectra directly computed with \textsc{CAMB} at different redshifts.
\label{fig:massive_nu}
}
\end{figure}

We give a validation of our treatment using the linear theory.
Fig.~\ref{fig:massive_nu} shows the ratio of the matter power spectrum calculated by two methods. The numerator is the one computed by the linear Boltzmann solver \textsc{CAMB} at the redshift indicated by the figure legend. On the other hand, the denominator is the one computed similarly by \textsc{CAMB} but at $z=0$ and then scaled to the redshift of interest by multiplying the square of the linear growth factor computed without massive neutrinos. This later one is effectively the underlying linear power spectrum for our simulations. The ratio is by definition unity at $z=0$ and grows with increasing the redshift, reaching a $\sim 3\%$ deviation at $z\sim30$. Our target redshifts are rather low, $z\lesssim1.5$, and the deviation stays well below $1\%$ level. While nonlinearity can in principle brings the sizable difference at earlier epochs to later epochs through mode coupling, it would be a higher order effect as the difference is at most a few percent level from the beginning.

\section{Dependence on the halo finder}
\label{app:halo}

\begin{figure}[h]
\begin{center}
\includegraphics[height=8cm,angle=0]{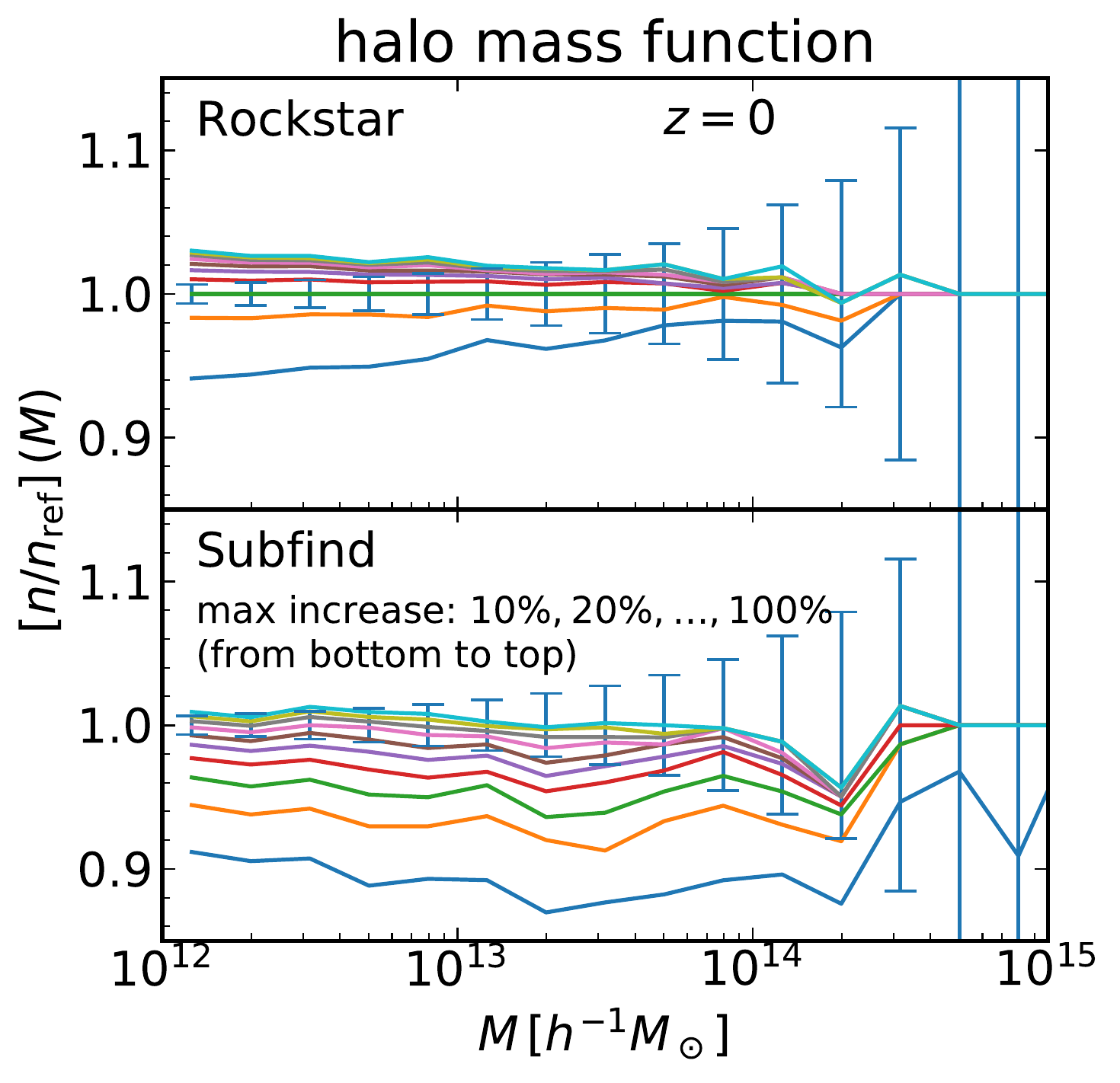}
\end{center}
\caption{Dependence of the halo mass function on the maximum allowed mass increase by including particles not dynamically associated to the halo of interest and on the halo finder (upper: \textsc{Rockster}, lower: \textsc{Subfind}). We normalize the mass function by that for the reference catalog based on the \textsc{Rockstar} finder with the fraction parameter $0.3$. We also show by the error bars the Poisson noise level for the reference halo catalog.
\label{fig:hmf_halodef}
}
\end{figure}

\begin{figure}[h]
\begin{center}
\includegraphics[height=8cm,angle=0]{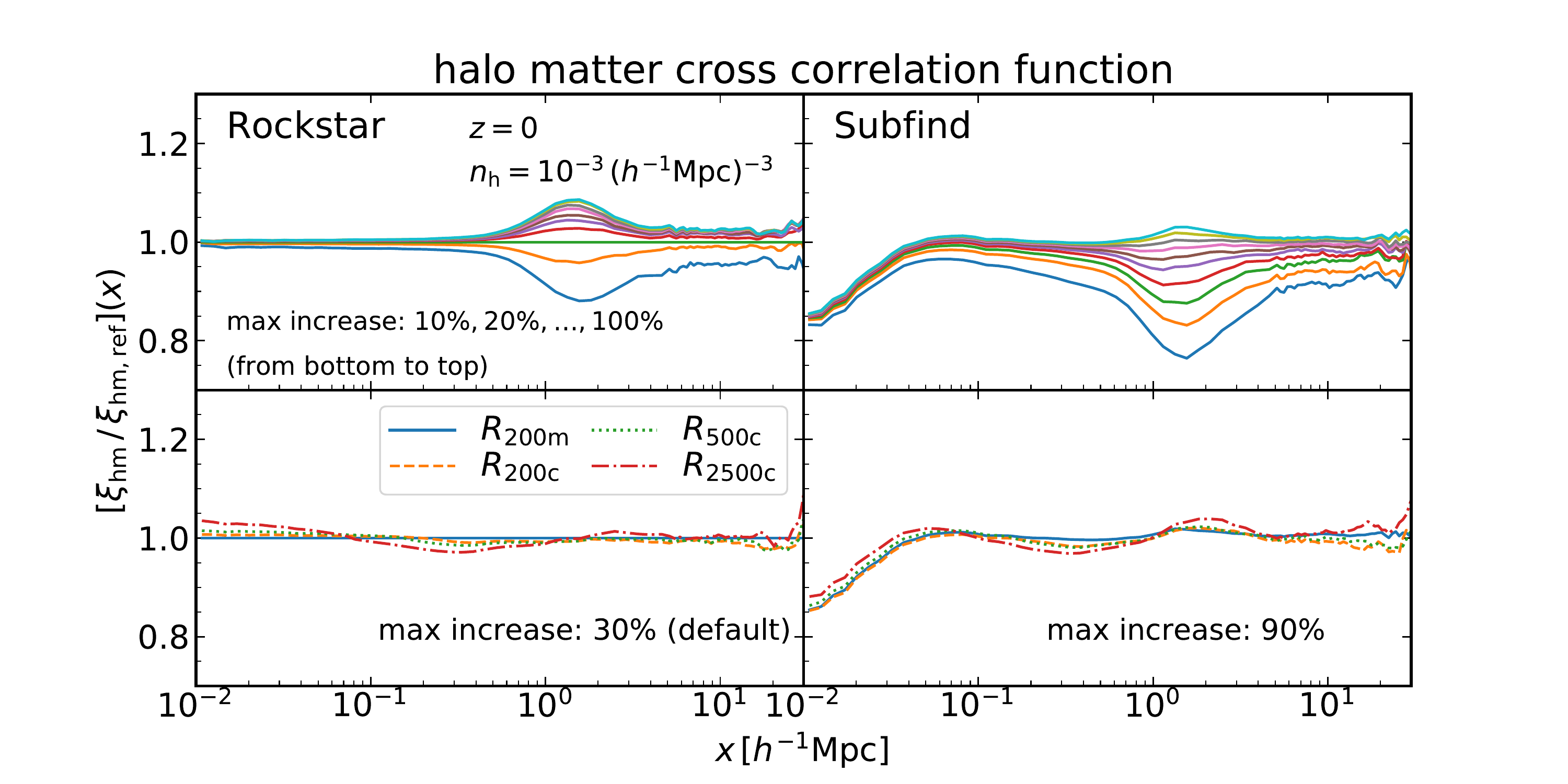}
\end{center}
\caption{Dependence of the halo matter cross correlation function on the maximum allowed mass increase by including particles not dynamically associated to the halo of interest and on the halo finder (left: \textsc{Rockster}, right: \textsc{Subfind}). We normalize the mass function by that for the reference catalog based on the \textsc{Rockstar} finder with the fraction parameter $0.3$.
\label{fig:cross_halodef}
}
\end{figure}

\begin{figure}[h]
\begin{center}
\includegraphics[height=8cm,angle=0]{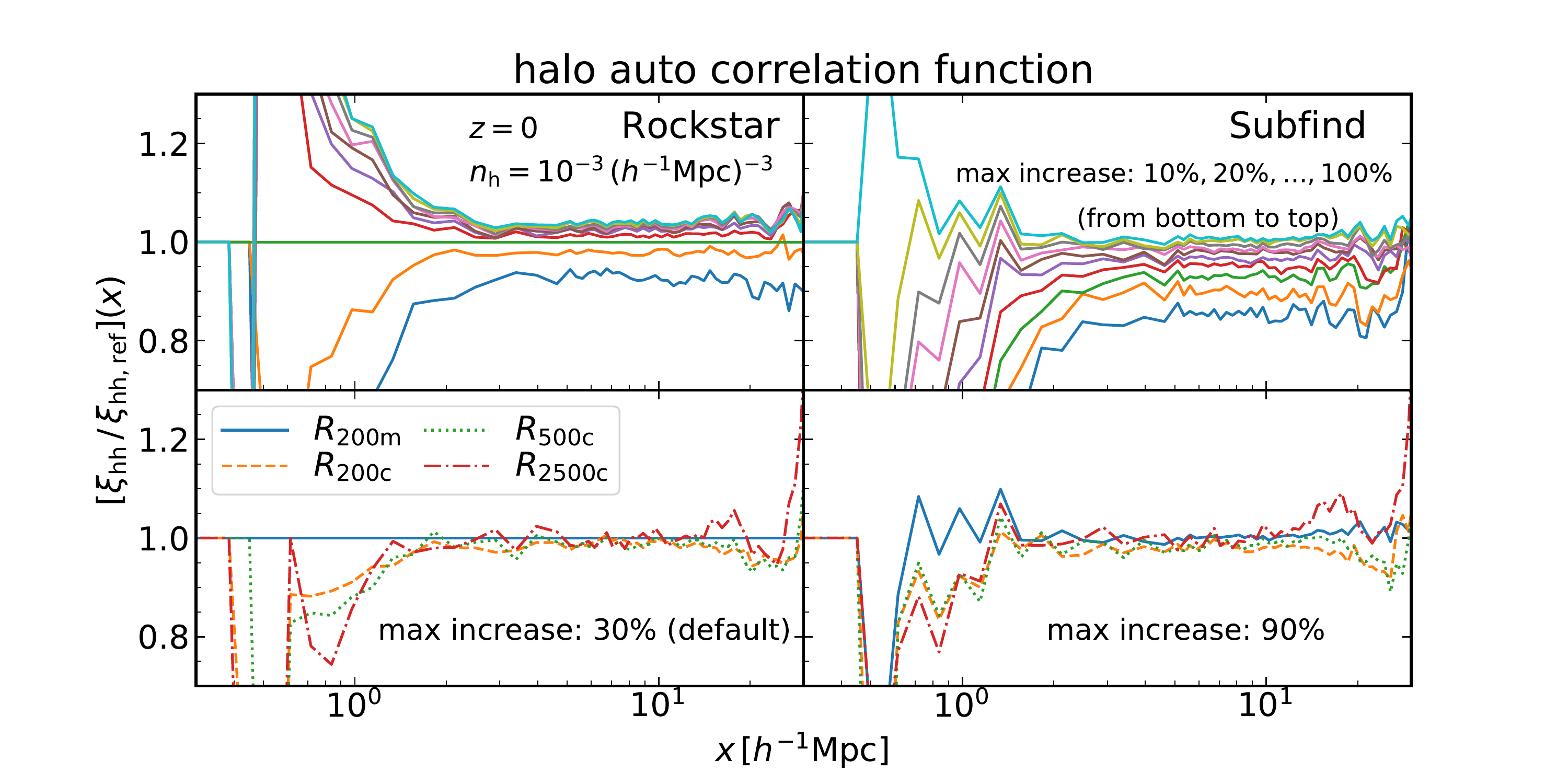}
\end{center}
\caption{Same as Fig.~\ref{fig:cross_halodef}, but for the halo auto correlation function. \label{fig:auto_halodef}
}
\end{figure}

The main target of this paper is to present the statistics of \textit{central} halos after removing substructures. However, the definition of central halos is rather ambiguous. We here examine two things: first, the dependence on the criterion to separate substructures and second the algorithm to identify a list of possible central-halo candidates. As discussed in the main text, we remove a halo if it is within the radius $R_{200\mathrm{m}}$ of a bigger halo in our default setting. In addition, we discard a halo as a fake central halo if the exact spherical mass withing $R_{200\mathrm{m}}$ is larger than that determined by \textsc{Rockstar} by $30\%$. This fraction
is a parameter that can alter the properties of the central halos remaining after these screening procedures. We also examine the \textsc{Subfind} algorithm \citep{subfind} in addition to \textsc{rockstar} employed in the main text.

We first examine in Fig.~\ref{fig:hmf_halodef} the halo mass function after removal of substructures. We show in the upper panel the halo mass function obtained by \textsc{Rockstar} while the lower panel shows that by \textsc{Subfind} at $z=0$. In both cases, we use the test simulation with $512^3$ particles in $(250\,h^{-1}\mathrm{Mpc})$ and divide the results with the reference result based on \textsc{Rockstar} with the maximum allowed mass increase of $30\%$. The upper panel shows that the parameter can change the mass function more severely near the low mass end. The change can reach $\sim 5\%$ level in the worst case with the fraction of $10\%$. When this fraction is larger, the change is only moderate, $\sim 3\%$ maximum at the low mass end. The same exercise is presented for \textsc{Subfind} in the lower panel. Here, the mass increase is based on the change of the mass from the bound mass determined by \textsc{Subfind}. Note that the reference mass function in the denominator is still the one with the \textsc{Rockstar} finder. Interestingly, we can match the \textsc{Subfind} mass function with \textsc{Rockstar} by adjusting the parameter. A maximum mass increase of $\sim 70 \text{--} 80\%$ with \textsc{Subfind} gives almost identical result to \textsc{Rockster} with the default parameter.

We perform similar tests for the correlation functions in Figs.~\ref{fig:auto_halodef} and \ref{fig:auto_halodef}, respectively for the halo matter cross and the halo halo auto correlation function in the upper panels for the two finders. The most significant effect on the cross correlation function appears at $\sim1\,h^{-1}\mathrm{Mpc}$ near the halo boundary. This is natural because our parameter controls the exclusion of substructures near the outskirt of a halo. Another notable thing is that the innermost part (i.e., $x \lesssim 0.05\,h^{-1}\mathrm{Mpc}$) for the \textsc{Subfind} finder. This is because of a different algorithm employed to define the halo center (the center of mass of particles in the core region versus the most bound particle). We, however, do not pay much attention here because the scale is close to the softening length ($0.024\,h^{-1}\mathrm{Mpc}$) and the main target of the emulator is on somewhat larger scale ($\gtrsim\,0.1\,h^{-1}\mathrm{Mpc}$). In case of the auto correlation function, the parameter can alter the overall bias factor. This acts in a sense as an assembly bias effect, as the halo population, recent merger history especially, is altered by changing the criterion to regard a structure at the outskirt of a halo as a substructure or not.

An important message here is again that the correlation functions from the \textsc{Subhalo} groups can be matched to those from \textsc{Rockstar} by adjusting the parameter that separates a subgroup from the central halo. In other words, the dependence of the result to the halo finding algorithm is subdominant and can be absorbed by a parameter that defines the central halos. While one has to beer in mind the possible dependence of the correlation functions on the precise definition of the central halos, we speculate that such a dependence would also be absorbed by HOD parameters when \textit{galaxy} clustering is considered. A structure discarded as a substructure in one algorithm but treated as a central halo in another might be accounted by populating a satellite galaxy there. We postpone further explicit tests of this point to a future investigation.

Finally, we examine the dependence of the cross and auto correlation functions on the outer boundary of the central halos that defines substructures. We consider $R_\mathrm{200c}$, $R_\mathrm{500c}$ and $R_\mathrm{2500c}$, in addition to the default choice of $R_\mathrm{200\mathrm{m}}$. Here, the numbers in the subscript before $\mathrm{c}$ indicate that the interior density is that number times the critical density. We show in the lower panels of Figs.~\ref{fig:cross_halodef} and \ref{fig:auto_halodef} the results normalized by the default setting. One can see trends similar to the one when we vary the maximum mass to increase parameter with a smaller variation. The typical change is within the target accuracy of this study (i.e., a few percent) except the case with a rather extreme choice of $R_\mathrm{2500c}$.

From the analyses presented in this Appendix, we conclude that the clustering properties of halos predicted by \textsc{Halo Modules} are robust against halo finding algorithms, but the parameter that determines the central-satellite separation can affect the results.

\section{Dependence of the performance on redshift and halo number density}
\label{sec:extra_dependence}
We have focused on how the emulator perform against simulations at $z=0.55$ and for halos with number density $10^{-4}\,(h^{-1} \mathrm{Mpc})^{-3}$ in the main text. We summarize our findings at different redshifts and for different halo samples in this appendix.

\begin{figure}[h]
\begin{center}
\includegraphics[width=8cm,angle=0]{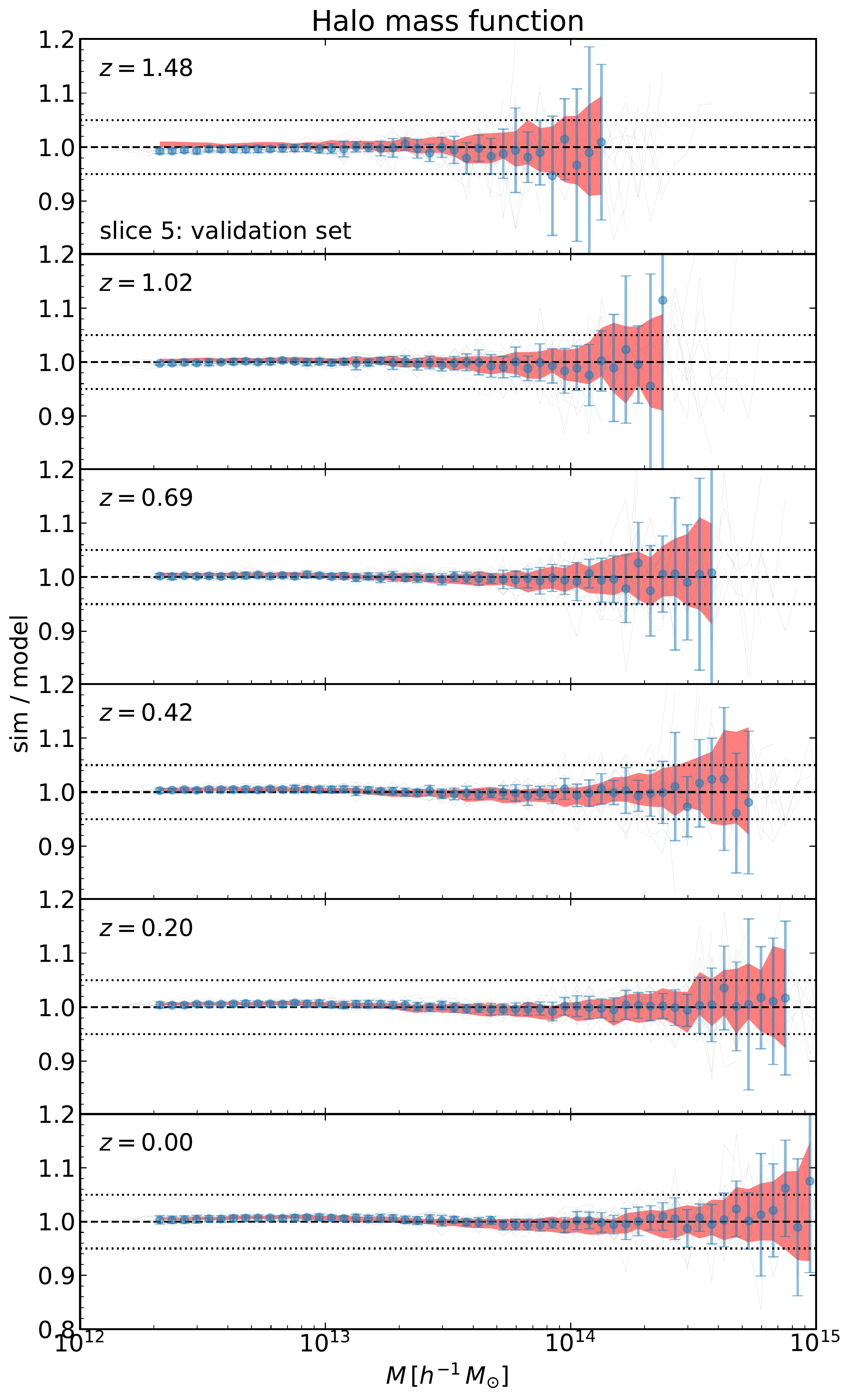}
\end{center}
\caption{Redshift dependence of the cross validation test for the halo mass function.\label{fig:zdep_hmf}
}
\end{figure}
We show first a cross validation study of the halo mass function
in Fig.~\ref{fig:zdep_hmf}. Each panel corresponds to the rightmost lower panel of Fig.~\ref{fig:HMF_model} where the accuracy of the emulator is tested for the $20$ cosmologies in Slice~$5$, which are not used in the Gaussian Process regression. We can confirm that the scatter among thin solid lines (i.e., different cosmologies) scales similarly to the width of the red shades (the scatter of the halo mass function for the different realizations of the fiducial {\it Planck} cosmology). Thus we conclude that the modeling is reasonably accurate given the uncertainties in the simulation data.

\begin{figure}[h]
\begin{center}
\includegraphics[width=18cm,angle=0]{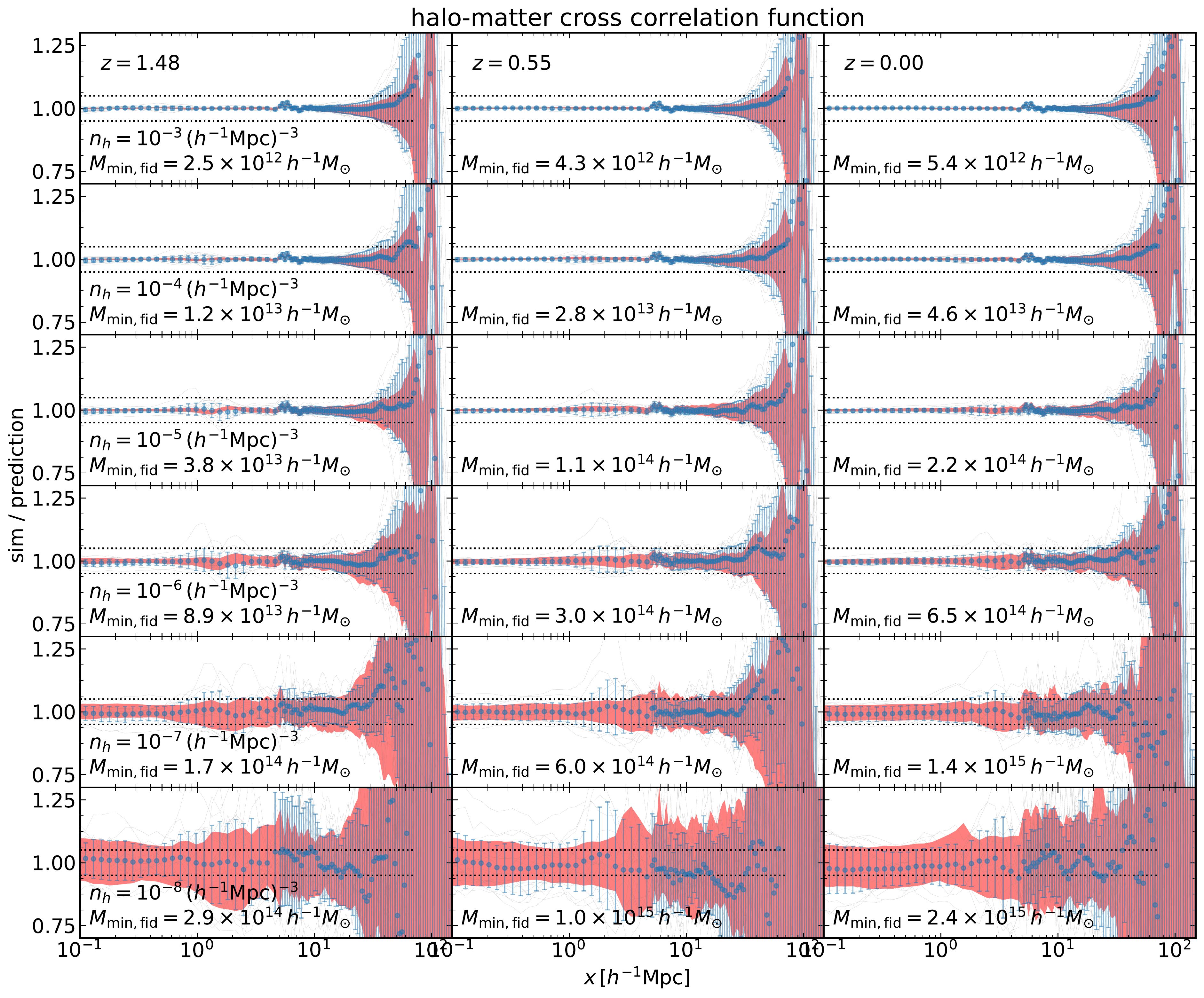}
\end{center}
\caption{Redshift and number density dependence of the cross validation test for the halo matter cross correlation function.\label{fig:dep_xihm}
}
\end{figure}
Next, we show in Fig.~\ref{fig:dep_xihm} a similar cross validation study for the halo matter cross correlation function at various number densities (rows) and at different redshifts (columns). The overall trend is that the accuracy is degraded as decreasing the number density reflecting the bigger uncertainties in the simulation data due to a larger noise. On the other hand, no clear dependence on redshift is found.
\begin{figure}[h]
\begin{center}
\includegraphics[width=18cm,angle=0]{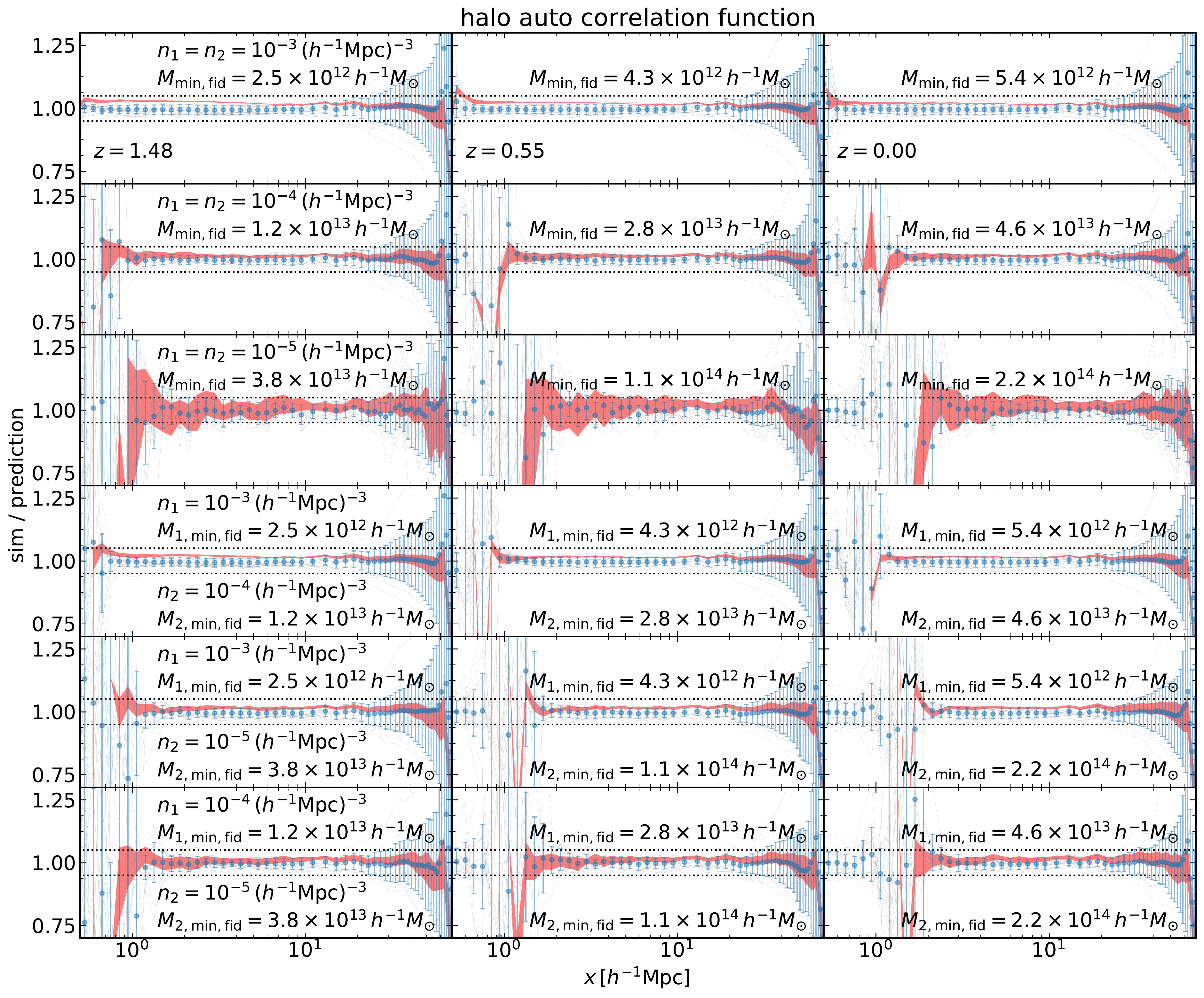}
\end{center}
\caption{Redshift and number density dependence of the cross validation test for the halo auto correlation function.\label{fig:dep_xihh}
}
\end{figure}
The halo-halo correlation function is tested in Fig.~\ref{fig:dep_xihh}. This time, the upper three rows are for the auto correlation function of the same halo samples and the remaining three rows are for two halo samples with different number densities as indicated in the figure legend. A similar trend, a bigger scatter for low density samples can be found.
\begin{figure}[h]
\begin{center}
\includegraphics[width=18cm,angle=0]{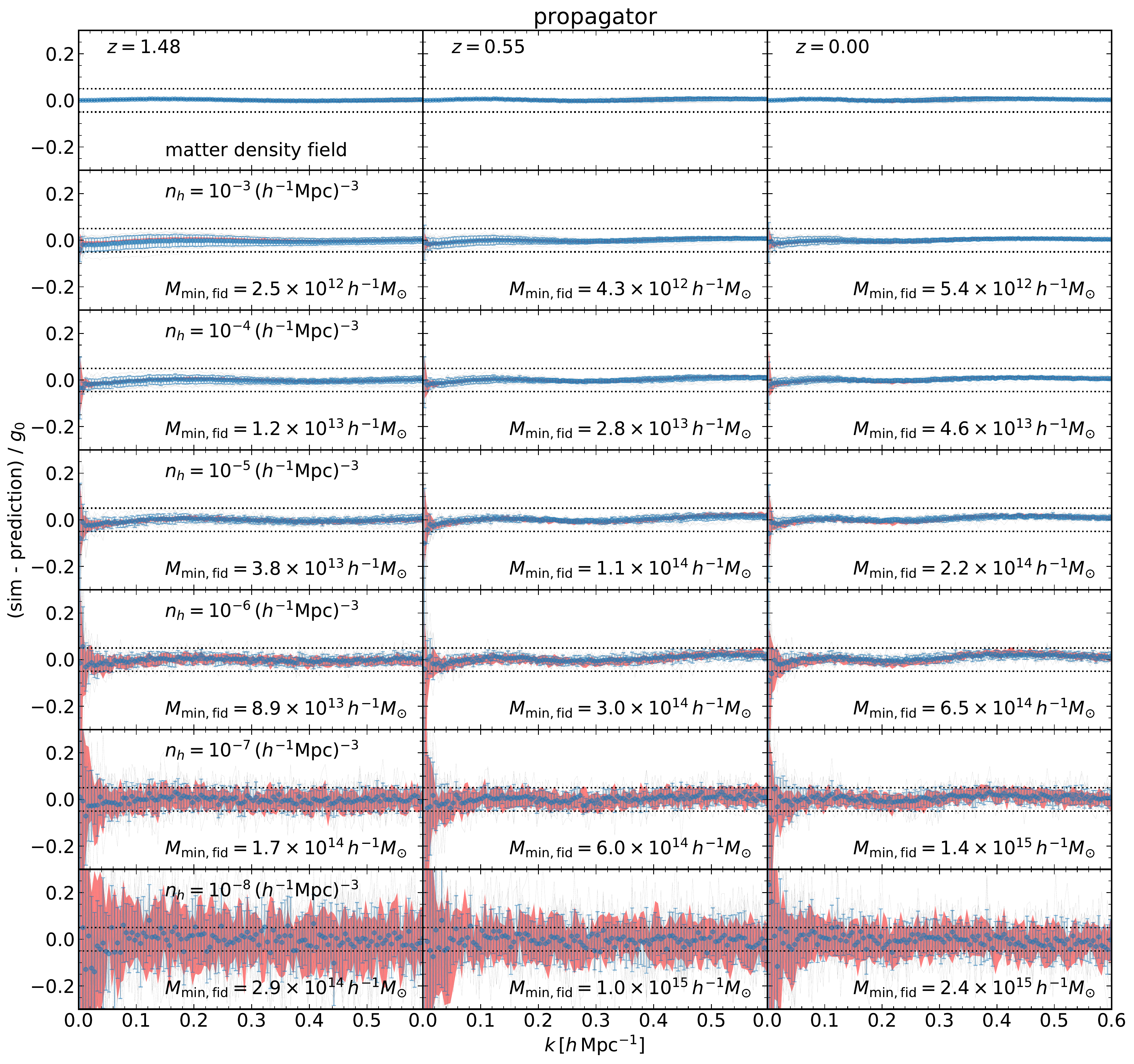}
\end{center}
\caption{Halo mass and redshift dependence of performance of the propagator module. We also show the matter propagator in the top row. \label{fig:dep_gamma}
}
\end{figure}
Finally, the propagators are shown in Fig.~\ref{fig:dep_gamma}.
These plots are a practical guide on the accuracy of the current code depending on the halo number densities (or halo masses) in actual use.

\section{Halo Occupation Distribution model}
\label{sec:HOD}

As we stressed in the main text, one can insert one's own module to \textsc{Dark Emulator}
to model how halos
are related to galaxies under consideration. Here, as a working example, we show a halo occupation distribution
model \citep[hereafter HOD;][]{1998ApJ...494....1J,peacock:2000qy,seljak:2000uq,scoccimarro:2001fj,2005ApJ...633..791Z}
to predict the abundance, the clustering and the lensing signal of galaxies. In particular we employ the model
in \citet{2015ApJ...806....2M}. A module based on this HOD prescription is provided in \textsc{Dark Emulator} as a default package.

We adopt the HOD model with an explicit split of the halo occupation into central and satellite galaxies:
\begin{equation}
\avrg{N}_M=\avrg{N_{\rm c}}_M+\avrg{N_{\rm s}}_M,
\label{eq:hod}
\end{equation}
where the average $\avrg{\dots}_M$ is taken for halos with mass $M$.
The mean HOD for central galaxies is given by
\begin{equation}
\avrg{N_{\rm c}}_M=f_{\rm inc}(M)\frac{1}{2}
\left[1+{\rm erf}\left(\frac{\log M-\log M_{\rm min}}{\sigma_{\log M}}\right)\right],
\label{eq:hodc}
\end{equation}
where ${\rm erf}(x)$ is the error function and
$M_{\rm min}$ and $\sigma_{\log M}$ are model parameters. The function $f_{\rm inc}(M)$ accounts for
potential incompleteness of central galaxies which models a possibility that a central galaxy in some halos can be missed
due to an imperfect selection effect or galaxies under consideration do not
necessarily occupy halos at the center even for sufficiently massive halos \citep[e.g.][]{masaki13}.
We assume a log-linear functional form given by
\begin{equation}
f_{\rm inc}(M)={\rm max}\left[0,{\rm min}[1,1+\alpha_{\rm inc}(\log M-\log M_{\rm inc})]\right],
\label{eq:finc}
\end{equation}
where $\alpha_{\rm inc}$ and $M_{\rm inc}$ are model parameters.
The mean HOD for satellite galaxies is given by
\begin{equation}
\avrg{N_{\rm s}}_M\equiv \avrg{N_{\rm c}}_M\lambda_{\rm s}(M)=\avrg{N_{\rm c}}_M\left[
\frac{M-\kappa M_{\rm min}}{M_1}
\right]^\alpha,
\label{eq:hods}
\end{equation}
where $\kappa$, $M_1$, and $\alpha$ are model parameters, and we have defined the notation $\lambda_{\rm s}(M)\equiv
[(M-\kappa M_{\rm min})/M_1]^\alpha$ for convenience in the following discussion. We assume that the distribution of central galaxies, $N_\mathrm{c}$, follows the Bernoulli distribution (i.e., can take only zero or one) with mean $\avrg{N_{\mathrm{c}}}_M$. On the other hand, we populate satellite galaxies to a halo only when a central galaxy exists. The conditional distribution of $N_\mathrm{s}$ in a halo with mass $M$ that has a central galaxy is given by the Poisson distribution with mean $\lambda_\mathrm{s}(M)$.
Our HOD model is fully specified by 7 parameters: $\{M_{\rm min},\sigma_{\log M},\alpha_{\rm inc},M_{\rm inc},\kappa,M_1,\alpha\}$.

Once the HOD model is specified, the mean number density of galaxies in a sample is computed as
\begin{equation}
\bar{n}_{\rm g}=\int\!\!\mathrm{d}M~\frac{\mathrm{d}n}{\mathrm{d}M}\left[
\avrg{N_{\rm c}}_M+\avrg{N_{\rm s}}_M
\right].
\end{equation}
Here the halo mass function $\mathrm{d}n/\mathrm{d}M$ is given by \textsc{Dark Emulator} for
a given cosmological model.

Now we consider the galaxy-galaxy weak lensing that measures the excess surface mass density profile around
lensing galaxies. The galaxy-galaxy weak lensing profile for lens galaxies at redshift $z_l$
can be expressed in terms of the galaxy-matter
power spectrum \citep[e.g.][]{2018ApJ...854..120M} as
\begin{equation}
\Delta\Sigma(R; z_l)=\bar{\rho}_{\rm m}\int_0^\infty\!\frac{k\mathrm{d}k}{2\pi}~ P_{\rm gm}(k; z_l)\,J_2(kR),
\label{eq:DeltaSigma_power}
\end{equation}
where $\bar{\rho}_{\rm m}$ is the present-day mean matter density and $J_2(x)$ is the second-order Bessel function.
Under the HOD model described above, we further make some assumptions on the location of central and satellite galaxies within the host halo. First, we allow some fraction ($f_\mathrm{off}$) of central galaxies to be located off from the true halo center following a Gaussian distribution with width $R_\mathrm{off}$.
In this case, this off-centering effect can be expressed in terms of a kernel
in Fourier space \citep{2011PhRvD..83b3008O,hikage:2013kx}
\begin{equation}
{\cal H}_{\rm off}(k; M, f_{\rm off}, R_{\rm off})
\equiv 1-f_{\rm off}+f_{\rm off}\exp\left[-\frac{1}{2}(k R_{\rm off})^2\right].
\end{equation}
We then introduce a function $\tilde{u}_{\rm s}(k; M)$ for the normalized radial profile of satellite galaxies again in Fourier space.
For this, we assume an NFW profile which is specified by a given model of the halo matter-concentration
relation, denoted as $c(M,z)$, for halos of a given mass.
We adopt the fitting formula in \citet{2015ApJ...799..108D} to compute $c(M,z)$
for a given cosmological model, and $c(M,z)$ is not a free parameter in our default setting.
Alternatively, we provide an option to distribute satellite galaxies following the mass distribution given by $\propto 1+\xi_\mathrm{hm}(x)$, which can be computed by one of our \textsc{Halo Modules}.
With these assumptions, the galaxy-matter cross-power spectrum is given as
\begin{eqnarray}
P_{\rm gm}(k)&=&
\frac{1}{\bar{n}_{\rm g}}
\int\!\!\mathrm{d}M\frac{\mathrm{d}n}{\mathrm{d}M}
\left[\avrg{N_{\rm c}}_M{\cal H}_{\rm off}(k; M, f_{\rm off}, R_{\rm off})+\avrg{N_{\rm s}}_M\tilde{u}_{\rm s}(k;M)
\right]P_{\rm hm}(k;M),
\label{eq:Pkgm}
\end{eqnarray}
In Eqs.~(\ref{eq:DeltaSigma_power}) and (\ref{eq:Pkgm}), $\mathrm{d}n/\mathrm{d}M$
and $P_{\rm hm}(k; M)$ are given by \textsc{Dark Emulator}, and the radial profiles of off-centering
central galaxies and satellite galaxies are modeled by the parameters: $\{c(M,z), f_{\rm off}, R_{\rm off}\}$.
Thus one can compute the galaxy-galaxy weak lensing profile
from \textsc{Dark Emulator}
for a given cosmological model, e.g., once 9 parameters for connecting halos to galaxies are specified:
7 parameters for HOD plus 2 profile parameters.

Next we consider the projected correlation function of galaxies, $w_{\rm gg}(R)$, which is defined
in terms of the three-dimensional correlation function as
\begin{equation}
w_{\rm gg}(R; z) = 2\int_0^{\pi_{\rm max}}\mathrm{d}\pi~ \xi_{\rm gg}\!\left(\sqrt{R^2+\pi^2};z\right).
\label{eq:wpgg}
\end{equation}
Here
\begin{equation}
\xi_{\rm gg}(x)\equiv \int_0^\infty\!\!\frac{k^2\mathrm{dk}}{2\pi^2}~P_{\rm gg}(k)j_0(kx),
\label{eq:xigg}
\end{equation}
where $j_0(x)$ is the zero-th order spherical Bessel function and $P_{\rm gg}(k)$ is the galaxy auto-power spectrum.
Using the above model ingredients, we can express the galaxy auto-power spectrum as
\begin{eqnarray}
P_{\rm gg}(k)&=&\frac{1}{\bar{n}_{\rm g}^2}\int\!\mathrm{d}M\frac{\mathrm{d}n}{\mathrm{d}M}
\left[2
\avrg{N_{\rm s}}_M~{\cal H}_{\rm off}(k;M)
 \tilde{u}_{\rm s}(k;M)+
 \avrg{N_{\rm c}}_M\lambda_{\rm s}(M)^2
 \tilde{u}_{\rm s}(k;M)^2\right]\nonumber\\
&&+\frac{1}{\bar{n}_{\rm g}^2}
\left[
\int\!\mathrm{d}M\frac{\mathrm{d}n}{\mathrm{d}M}\left\{
\avrg{N_{\rm c}}_M{\cal H}_{\rm off}(k;M)+\avrg{N_{\rm s}}_M~ \tilde{u}_{\rm s}(k;M)
\frac{}{}\right\}
\right]\nonumber\\
&&\hspace{2em}\times
\left[
\int\!\mathrm{d}M'\frac{\mathrm{d}n}{\mathrm{d}M'}\left\{
\avrg{N_{\rm c}}_{M'}{\cal H}_{\rm off}(k;M')+\avrg{N_{\rm s}}_{M'}~ \tilde{u}_{\rm s}(k;M')
\frac{}{}\right\}
\right]P_{\rm hh}(k;M,M').
\end{eqnarray}
Note that in deriving the equation above, we have used the relations
\begin{eqnarray}
&&\avrg{N_{\rm c}N_{\rm s}}_M=\avrg{N_{\rm s}}_M, \nonumber\\
&&\avrg{N_{\rm s}(N_{\rm s}-1)}_M=\avrg{N_{\rm c}}_M\lambda_{\rm s}(M)^2,
\end{eqnarray}
which follow from our assumptions on the distribution of $N_\mathrm{c}$ and $N_\mathrm{s}$ described above.
Once again, the projected correlation function of galaxies can be computed for the same sat of model parameters
as those of the galaxy-galaxy weak lensing profile (i.e., 9 parameters for the halo-galaxy connection).

\end{document}